\documentclass{emulateapj}

\usepackage{graphicx}
\usepackage{txfonts}
\usepackage{lscape}
\usepackage{epsfig}
\usepackage{times}
\usepackage{graphicx}
\usepackage{natbib}
\bibpunct{(}{)}{,}{a}{}{}

\shorttitle{Galaxy population properties in a z$\sim$2 cluster}
\shortauthors{Strazzullo et al.}

\begin{document}


\title{Galaxy evolution in overdense environments at high redshift:
  \\ passive early-type galaxies in a cluster at redshift 2}


\author{V. Strazzullo$^1$, R. Gobat$^1$, E. Daddi$^1$, M. Onodera$^2$,
  M. Carollo$^2$, M. Dickinson$^3$,
  A. Renzini$^4$,\\ N. Arimoto$^{5,6}$, A. Cimatti$^7$,
  A. Finoguenov$^{8}$, R.-R. Chary$^{9}$ }

\altaffiltext{1}{CEA Saclay, Orme des Merisiers, 91191 Gif sur Yvette, France}
\altaffiltext{2}{Institute for Astronomy, ETH Z\"urich Wolfgang-Pauli-strasse 27, CH-8093 Z\"urich, Switzerland} 
\altaffiltext{3}{National Optical Astronomy Observatory, P.O. Box 26732, Tucson, AZ 85726, USA}
\altaffiltext{4}{INAF, Osservatorio Astronomico di Padova Vicolo dell’Osservatorio 5, 35122 Padova, Italy}
\altaffiltext{5}{Subaru Telescope, 650 North A’ohoku Place, Hilo, HI 96720, USA}
\altaffiltext{6}{Graduate University for Advanced Studies, 2-21-1 Osawa, Mitaka, Tokyo 181-8588, Japan}
\altaffiltext{7}{Universit\`a di Bologna, Dipartimento di Astronomia, via Ranzani 1, 40127 Bologna, Italy}
\altaffiltext{8}{Department of Physics, University of Helsinki, Gustaf H\"allstr\"omin katu 2a, FI-00014 Helsinki, Finland}
\altaffiltext{9}{Division of Physics, Mathematics and Astronomy, California Institute of Technology, Pasadena, CA 91125, USA}




\begin{abstract}

We present a study of galaxy populations in the central region of the
IRAC-selected, X-ray detected galaxy cluster Cl J1449+0856 at $z=2$.
Based on a sample of spectroscopic and photometric cluster members, we
investigate stellar populations and morphological structure of cluster
galaxies over an area of $\sim$0.7Mpc$^2$ around the cluster core. The
cluster stands out as a clear overdensity both in redshift space, and
in the spatial distribution of galaxies close to the center of the
extended X-ray emission. The cluster core region ($r<200$~kpc) shows a
clearly enhanced passive fraction with respect to field
levels. However, together with a population of massive passive
galaxies mostly with early-type morphologies, it also hosts massive
actively star-forming, often highly dust-reddened sources. Close to
the cluster center, a multi-component system of passive and
star-forming galaxies could be the future BCG still assembling. We
observe a clear correlation between passive stellar populations and an
early-type morphology, in agreement with field studies at similar
redshift. Passive early-type galaxies in this clusters are typically a
factor 2-3 smaller than similarly massive early-types at $z\sim0$, but
also on average larger by a factor $\sim2$ than their field analogs at
$z\sim2$, lending support to recent claims of an accelerated
structural evolution in high-redshift dense environments. These
results point towards the early formation of a population of massive
galaxies, already evolved both in their structure and stellar
populations, coexisting with still-actively forming massive galaxies
in the central regions of young clusters 10 billion years ago.

\end{abstract}





\section{Introduction}
\label{intro}

\setcounter{footnote}{0} 

In the nearby Universe and at least up to z$\sim$1, overdense
environments, and specifically galaxy cluster cores, are invariably
found to preferentially host galaxy populations dominated by massive,
passive early-type galaxies \citep[e.g., among many others,
][]{dressler1980,postman2005,baldry2006,vanderwel2007,patel2009,rosati2009,peng2010,wetzel2012}. The
way these galaxies are formed and evolve is a longstanding matter of
debate, with different pieces of their formation history, and in
particular peculiarities with respect to field galaxies, being put
together thanks to high-redshift observations
\citep[e.g.,][]{vandokkum2007,gobat2008,mei2009,rettura2010} as well
as fossil-record studies \citep[e.g.,][]{thomas2005,thomas2010}. Most
studies agree on the early formation of a population of massive
cluster early-types, with their stars formed at high redshift
($z\sim2$ or beyond), and their mass largely assembled before $z\sim1$
\citep[e.g.,][]{depropris2007,lidman2008,andreon2008,mancone2010,strazzullo2010}.

While detailed studies of cluster galaxy populations are relatively
common up to redshift one, they become increasingly rarer at higher
redshifts, and in particular beyond $z\sim1.5$, due to observational
challenges both in reliably identifying clusters, and in accurately
determining the properties of their galaxies. Nonetheless, the $z>1.5$
range is a crucial epoch to study massive cluster galaxies close to
their main formation epoch.  Indeed, recent observations of
$z\gtrsim1.5$ clusters started to show that massive galaxy populations
are often in a still active formation stage, even in the cluster core
\citep[e.g.,
][]{tran2010,hilton2010,hayashi2010,hayashi2011,santos2011,fassbender2011}.

From the theoretical point of view, current models, while invoking an
early formation for the stars ending up in massive early-type galaxies
today, maintain their hierarchical nature in predicting the late
assembly of their stellar mass from smaller, mostly passive
progenitors \citep[e.g.,][]{delucia2006,johansson2012}.  The relevance
of such merging events, as well as of other processes (e.g., AGN or
stellar feedback), possibly affecting both the star formation history
and the galaxy structure in the evolutionary path of these systems,
may be probed by the (albeit biased and complicated) comparison of
cluster galaxy samples at different redshifts.  

Reaching the cosmic
epochs when massive cluster galaxies are still forming is thus
fundamental in order to directly observe the formation of the bulk of
the stars, the way stellar mass is assembled, and their morphological
evolution, that together lead to the massive early-types dominating
cluster cores at later times.

Ideally, this kind of investigation is carried out in clusters which
are not pre-selected based on the characteristics of their galaxy
populations, but rather based on their mass or overdensity. At this
redshift, and with current facilities, X-ray selection becomes very
challenging for the identification of moderately massive systems
representative of the progenitors of typical lower-redshift
clusters. On the other hand, ``IRAC-selected'' clusters identified
based on overdensities of stellar mass-limited galaxy samples
\citep{eisenhardt2008,papovich2008}, ideally with a-posteriori
detection of a (generally faint) X-ray emission, offer a suitable
alternative for the identification of clusters beyond $z\sim1.5$.

In this work, we study galaxy populations in the IRAC-selected and
X-ray detected cluster Cl J1449+0856 at $z=2$ \citep[RA = 14h49m14s,
  Dec = $8^{\circ}56'21''$, ][2013]{gobat2011}. This is among the most
distant spectroscopically confirmed galaxy clusters discovered so far,
and the most distant with a detected X-ray emission. The first
spectroscopic investigation with VLT/VIMOS and FORS2 spectroscopy on a
wide field around the cluster, found a peak in the redshift
distribution of star-forming galaxies at $z\sim 2.07$
\citep{gobat2011}. 

Subsequent follow-up on the cluster center with
slitless HST/WFC3 spectroscopy, unveiled a much stronger peak in the
redshift distribution at z=2, which is the most prominent peak in the
area of the galaxy overdensity, and contains about 20 spectroscopic
cluster members to date, including spectroscopically confirmed massive
passive red galaxies in the cluster core (Gobat et
al. 2013). Cl~J1449+0856 is thus now spectroscopically confirmed to be
at z=2.  

Given the massive use of photometric redshifts required to
carry out this work, we are not able to distinguish galaxies at z=2
from galaxies at z=2.07, and we thus retain sources belonging to the
z=2.07 structure in our sample of candidate members (unless a
spectroscopic redshift is available). On the other hand, as discussed
in detail in the Gobat et al. (2013) companion paper, the z=2.07
redshift peak seems to be associated to a large scale, less prominent,
diffuse structure, which does not significantly contribute to the
overdensity in the central cluster region studied here, and is likely
not to significantly affect most of the results presented in this
work.

A wide multi-wavelength coverage, and high-resolution restframe
optical imaging, allow us to study in detail fundamental properties of
cluster galaxies already 10 billion years ago.  In particular, we
focus in this paper on the identification of a population of passive
candidate members, and on their structural properties. The trademark
cluster-core galaxies up to $z\sim1$, massive galaxies with low star
formation are in fact expected to be significantly rarer by redshift
two, approaching the epoch where not only they are still forming many
of their stars, but also when velocity dispersion in the cluster core
is still low enough that merging-driven mass assembly can play an
important role. Although conclusive evidence is still lacking, the
early merging of already gas-poor galaxies in an overdense environment
might also affect the structural properties of the resulting massive,
passive systems, producing galaxies which are structurally more
evolved than their field counterparts. While statistics are still poor,
Cl J1449+0856 offers one of the very rare chances of studying
passive early-types very close to their  formation epoch,
together with still actively forming galaxies, in an already
relatively evolved cluster core.

Throughout this work we assume $\Omega_{M}$=0.3,
$\Omega_{\Lambda}$=0.7, H$_{0}$=70~km~s$^{-1}$~Mpc$^{-1}$, and a
\citet{salpeter1955} IMF. Magnitudes and colors are in the AB system.

~\
\section{Data and sample selection}

\subsection{Catalogs and derived quantities}
\subsubsection{Photometry}

We use photometry measured on imaging in the U,V (VLT/FORS), B,R,i,z
(Subaru/Suprime-Cam), Y,J,H,K$_{s}$ (Subaru/MOIRCS, plus additional
VLT/ISAAC data for J and K$_{s}$), F140W (HST/WFC3), and 3.6,4.5$\mu$m
(Spitzer/IRAC) bands. Sources were extracted with SExtractor
\citep{sextractor} on the F140W image, and photometry was measured in
two ways, producing two independent multi--wavelength catalogs. One
catalog is based on aperture photometry (1.5'') measured with
SExtractor, corrected for the different resolution of the images by
using aperture corrections estimated on each image from the growth
curve of point--like sources. The other catalog is based on photometry
measured on each image by fitting with Galfit \citep{galfit1,galfit2}
Gaussian profiles, convolved with the image PSF, at the position of
the F140W sources. While the two approches yield broadly consistent
measurements, in most cases aperture photometry will be more
accurate\footnote{Note that this is not a general statement: it
    is true, as results from simulations, given the specific approach
    and settings we use. A further indication of the generally more
    accurate aperture photometry in our case, comes from photo-z
    performance as discussed in section \ref{sec:zphots}.}. On the
other hand, and especially in crowded fields typical of a cluster
core, blending may be a severe issue and the second approach offers a
way to deal with it. In order to take this into account, in the
  following we will use both catalogs as described in detail in
  section \ref{sec:zphots}.

In the following we select a sample with $m_{140}<25.7$ (corresponding
to the $10\sigma$ limit in a 1'' aperture), within an area of
$\sim$3.3 square arcminutes uniformly covered in the WFC3/F140W
imaging. This catalog contains $\sim$370 objects. Seven point-like
sources (in the F140W image) down to $m_{140}\sim22$ were removed; at
fainter magnitudes, we further removed $\sim30$ sources which may be
stars based on their BzK colors \citep{daddi2004}. The inclusion or
removal of these sources has no impact on the results of this
work. The galaxy sample we use in the following thus contains
$\sim$330 galaxies down to $m_{140}=25.7$.

\subsubsection{Photometric redshifts}
\label{sec:zphots}

From the 13-bands photometry, we estimated photometric redshifts
(photo--zs) with EAZY, using the standard set of templates
\citep{eazy,brammer2011,whitaker2011}. Before the actual photo--z
estimation, we determined possible offsets in the photometry in
different bands by iteratively comparing the best-fit vs. measured
photometry at fixed (spectroscopic) redshift \citep[e.g.,
][]{capak2007,ilbert2009} for a sample of $\sim110$ sources with
redshifts measured from WFC3 grism spectroscopy over the whole
WFC3-covered field\footnote{Out of these, 94 are in the uniformly
  covered WFC3 image area used in the following analysis of galaxy
  populations, the remaining being located in the external part of the
  WFC3 image.} (Gobat et al. 2013). Systematic offsets between
measured photometry and model SED for spectroscopic sources can be
attributed to different causes including zero-point and/or aperture
correction errors, as well as model uncertainties\footnote{For our
  catalogs and with our settings, offsets were in most cases
  $<$10-20\%, except for bands with large uncertainties in the
  determination of the photometric zero-point -- as often suggested
  also by comparison of SEDs of stars with stellar templates -- and/or
  in the instrument response function.}. In the following, we use
photometry corrected for these systematic offsets; we note that the
offsets determined for the two independent photometric catalogs
(SExtractor- and Galfit-based) are generally consistent within
$\sim$10\%.

\begin{figure}[t]
\includegraphics[width=.485\textwidth,bb=92 408 362 718,clip]{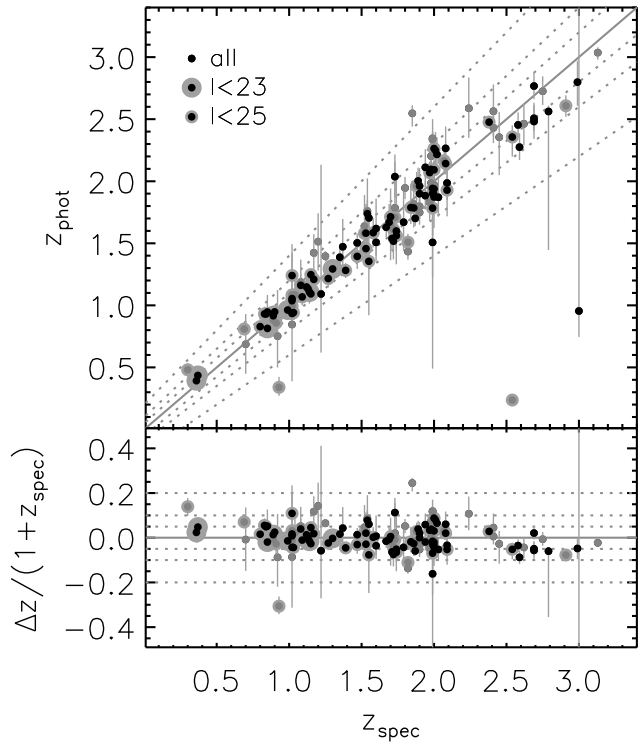}
\caption{The comparison of photometric and spectroscopic redshifts for
  the available spectroscopic sample. All available spectroscopic
  redshifts are shown, but secure and less reliable $z_{spec}$
  determinations are shown as black and gray dots, respectively. The
  lower panel shows $\frac{z_{phot}-z_{spec}}{1+z_{spec}}$
  vs. $z_{spec}$.  In all panels, the solid line traces
  $z_{phot}=z_{spec}$ and the dotted lines show a $\Delta z/(1+z)$ of
  5, 10, and $20\%$. Thin and thick gray circles around symbols
  highlight sources brighter than I$<$25 and 23 mag, respectively, as
  indicated in the legend. \label{fig:zz}}
\end{figure}

The interquartile redshift range of the spectroscopic sample is
z=[1.1-2] (with redshifts up to z$\sim$3), and its interquartile
$m_{140}$ range is $m_{140}$=[22.7-24.3] (reaching up to
$m_{140}\sim25.5$). Therefore, while compared to the whole sample of
$m_{140}<25.7$ sources the spectroscopic sample is obviously typically
brighter, it can still be considered generally representative of
sources in the magnitude and redshift ranges that are the focus of
this study.

Photo--zs were determined for both (SExtractor- and Galfit-based)
catalogs, and for the SExtractor catalog they were also determined
excluding the IRAC bands. The comparison of these three different
photo-z determinations was used to improve the photo--z accuracy,
selecting for each source the best estimate to be used, as follows. By
comparison with the spectroscopic catalog, we find that photo--zs
determined from aperture photometry including IRAC bands show the
lowest scatter, but also a significant fraction of outliers (almost
10\%). For sources in the magnitude range typical of our spectroscopic
sample, outliers may be due to fatal errors and/or degeneracies (e.g.,
a double-peaked photo--z probability distribution), with these
conditions worsened by systematic offsets in the photometry (of some
bands) due to bad resolution (and thus contamination by neighboring
sources).  In such cases, the Galfit-based photometry -- and thus 
the derived photo-z --  may be more accurate than SExtractor aperture 
photometry, as discussed above. 

We identified sources potentially affected by neighbor's
contamination by selecting in the K$_{s}$ and 3.6$\mu$m bands objects
which were falling in the circle contaning 99\% of the flux of a
different source\footnote{This approach is quite conservative in that
  it does not make assumptions on the relative flux of the neighbors
  and includes potential contamination also from much fainter
  sources.}. In the area and magnitude range that we use in this work,
about 10\% of our sample is classified as potentially contaminated at
the K-band resolution ($\sim$0.65''), and $\sim$50\% at the IRAC
resolution ($\sim$2''). 

For uncontaminated sources we use photo-zs from the SExtractor catalog,  
as well as for sources where only the IRAC photometry is flagged as
potentially contaminated and the SExtractor-based photo-zs 
with and without IRAC bands are consistent. For the remaining sources 
(about 20\% of the sample) photos-zs from the Galfit catalog are used.

Applying this approach and comparing to the spectroscopic sample, and
in spite of the comprehensibly smaller fraction (30\%) of potentially
contaminated sources in the spectroscopic vs. the full sample (50\%),
we find that we can significantly reduce the fraction of catastrophic
outliers, while retaining the higher photo-z accuracy obtained with
aperture photometry for the majority of the sample. The final photo-z
catalog we use in the following has a scatter \citep[as estimated with
  the normalized median absolute deviation (NMAD),][]{hoaglin1983} of
5.7\% in $\Delta z/(1+z)$, and 3\% catastrophic outliers ($|\Delta
z|/(1+z)>0.2$) (at least half of these have a ``less reliable''
  spectroscopic redshift determination, see Figure \ref{fig:zz}).
The comparison of photometric redshifts with the available
spectroscopic sample is shown in Figure \ref{fig:zz}. For comparison
with other studies, we note that, thanks to the WFC3 slitless
spectroscopy, the spectroscopic sample we use is considerably deeper
than those generally obtained from ground-based spectroscopy, with the
median I band magnitude of our spectroscopic sample $\sim$25, and
almost 80\% of the sample fainter than I=24. For instance, considering
only spectroscopic sources brighter than I=25 (23) the NMAD scatter of
$\Delta z/(1+z)$ would be $<$4.5\% ($<2.5\%$).

\subsubsection{Stellar masses}

Stellar masses were determined with FAST \citep{fast} on the 13-bands
U to 4.5$\mu$m photometry, using \citet[][hereafter, BC03]{bc03}
delayed exponentially declining star formation histories (SFHs,
$\psi(t) \propto \frac{t}{\tau^2} exp(-t/\tau)$) with 0.01$<\tau<$10
Gyr, solar metallicities, Salpeter IMF, and the \citet{calzetti2000}
reddening law with E(B-V) up to 1 mag. Stellar masses were
independently derived for both the SExtractor and the Galfit catalogs:
for sources where contamination was expected to significantly
  affect the aperture-based SED, as discussed concerning photo-zs in
  section \ref{sec:zphots}, stellar masses from the Galfit catalog
  were used.

Masses from the SExtractor catalog were corrected to "total'' masses
using the ratio between AUTO and aperture flux in the detection image
(F140W). For the $m_{140}<25.7$ sample, more than 90\% of the objects
have a correction lower than $50\%$, and only $\sim$3\% of sources
have correction factors $\gtrsim2$. While this approach corrects for
the bulk of the flux loss, we note that it still relies on
approximating the total flux with FLUX\_AUTO, and it is based on just
one band thus neglects any color gradient within the galaxy. In this
respect, we note that the systematic offset between stellar masses
from the Galfit catalog and the two SExtractor catalogs (with and
without IRAC, corrected to total masses) is less than 0.1dex, with a
scatter of up to $\sim$0.3dex.  Leaving the metallicity free, and
using exponentially declining SFHs rather than delayed exponentials,
would introduce no systematics for the overall sample and a further
scatter of less than 0.1dex in stellar mass (which is small compared
to the scatter estimated above at fixed metallicity and SFH). With
respect to the choice of the SFH, we note that it has been shown how
other forms of SFHs might be more appropriate for star-forming
galaxies at high redshift \citep{maraston2010,papovich2011}. For what
directly concerns this work, rising or constant (possibly truncated)
SFHs would change the stellar masses of our targets negligibly, and in
any case well within the estimated errors. Other parameters which may
be more affected by the SFH choice (like, notably, star formation
rates and ages of star-forming galaxies) are not used in this paper.

On the other hand, using \citet[][hereafter, M05]{maraston2005} rather
than BC03 models would produce stellar masses systematically smaller
(overall for the sample of interest) by a factor $\sim$0.15dex, with a
scatter of $\sim$0.15dex; this is discussed in detail below, where
relevant. After accounting for the overall 0.15dex systematic offset,
the stellar mass determinations with BC03 and M05 models (with
metallicity either fixed to Z$_{\odot}$ or allowed to vary within a
factor 2 from a Z$_{\odot}$) are consistent within a factor of at most
2 for $>$90\% of the sample of interest.  

Finally, the median formal error on stellar masses for our sample of
candidate members is 0.2-0.3 dex (or 0.1-0.15 dex for
m$_{140}<24.5$). In summary, we thus estimate a typical accuracy of
about a factor 2 for the stellar mass determination for our targets.

\subsection{Morphological analysis}
\label{sec:morph}

A rough indication of galaxy structure (early type vs. late type),
effective radius and ellipticity, was obtained by modeling of the 2D
surface brightness distribution, carried out with Galfit (version 3)
assuming a single Sersic profile for each F140W-detected source. The
modeling was performed on the WFC3 F140W image, which has the best
resolution in our data set, and probes the restframe optical light
(approximately B band) at the cluster redshift. We used a PSF built
from the data by using median stacking of 6 high S/N stars in the
field. The background was measured and subtracted locally over the
whole image, and was fixed to zero in the fit. The whole image (and
thus each source) was fitted multiple times, split in overlapping
cutouts of different sizes\footnote{For each source, the final
  estimate of each parameter was calculated as the median among all
  fits with residuals of $<$25\% on at least 90\% of the S/N$>$10
  pixels. Overall, 3/4 (1/2) of the sources in the magnitude range of
  interest ($m_{140}\lesssim 24.5$, see below) has results derived
  from the median of at least 5 ($>$10) different fits,
  respectively.}, modeling simultaneously all sources in the cutout.

In order to estimate the reliability of the results as a function of
magnitude and profile type, specifically for the image and fitting
settings that we used, we carried out simulations of the fitting
procedure by adding synthetic sources in blank parts of the
image. Sources with a range of magnitudes, $n_{Sersic}$, radius,
ellipticity and position angle were added and then fitted with Galfit
using the same procedure used for real objects. These simulations
provide an estimate of the reliability of our analysis in somewhat
``optimistic conditions", since they assume that sources are
relatively isolated, regular Sersic profiles, convolved with the same
PSF that we use for the actual fitting. The input flux is recovered
within $10\%$ down to $m_{140}=24.5$ (corresponding to $\sim$30
times the noise in a 1'' aperture). At this magnitude, these
simulations suggest that the error on the semi-major axis is $\sim$10,
15, 25\% for profiles with $n_{Sersic}<$1.5, 1.5-3, $>$3,
respectively, while the error on the Sersic index is between
$\sim25\%$ for late-type profiles and $\sim30\%$ for early-type
profiles. In addition, as it is known from previous work
\citep[e.g.,][]{trujillo2006a,sargent2007,pannella2009b} at faint
magnitudes the Sersic index of high-Sersic profiles tends to be
underestimated; with our settings, at $m_{140}=24.5$, these
simulations find a median offset in the Sersic index of about $-10\%$
for early-type profiles (disk-like profiles are unaffected). This
systematic underestimation is negligible down to $m_{140}\sim24$,
where errors on semi-major axis are $<$5\%, 20\%, and errors on
$n_{Sersic}$ $<$15\%, 20\%, for low- and high-Sersic profiles,
respectively. All parameters, for all kinds of profiles, are retrieved
at better than 10\% down to $m_{140}\sim23$ (corresponding to
S/N$\sim$100 in a 1'' aperture).

Based on these simulations we set $m_{140}=24.5$ as the limiting
magnitude where we consider our surface brightness modelling
reliable. Beside the general $m_{140}<25.7$ sample, we will thus
consider this $m_{140}<24.5$ sample for the
morphological analysis.  Galaxy sizes quoted in the following are the
circularised half-light radii, calculated from the Galfit-based
parameters as the effective semi-major axis times the square root of
the axis ratio.

\subsection{The candidate member sample}
\label{sec:candidatemembersample}

A spectroscopic redshift is measured for about one fourth of the
$m_{140}<25.7$ sample, and for $\sim$45\% of the $m_{140}<24.5$
sample. Based on the available spectroscopy, and otherwise on the
photo-z analysis, we thus determine which sources in our sample are
(candidate) cluster members.

We select as spectroscopic members all sources with spectroscopic
redshift $1.97<z<2.01$. All other spectroscopic sources are considered
interlopers. From the spectroscopic sample of Gobat et al. (2013), we
retain 14 cluster members in the area studied in this work. One source
in the $1.97<z<2.01$ range, close to the edge of the studied area, is
not in our catalog being close to a bright star. From its spectrum it
is classified as a star forming source, and given its position more
than 600~kpc away from the cluster center it would not alter (if
anything it would rather reinforce) the conclusions of this
study. Furthermore, an X-ray detected AGN classified as a cluster
member in Gobat et al. (2013) is not included in our member
sample. This source appears to have a very close neighbor at $\sim
0.5''$ distance (with undetermined redshift), producing a likely
significant contamination to its observed SED. Assuming that both
sources belong to the cluster, and that the measured photometry is not
significantly altered by the emission of the AGN, the total observed
SED produced by both sources would suggest a dusty star-forming
stellar population. The morphology of the AGN host appears very
compact, essentially unresolved (although its magnitude is close to
the limit where we can carry out a reliable morphological
analysis). Because of such considerable uncertainties in determining
the properties of this source, we will not consider it in the
following analysis of galaxies in Cl J1449+0856.

For all sources without an available spectroscopic redshift, we rely
on photo-zs. In determining membership by photo-zs, we decided not to
purely rely on the redshift probability distribution estimated by EAZY
for each object. We adopted instead an hybrid approach, taking into
account also the ``empirical'' photo-z uncertainty as independently
determined by comparison with the spectroscopic sample. By analysing
photo-z results for spectroscopic sources (for the full spectroscopic
sample, not just cluster members) as a function of photo-z probability
distribution and quality of the best fit, we first estimated the
likely reliability of a photo-z given its odds and $\chi^2$ calculated
by EAZY (within our catalog). Based on such comparison, we deemed as
``most reliable" photo-zs with odds$>$98\% and a $\chi^2$ within three
times the median $\chi^2$ in our catalog. To the other extreme, we
defined as ``unreliable" photo-zs with odds lower than 95\%, or with
$\chi^2$ worse than six times the median $\chi^2$.

For the purpose of membership assignment, the full sample was then
split in three classes of objects, identified as interlopers,
``likely'' candidate members, and ``possible'' candidate members
(which are lower-priority candidates, mainly due to a more uncertain
photo-z determination), as follows. Based on the observed scatter of
$\Delta$z/(1+z) (Sec.~\ref{sec:zphots}), and considering the estimated
low fraction of catastrophic failures, all galaxies with a photo-z
beyond 2$\sigma$ from the cluster redshift are considered interlopers,
except those within 3.5$\sigma$ and with a photo-z deemed unreliable,
which are retained as possible members. All galaxies with a photo-z
within 2$\sigma$ from the cluster redshift are considered as possible
members, unless they have a highly reliable photo-z and an integrated
photo-z probability distribution within 1$\sigma$ of the cluster
redshift\footnote{That is, $P(z_{cl}) =
  \int_{2-3\times0.057}^{2+3\times0.057} p(z) dz$.}  $P(z_{cl})>$30\%,
which classifies them as likely members. All galaxies with a photo-z
within 1$\sigma$ of the cluster redshift are selected as likely
members.

Likely members make up $\sim$15\% of the full sample, and a further
15\% is made of possible members. The remaining $\sim70$\% of the full
sample is rejected as foreground ($\sim$55\%) or background
($\sim$15\%) interlopers.

For comparison with other studies, we note that $\sim$3/4 of galaxies
selected as likely members have $P(z_{cl})>$30\% (with
$P(z_{cl})>$20\% for almost all likely members), and in turn $>$80\%
of galaxies with $P(z_{cl})>$30\% are classified as likely members,
making the likely-member selection roughly similar to that used by
e.g., \citet{tran2010,papovich2012}. On the other hand, possible
members have $P(z_{cl})$ as low as $\sim1\%$, with an interquartile
range of $P(z_{cl})$ of about 6-22\%. The inclusion of possible
members thus brings our selection closer to more conservative
$P(z_{cl})$-based criteria adopted by other studies as
e.g., \citet{tanaka2012,raichoor2012}. In fact, in this work galaxies
with a $P(z_{cl})$ larger than 10\% are essentially always included in
the candidate sample (as either likely or possible members), unless
they are spectroscopic interlopers.  Overall, the selection we use is
thus quite conservative, which is reflected in the considerations
about completeness and contamination discussed below.

\begin{figure}
\includegraphics[width=.5\textwidth, bb= 61 350 363 730, clip]{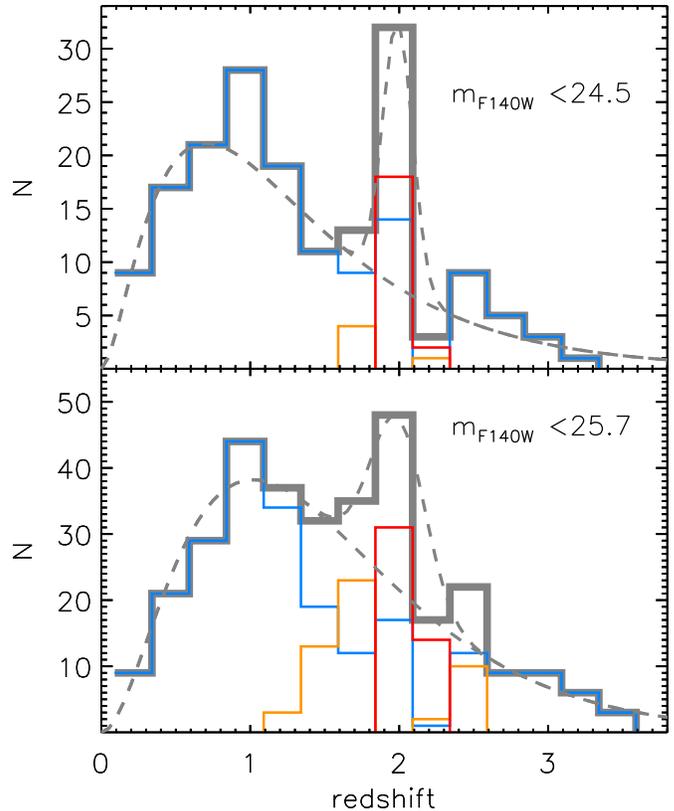}
\caption{The redshift distribution in the cluster field. Gray lines
  show the distribution (of photometric redshifts, or spectroscopic
  where available) of all sources in the target area down to
  $m_{140}=$24.5 and 25.7 (upper and lower panel, respectively). The
  contributions of galaxies identified as interlopers, "possible'' and
  ``likely" cluster members, as described in the text, are shown as
  blue, orange and red lines, respectively. Dashed gray lines in both
  panels show a modeling of the redshift distribution in this field,
  with and without the cluster contribution (see
  text). \label{fig:zdist}}
\end{figure}

\subsubsection{Completeness and contamination of the candidate member sample}
\label{sec:complcont}

Figure \ref{fig:zdist} shows the redshift distribution in the
considered area, down to $m_{140}=$24.5 and 25.7, highlighting the
different contribution of interlopers vs. possible and likely
candidate cluster members. The presence of the cluster on the
underlying field redshift distribution is clearly visible, even in
this distribution largely based on photo-zs. By comparison with
comoving number densities measured in wide fields
\cite[e.g.][]{muzzin2013} the comoving number density of massive
($>10^{10}$M$_{\odot}$) galaxies at $1.7<z<2.1$ (the photo-z range
most affected by cluster members) in the area within 150~kpc from the
cluster center, is about 30 times larger than in the general field
($\Delta log(n/Mpc^{3}) = 1.49\pm0.15$~dex)\footnote{We verified that
  the comoving number density measured in our field in the $1<z<1.5$
  range, which is not affected by the cluster, is in excellent
  agreement with what measured by \citet{muzzin2013}. Also note that,
  even in the $1.7<z<2.1$ range affected by the cluster, the comoving
  number density of massive galaxies beyond 200~kpc from the cluster
  center is only a factor $\sim$4 times the density in the general
  field ($\Delta log(n/Mpc^{3}) = 0.58\pm0.15$~dex).}.

We note that, by comparison with the spectroscopic sample, our
selection of candidate cluster members is highly complete (all the 19
spectroscopic cluster members in the area probed by WFC3 grism data
would be classified as likely candidate members based on their photo--zs).
On the other hand, as a
tradeoff for completeness, the sample of candidate members is
significantly contaminated by interlopers. 

As a rough estimate of such contamination, we find that by comparison
with the spectroscopic sample about half of the sample of ``likely''
cluster members would be interlopers. We note that this fraction
increases significantly, possibly to $\sim$80\%, for ``possible''
members, for the obvious reason that, by selection, this sample is
made of objects with a photo-z more distant from the cluster redshift,
and in most cases poorly constrained. For such (typically faint)
sources, not only it is difficult to obtain a reliable redshift
estimate, but also to estimate photo-z accuracy and contamination. On
the other hand, as we show in the following, the vast majority of
these uncertain candidates is made of low-mass star-forming galaxies
that do not enter our mass-complete samples, and have generally little
weight in our conclusions. We finally note that spectroscopic
interlopers classified as likely and possible members are generally
close to the cluster redshift ($\gtrsim90$\% at $1.8<z<2.1$ and
$1.4<z<2.7$, for likely and possible members, respectively).

As a further check of the relevance of contamination, we model the
redshift distribution in Figure \ref{fig:zdist} with a $f(z)=C \times
\frac{\beta z^2}{\Gamma(3/\beta)z_0^3} e^{-(z/z_{0})^{\beta}}$
\citep{brainerd1996} for the field plus a Gaussian centered at z=2 for
the cluster (dashed gray lines in the figure). The modeling is only
done with the purpose of estimating the cluster and field
contributions in the $1.5<z<2.5$ redshift range\footnote{We use
  $\beta$=0.72, $z_{0}$=0.17 for $m_{140}<24.5$, and $\beta$=1,
  $z_{0}$=0.51 for $m_{140}<25.7$, but note that given the very small
  area we probe, and the contamination from the cluster itself, these
  data are not ideal for modeling $f(z)$, and thus the parameters
  determined here should not be considered for general purposes.}, as
relevant to membership determination.
 
For the $m_{140}<25.7$ sample (lower panel of Figure \ref{fig:zdist}),
the $\sigma$ of the Gaussian is fixed to the scatter estimated for the
photo-zs, and from the modeling we estimate that, in the $1.5<z<2.5$
range, about one third of this sample should be made of cluster
galaxies. Therefore, statistically we should have in our catalog, in
this redshift and magnitude ranges, about 40 cluster members: we have
14 spectroscopic members, and then 31 likely and 40 possible
candidates. Assuming our estimated 50\% contamination for likely
members and 80\% contamination for possible members, yields $\sim24$
members in very close agreement with the statistical estimate.

For the $m_{140}<24.5$ sample, the photo-z scatter is smaller
($\sim5\%$), and furthermore many cluster members are
spectroscopically confirmed, resulting in a tighter Gaussian in the
upper panel of Figure \ref{fig:zdist}, thus for the bright sample we
use $0.1<\sigma<0.15$. For this sample, we estimate that about 50\%
(and at least $40$\%) of the galaxies should be cluster members, thus
statistically $\sim 29$ (and at least 23) cluster members in this
magnitude and redshift range in our catalog. Since we have 13
spectroscopic members and 12 candidates, this could suggest that, for
the bright sample, our membership determination is less affected by
contamination (as could be expected). Overall, this check confirms
that our estimate of contamination for the whole sample is realistic,
if anything somewhat too high for bright sources.

\subsubsection{Final samples}
\label{sec:finalsamples}

In the end, we have a sample of 96 candidates, with 14 spectroscopic
members, 31 ``likely'' and 51 ``possible'' candidates\footnote{Four
  more sources classified as possible members are embedded in the
  halos of bright objects and were excluded, because of their unclear
  nature and severely corrupted photometry.}  down to
$m_{140}=25.7$. Based on the considerations discussed above, we expect
the whole sample of candidate members to include about 50 interlopers,
for the most part (3/4) selected as possible members.

We stress that, because of the significant contamination of our
candidate member sample by field galaxies at similar redshift, in most
of this paper we will not be able to investigate the detailed
comparison of galaxy properties in cluster and field environment at
$z\sim2$. On the other hand, in spite of the significant contamination
or in some cases thanks to the extensive WFC3 spectroscopy, some
properties of the cluster galaxy populations are clearly visible, even
after dilution of their signal with field galaxies, as discussed
below.

As discussed in section \ref{sec:morph}, for all analyses involving
the characterisation of galaxy morphological structure, we limit our
sample to objects brighter than $m_{140}=24.5$. The sample is thus
reduced to $\sim170$ sources, out of which 13 and 12 galaxies are
spectroscopic and candidates members, respectively.

The mass completeness limits corresponding to the magnitudes
$m_{140}=$25.7 and 24.5, estimated for an SSP formed at $z_f= 5$, with
solar metallicity and no dust reddening, are $\sim
8.5\times10^9$M$_{\odot}$ and $2.5 \times 10^{10}$M$_{\odot}$,
respectively (Salpeter IMF).

\subsection{Passive and star-forming galaxies}
\label{sec:stellarpop}

In order to broadly characterize the stellar population properties of
individual sources, we split the sample in two classes of galaxies
which are either essentially passively evolving, or still actively
forming stars. We initially assign galaxies to one class or the other
based on their restframe U-V and V-J colors \citep[e.g.,][hereafter
  UVJ classification]{wuyts2007}, as calculated based on its redshift
and the appropriate observed SED (as selected in
Sec.~\ref{sec:zphots}; spectroscopic redshifts are used where
available). We use here the division between passive and active
galaxies in the U-V vs V-J plane as adopted in \citet{williams2009}.

\begin{figure}
\includegraphics[width=.49\textwidth,bb=70 360 522 720,clip]{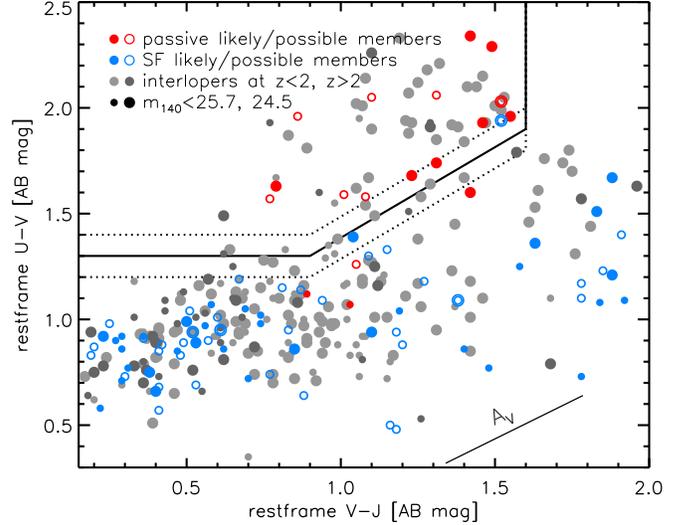}
\caption{The restframe U-V vs V-J color-color plot. Larger/smaller
  symbols show galaxies brighter than the two limits used
  ($m_{140}<$24.5 and 25.7, respectively). The solid line shows the
  separation criterion between passive and star-forming galaxies at
  $z\sim2$ as adopted in \citet{williams2009} -- dotted lines are, as
  a reference, $\pm$0.1 mag around this limit. Interlopers in the
  fore- and background are shown as light and dark gray symbols.
  Coloured symbols show cluster candidate members (``likely" and
  "possible'' as full and empty circles), red for passive and blue for
  star-forming sources according to the SED classification -- see text
  for details. \label{fig:uvjall}}
\end{figure}

We then re-fit the observed SEDs of candidate members, at fixed
redshift (photo-z or spectroscopic value), using combinations of
templates from two different libraries. The first library includes
only BC03 passive templates (age/$\tau >$4.5 and age $\ge$ 0.6 Gyr),
with different metallicities, no dust attenuation, and a range of ages
appropriate for the redshift range of the candidate members. This
library is thus only appropriate for passive galaxies, in the relevant
redshift range, with little dust attenuation. The second library
includes only BC03 templates with constant SFH, attenuated by dust
with E(B-V) up to 1.2, and with a range of ages appropriate for the
redshift considered. This library may thus only describe actively
star-forming (SF) populations with a broad range in dust attenuation,
including highly reddened sources.

\begin{figure*}
\includegraphics[height=.32\textheight,bb=104 390 456 720,clip]{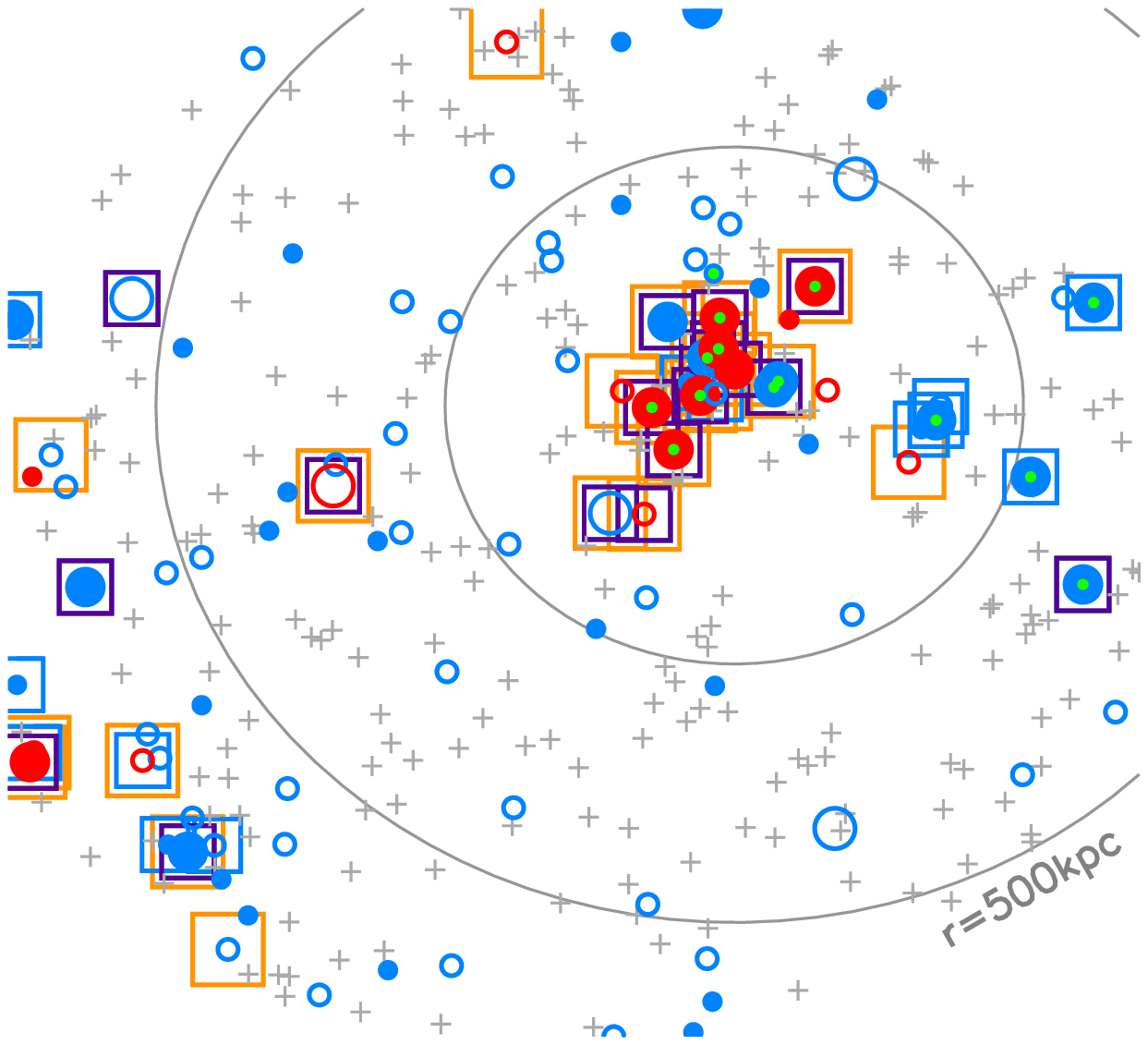}%
\hspace{.6cm}
\includegraphics[height=.32\textheight,bb=104 390 513 720,clip]{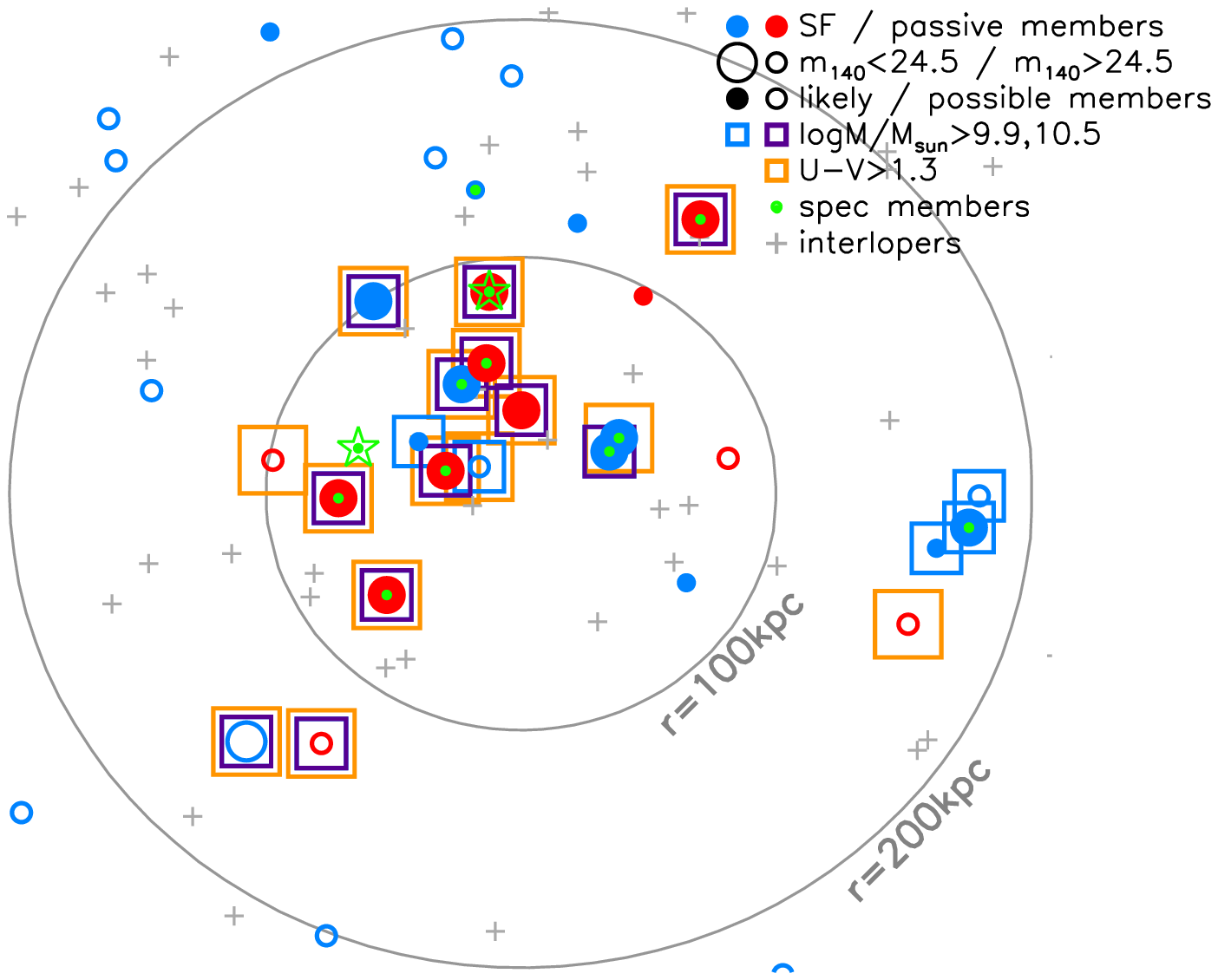}
\caption{ {\it Left:} The distribution of galaxies brighter than
  $m_{140}=25.7$ in the studied field. Interlopers are plotted as gray
  crosses, while passive and star-forming candidate members are
  highlighted in red and blue, respectively.  Filled and empty circles
  show likely and possible candidates, and spectroscopically confirmed
  members are marked with a small green point in the center. Large and
  small circles show sources brighter than $m_{140}=24.5$ and
  $m_{140}=25.7$. Yellow squares mark candidate members with restframe
  U-V$>1.3$, while blue and purple squares show the {\it mass
    complete} samples of members more massive than
  logM/M$_{\odot}$=9.9 and 10.4, respectively. Solid gray circles show
  clustercentric radii of 250 and 500 kpc (proper) at the cluster
  redshift. North is up, East to the left. {\it Right: } A close-up of
  the left-hand panel in the cluster center. Symbols are the same,
  gray circles mark clustercentric radii of 100 and 200 kpc (proper)
  at the cluster redshift. Two AGNs spectroscopically confirmed to
  belong to the cluster (Gobat et al. 2013) are marked by green
  stars. \label{fig:panoramixall}}
\end{figure*}

If we compare the $\chi^2$ of the best-fits for cluster members
obtained with these two libraries and with the EAZY standard
templates, we generally find that more than half of the galaxies
UVJ-classified as star-forming are best-fitted with the EAZY library
(lowest $\chi^2$ in 55\% of cases), 40\% are best-fitted by the
constant SFH library, and only 4\% have the lowest $\chi^2$ with the
passive library. Instead, candidate members classified as UVJ-passive
are essentially never best-fitted by constant SFH templates (with the
exception of a source close to the dividing line), and in 90\% of
cases have $\chi^2_{PASSIVE} \lesssim \chi^2_{EAZY}$ (we note that the
best-fit EAZY SED can also be essentially passive, with the greatest
contribution coming from templates of evolved
populations)\footnote{When a spectroscopic redshift is not
    available, as discussed above we fix the redshift to the
    photometric value, which is a sensible choice given that a
    photometric redshift is incomparably better constrained with the
    general template library. Nonetheless, we note that for most
    (70\%) of the passive sources, constant SFH templates give a
    poorer fit even if redshift is left as a free parameter.}. In the
great majority of cases, this alternative classification based on the
comparison of SED fit with different libraries agrees with the UVJ
selection.  In very few cases the two classifications do not agree
(see Figure \ref{fig:uvjall}): these are close to the dividing line of
the UVJ plot, where it is thus particularly useful to also have a
different approach, and/or faint sources whose SED is poorly
constrained due to large photometric errors. In these few cases we
adopt the SED-based classification.

Our sample of candidate members is thus ultimately divided in 18
passive (6 secure members plus 4 likely and 8 possible candidates) and
78 actively star-forming galaxies (8 secure members plus 27 likely and
43 possible candidates).

\begin{figure*}
\centering
\includegraphics[width=.6\textwidth,bb=59 640 404 703,clip]{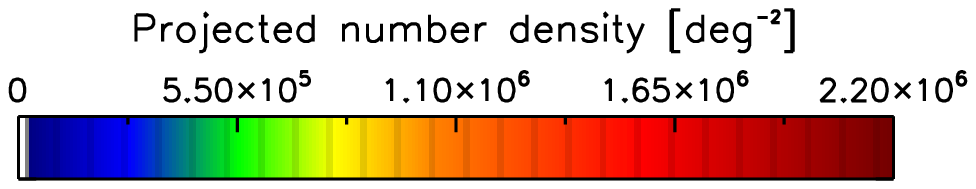}
\includegraphics[width=.33\textwidth,bb=100 382 351 661,clip]{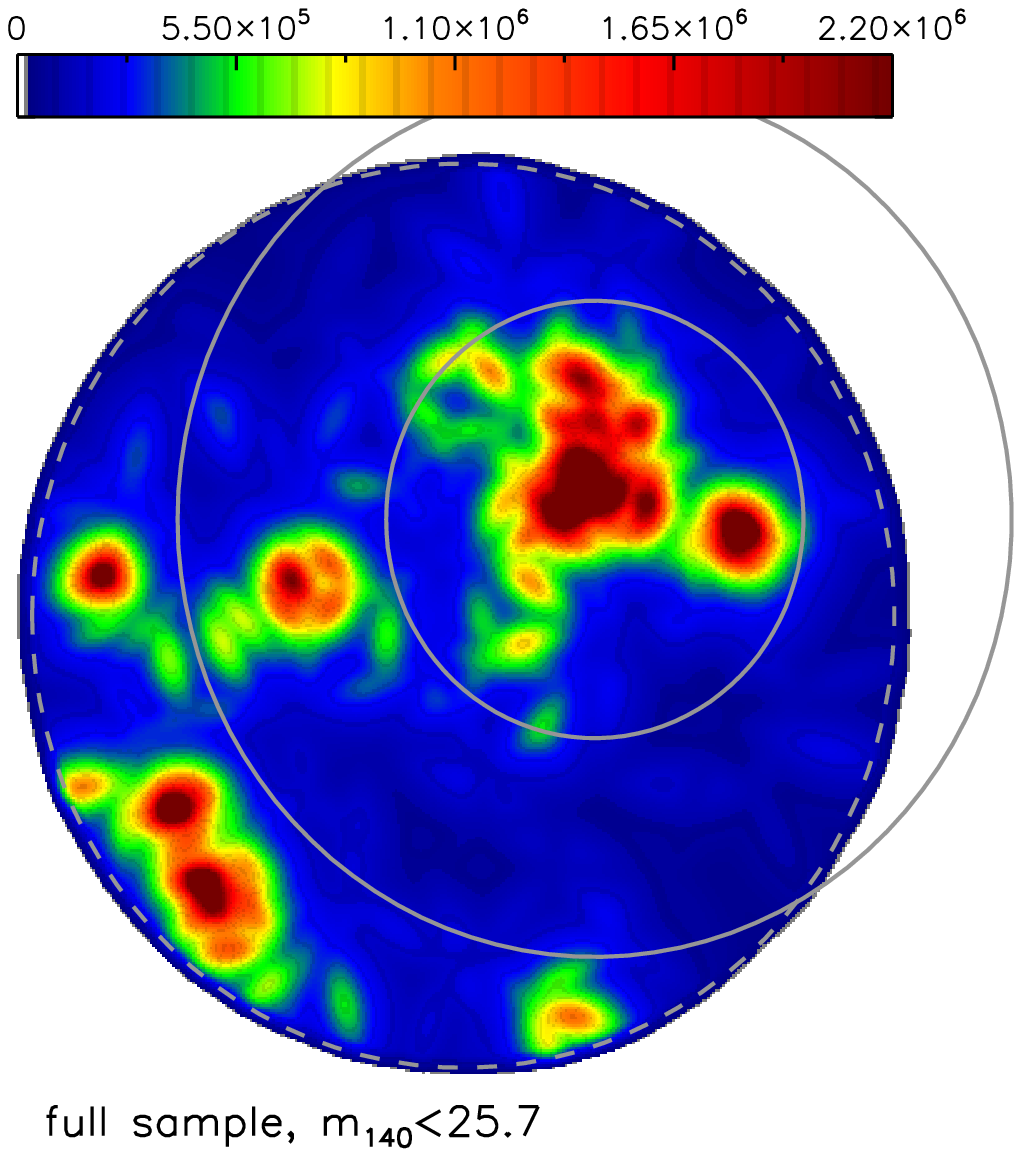}%
\includegraphics[width=.33\textwidth,bb=100 382 351 661,clip]{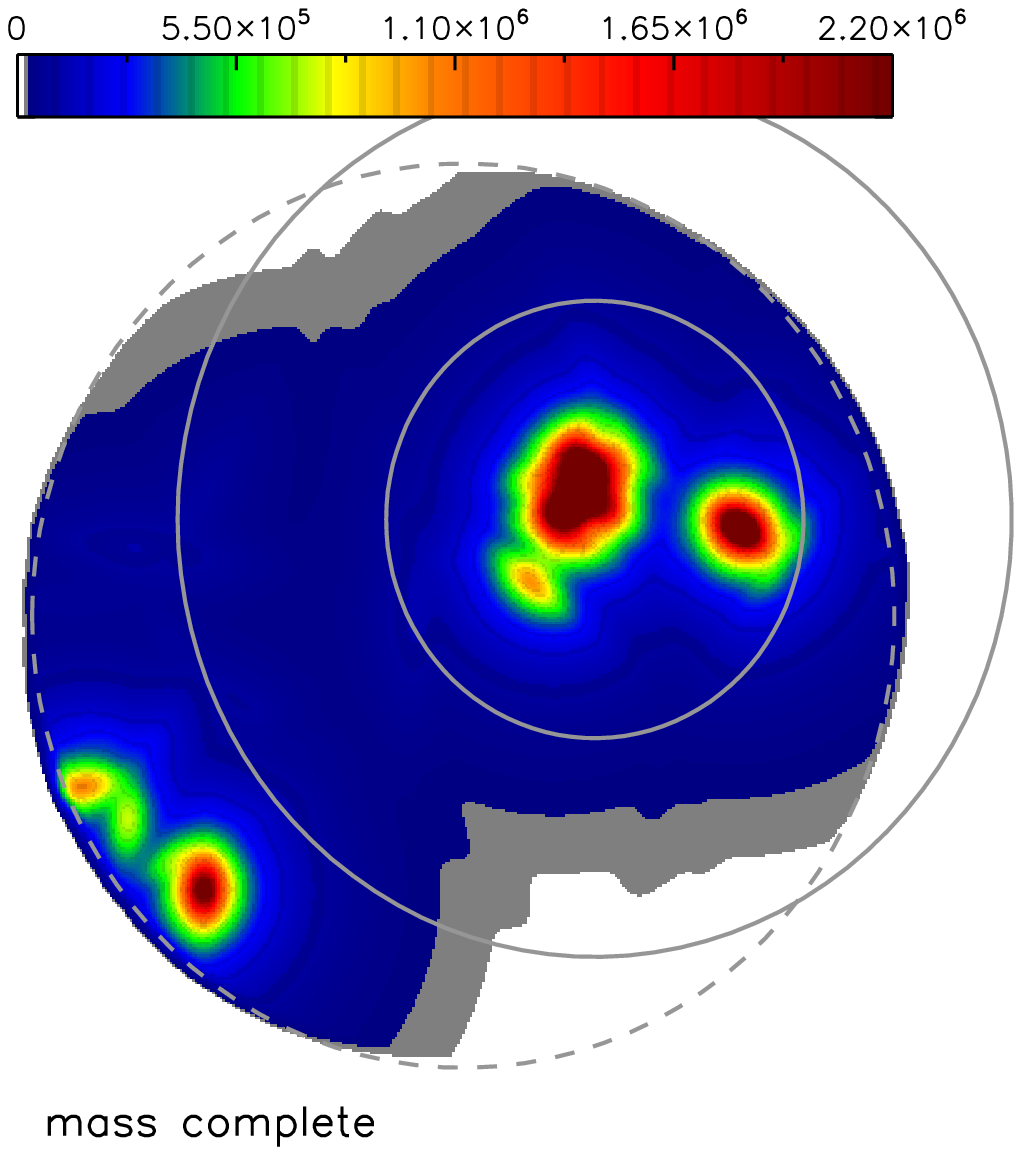}%
\includegraphics[width=.33\textwidth,bb=100 382 351 661,clip]{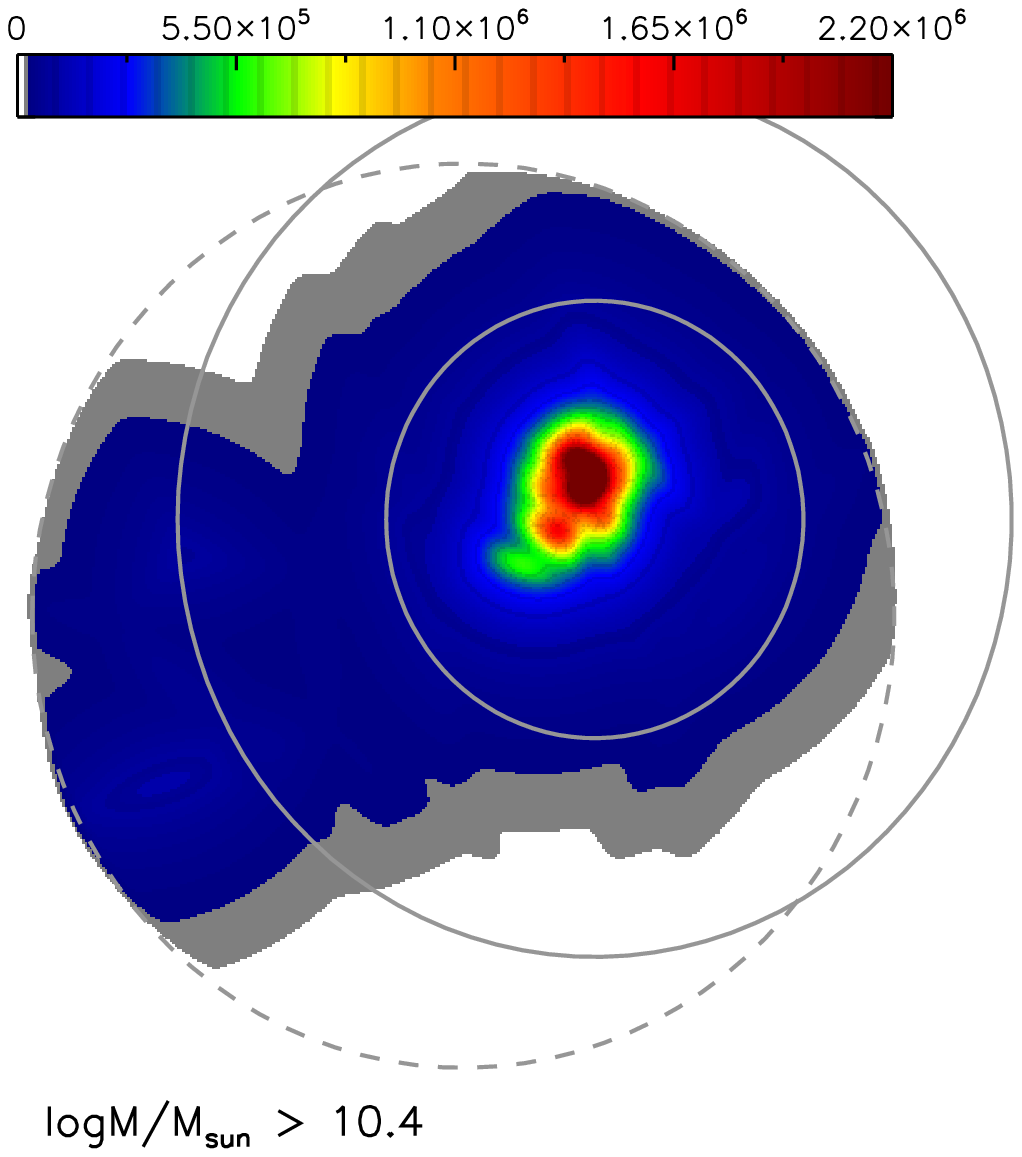}

\includegraphics[width=.25\textwidth,bb=100 382 351 661,clip]{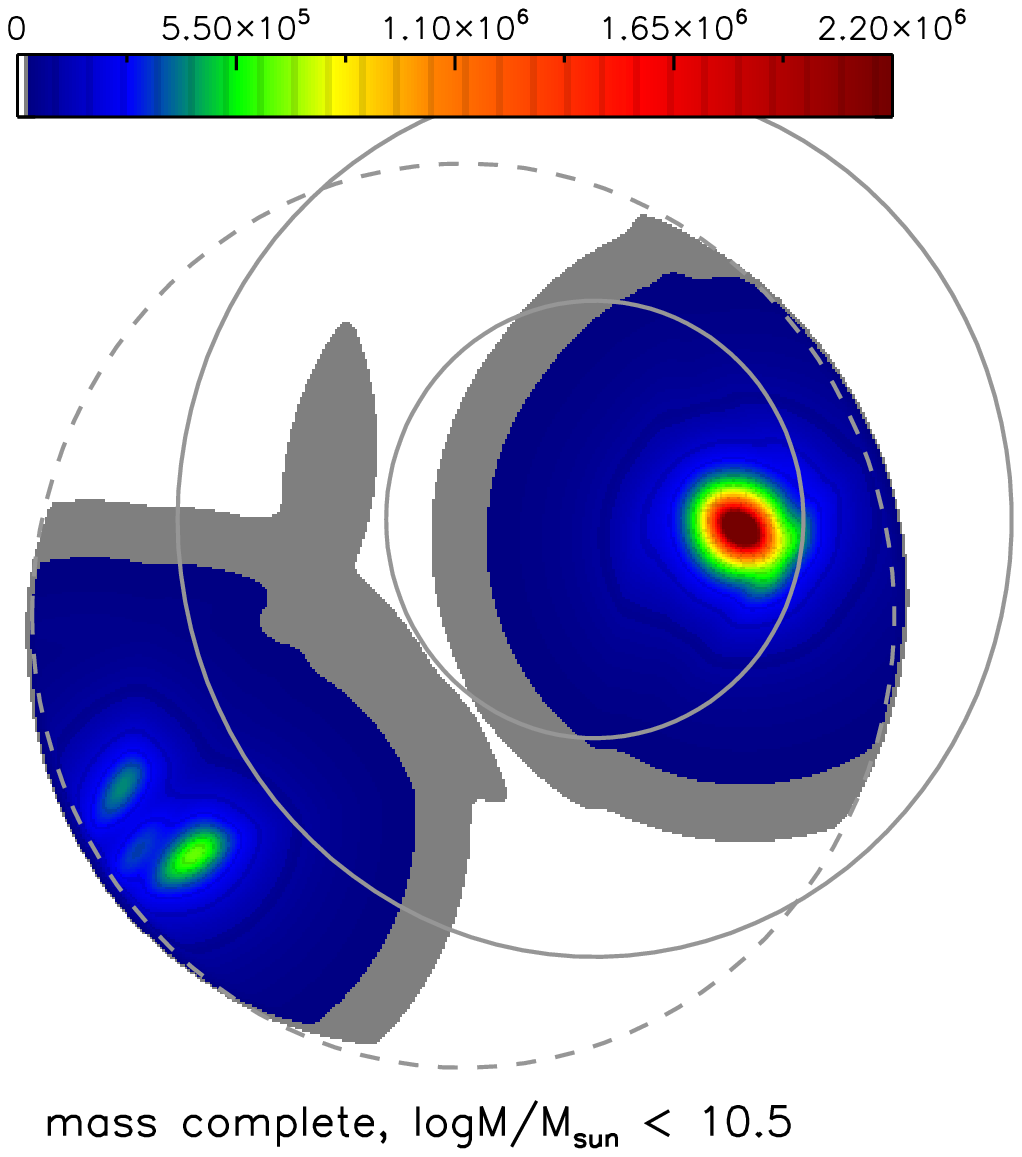}%
\includegraphics[width=.25\textwidth,bb=100 382 351 661,clip]{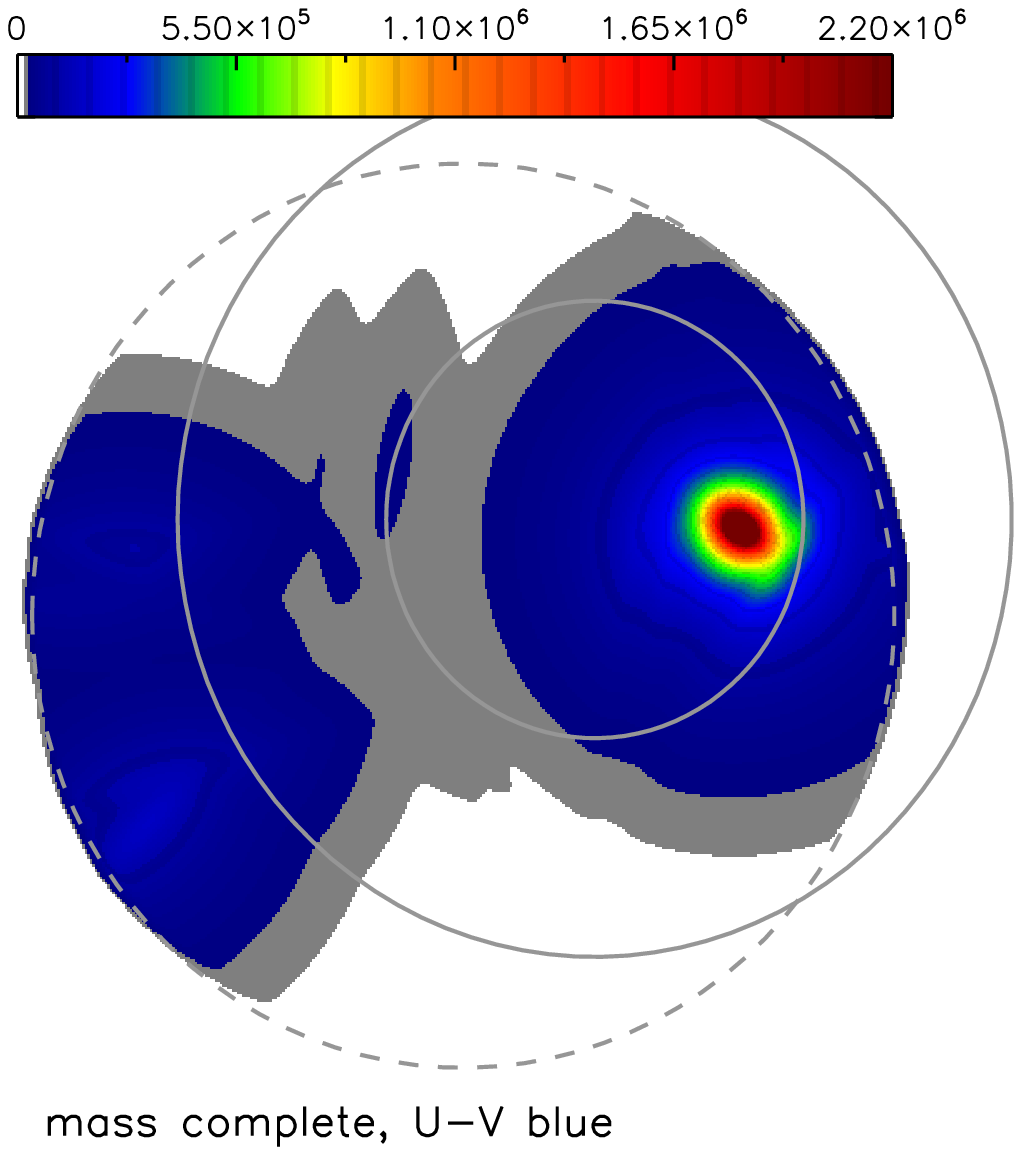}%
\includegraphics[width=.25\textwidth,bb=100 382 351 661,clip]{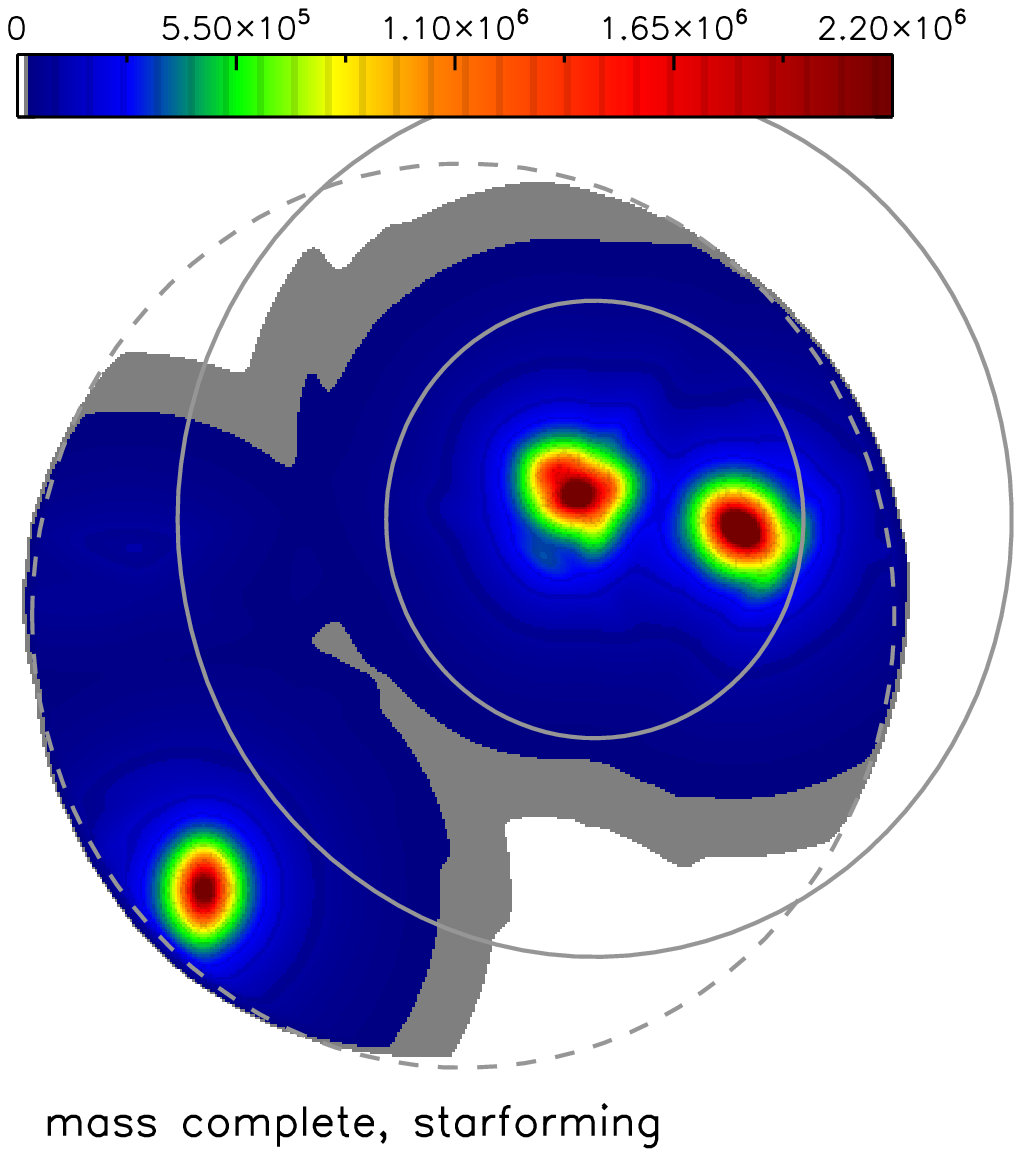}%
\includegraphics[width=.25\textwidth,bb=100 382 351 661,clip]{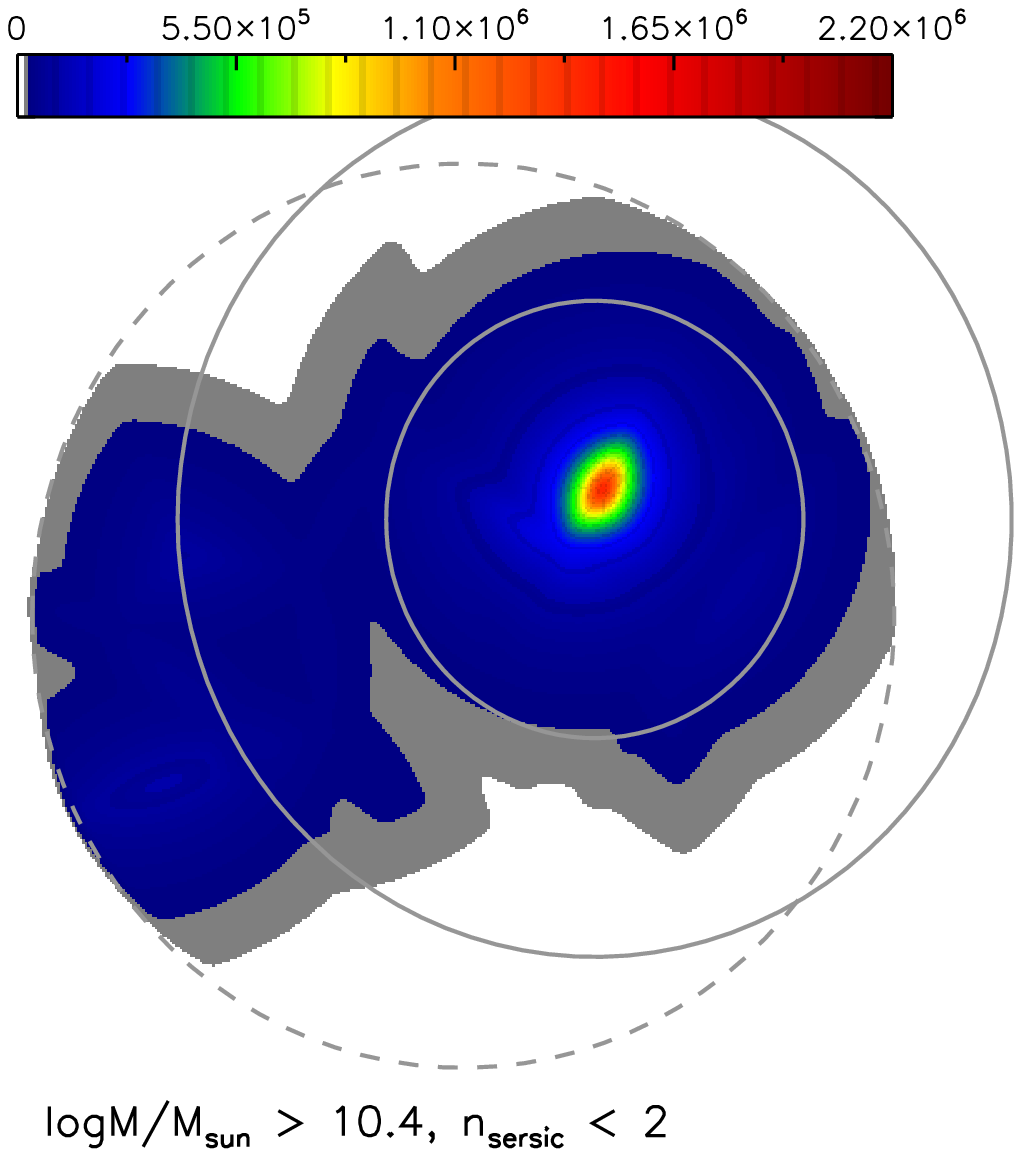}

\includegraphics[width=.25\textwidth,bb=100 382 351 661,clip]{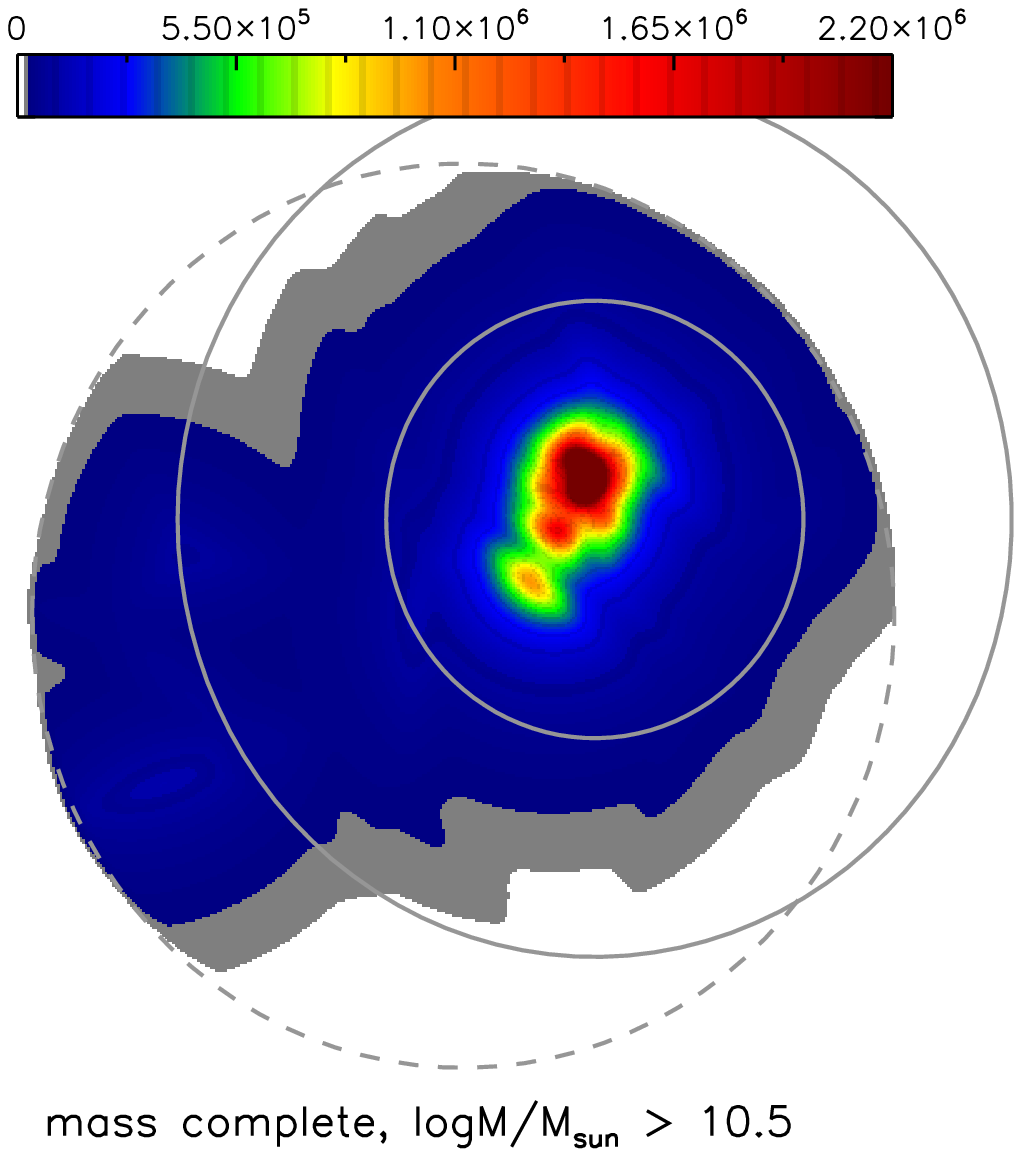}%
\includegraphics[width=.25\textwidth,bb=100 382 351 661,clip]{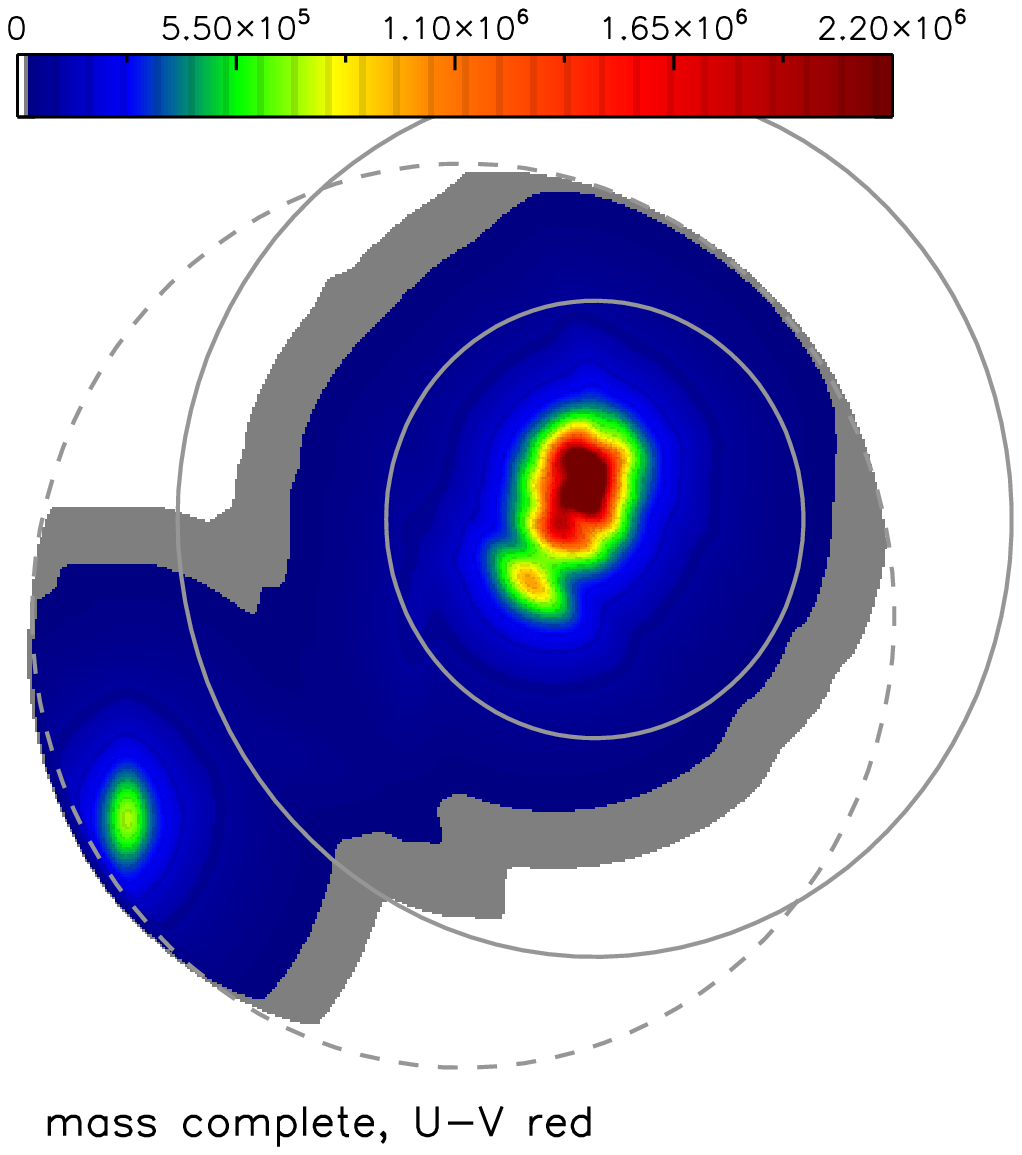}%
\includegraphics[width=.25\textwidth,bb=100 382 351 661,clip]{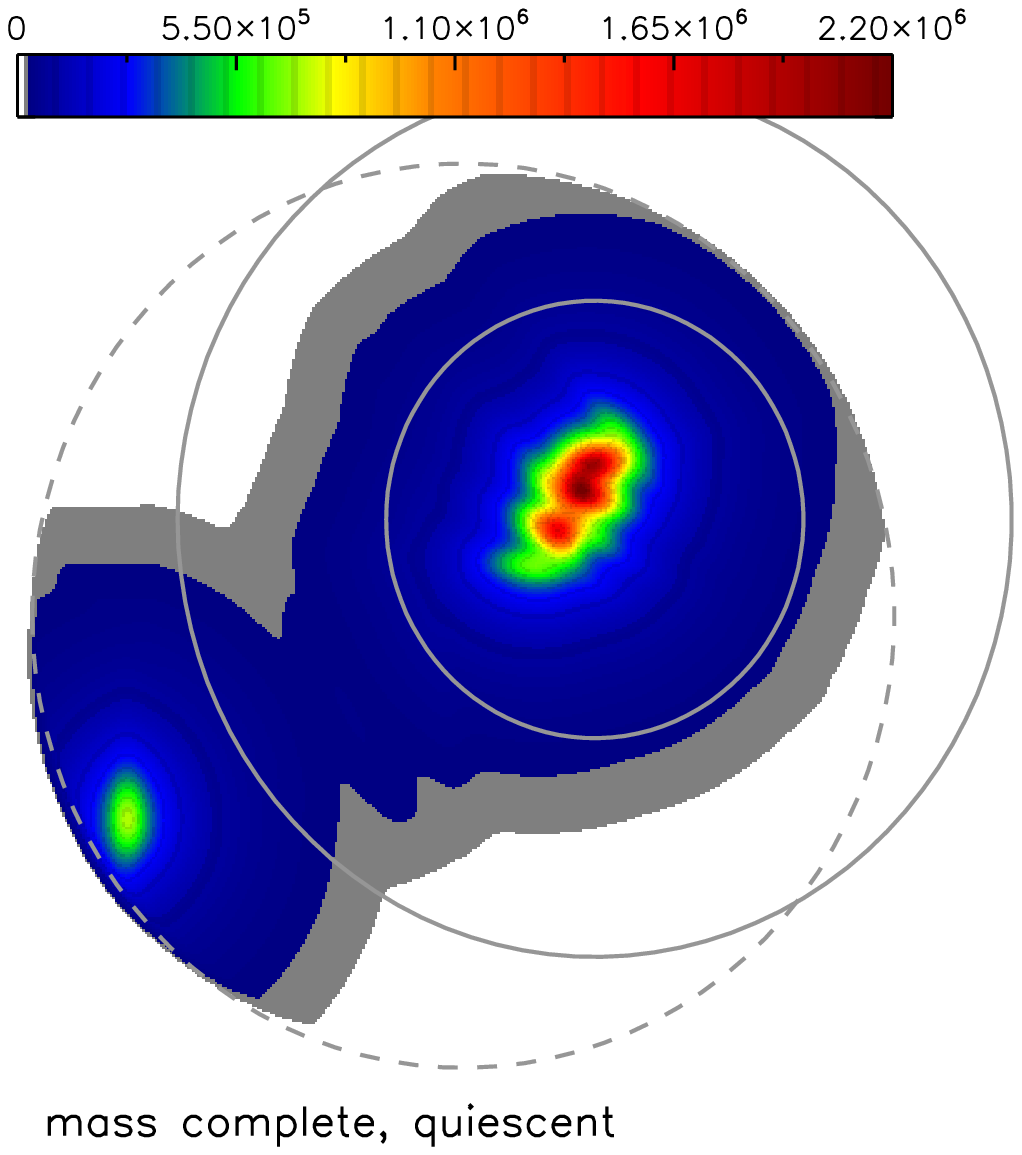}%
\includegraphics[width=.25\textwidth,bb=100 382 351 661,clip]{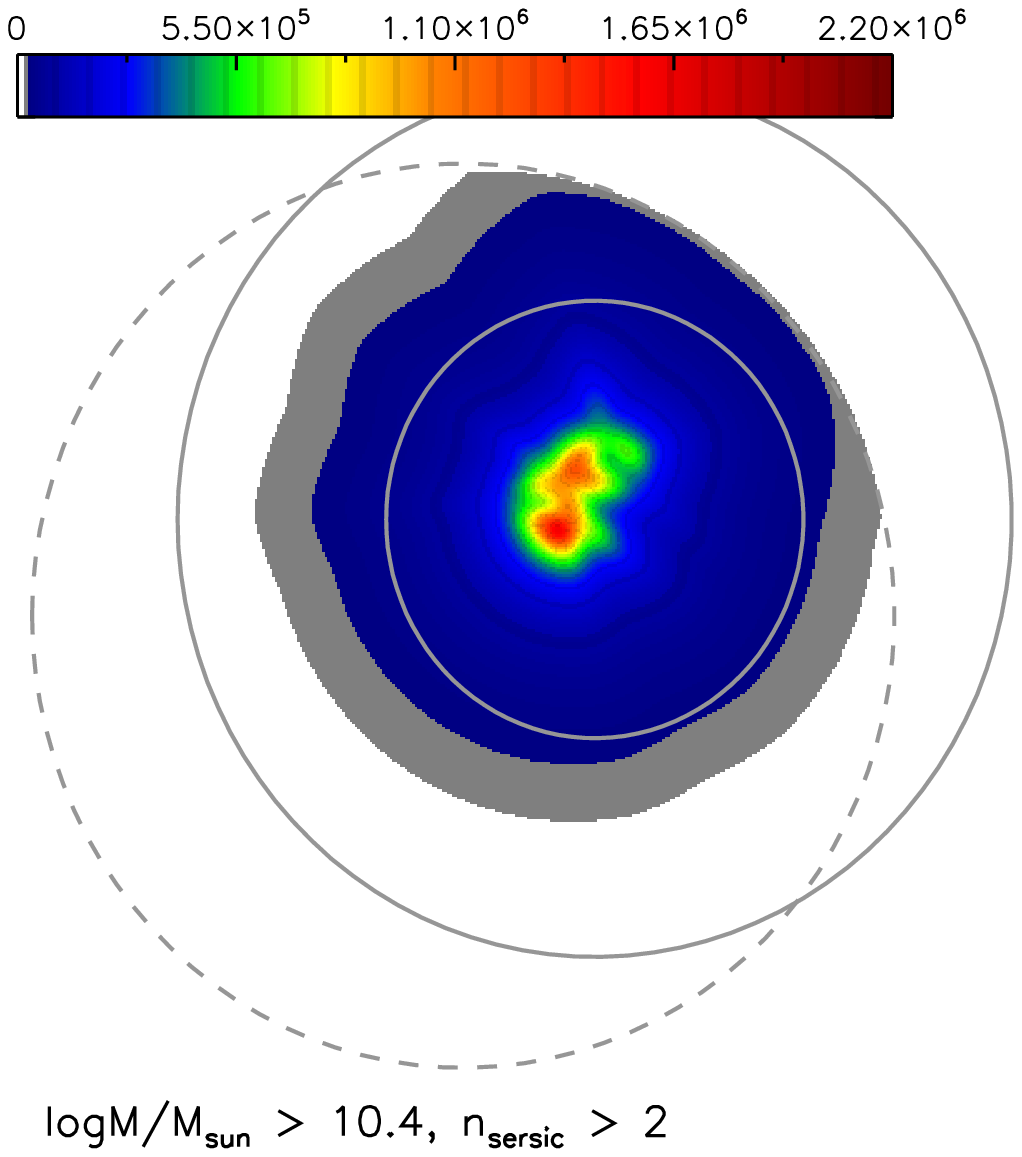}

\caption{{\it Top panels:} The projected density of candidate members
  around the center of the extended X-ray emission. A smoothed map of
  $\Sigma_{3}$ for the whole flux-limited sample of candidate cluster
  members down to $m_{140}=25.7$ (left panel, same sample as in Figure
  \ref{fig:panoramixall}), and for the mass complete samples of
  candidate members with logM/M$_{\odot}>$9.9 and 10.4, respectively
  (middle and right panel). {\it Middle and lower panels:} the two
  rightmost panels show the $\Sigma_{3}$ maps for high and low Sersic
  index candidate members, for the logM/M$_{\odot}>$10.4 sample shown
  in the top right panel. All other panels, which do not rely on
  morphological analysis, show $\Sigma_{3}$ maps for different
  sub-populations of the full mass complete (logM/M$_{\odot}>$9.9)
  sample shown in the top middle panel. These sub-populations are
  selected in stellar mass, restframe U-V color, or star formation
  classification (thus essentially specific star formation rate), as
  indicated at the bottom of each map. In all panels, the dashed
  circle shows the footprint of the catalog we used, while the two
  solid gray circles show clustercentric distances of 250 and 500 kpc
  at the cluster redshift, from the center of the extended X-ray
  emission. North is up, East to the left. Note that these maps refer
  to the full sample of candidate members, with no correction for
  contamination by interlopers. \label{fig:mapall}}
\end{figure*}

\section{Galaxies associated with the Cl1449+0856 structure}
\label{generalgalaxyprop}
As discussed above, our selection of candidate members should be
highly complete but also significantly affected by contamination from
interlopers. While it is impossible to remove this contamination based
on photometric redshifts, we can at least statistically investigate
some properties of the cluster galaxy populations which are strong
enough not to be diluted by the significant presence of
interlopers.

\subsection{Projected distribution of candidate members}

We show in Figures \ref{fig:panoramixall} and \ref{fig:mapall} the
projected distribution of candidate cluster members in the field.
Note that both figures show all candidate members -- according to the
specific selection as labeled in individual panels -- and are thus
affected by interloper contamination (as discussed in sections
\ref{sec:complcont} and \ref{sec:finalsamples}). While the more
uncertain "possible'' members make up about half of the full
$m_{140}<25.7$ candidate sample, their contribution is higher at low
masses, and goes down to $<$30\% and 25\% for the logM/M$_{\odot}>$9.9
and logM/M$_{\odot}>$10.4 mass-complete samples highlighted in the
figures. Accounting for spectroscopic members and the estimated
contamination for ``likely'' and ``possible" candidates, we estimate
that these mass-complete samples are affected by an overall
contamination of $<40$\% and $\sim30\%$, for logM/M$_{\odot}>$9.9 and
logM/M$_{\odot}>$10.4, respectively.

Figure \ref{fig:panoramixall} shows individually all galaxies in the
field, highlighting "possible" and "likely" candidate members, as well
as the nature of their stellar populations as estimated from their SED
(section \ref{sec:stellarpop}).

\begin{figure}
\centering
\includegraphics[width=.46\textwidth, bb = 15 16 215 225 , clip]{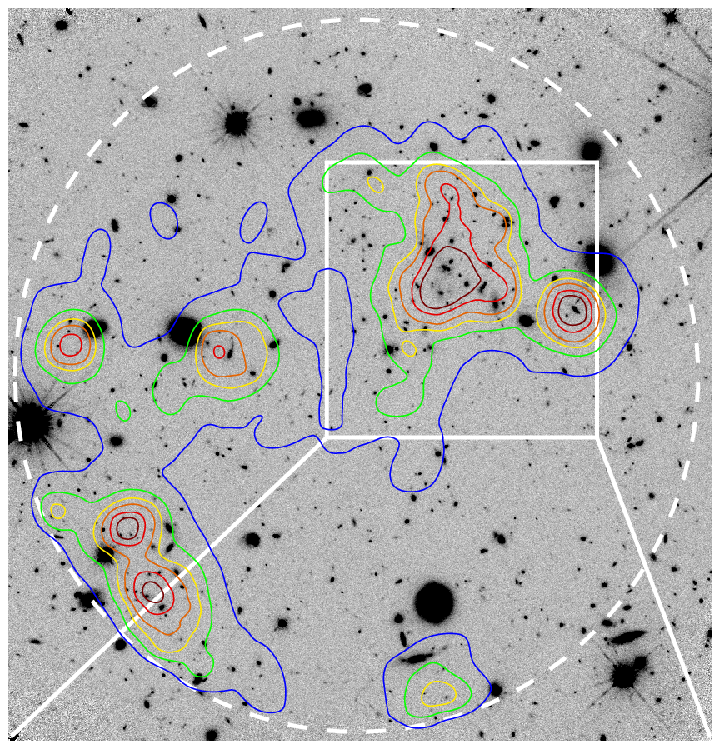}
\includegraphics[width=.46\textwidth,bb = 15 16 91 93, clip]{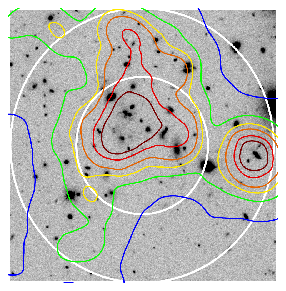}
\caption{The WFC3 F140W image of the studied area. Contours correspond
  to the density map of the full flux-limited sample of candidate
  members shown in Figure \ref{fig:mapall} (top-left panel), with
    colors corresponding to projected number density levels in the
    same color scale. The dashed circle shows the footprint of the
    catalog we used, as in Figure \ref{fig:mapall}. In the bottom
    panel, a close-up of the inner cluster region (white square in
    upper panel) is shown. White circles show radii of 100 and 200 kpc
    (proper) at the cluster redshift.  \label{fig:contoursonwfc3}}
\end{figure}

Figure \ref{fig:mapall} shows local density maps of the same sample of
candidate cluster members ($m_{140}<25.7$), as well as of the
mass-complete sample (logM/M$_{\odot}>$9.9), to picture more clearly
their projected distribution, the density enhancement around the
cluster center \citep[taken as the center of the extended X-ray
  emission, as quoted in][]{gobat2011}, and possible surrounding
structures.  As an estimator of local (projected) density, we used the
density based on the distance to the 3$^{rd}$ nearest neighbour,
$\Sigma_{3}$. We correct for edge effects by accounting for uncovered
area within the distance to the 3$^{rd}$ nearest neighbour, however
minor edge effects may still persist. Contours from the density map of
the full flux-limited $m_{140}<25.7$ sample of candidate members
(top-left panel of Figure \ref{fig:mapall}) are shown overlaid on the
WFC3 F140W image in Figure \ref{fig:contoursonwfc3}. Figure
\ref{fig:mapall} also shows local density maps of different
sub-populations of the mass-complete sample, to highlight their
different (projected) distribution.

Figures \ref{fig:panoramixall} and \ref{fig:mapall} clearly show the
characteristic nature of the galaxies in the central concentration,
including many massive, red, passive sources within 100-150 kpc of the
cluster center.  West of the cluster center, these figures show an
overdensity of galaxies which seem distinct in nature, less massive,
star-forming, and (whenever the measurement is possible) with a
late-type structure (Figure \ref{fig:uvjmorphmembers}).  Half of these
are spectroscopically confirmed to be cluster members.

Another overdensity in the projected distribution of candidate members
is located south-east of the cluster center.  However, as
Figure \ref{fig:panoramixall} shows, it is made in large part of
candidates less likely to be at the cluster redshift, and it contains
no spectroscopically confirmed member. The reality of this structure
could not be confirmed with the current spectroscopic coverage, partly
due to observational issues (being located at the edge of the field
fully covered by WFC3 grism spectroscopy with all 4 orientations, and
being made in large part of faint sources).

Considering the mass-complete sample (blue and purple squares in
Figure \ref{fig:panoramixall}), and in spite of dilution due to
interloper contamination, a concentration of massive, of optically red
(restframe U-V$>$1.3\footnote{The restframe U-V$>$1.3 threshold, as
  used here when referring to optically red sources, is close to the
  observed Y-K color cut used in \citet{gobat2011}.}), and of passive
galaxies in the cluster core is evident. This central concentration
appears to include mostly passive sources, but also some dust-reddened
star-forming galaxies. These results seem to be largely stable against
the inclusion of interlopers. In fact, due to the selection criteria a
large fraction of the less likely ("possible'') members is made of
low-mass star-forming galaxies below the mass completeness limit
($75\%$, vs $50\%$ for ``likely'' members). More specifically, the
concentration of massive, red, and passive galaxies in the cluster
core does not depend on the inclusion of less-likely members.

\subsection{Structural and stellar population properties}

Passive systems, as well as more massive galaxies
(logM/M$_{\odot}>$10.5), seem to be effectively segregated in the
central cluster region, with two thirds of these galaxies within
$<$200~kpc from the cluster centre.

\begin{figure}[b]
\centering
\includegraphics[width=.4\textwidth,bb=104 415 456 706,clip]{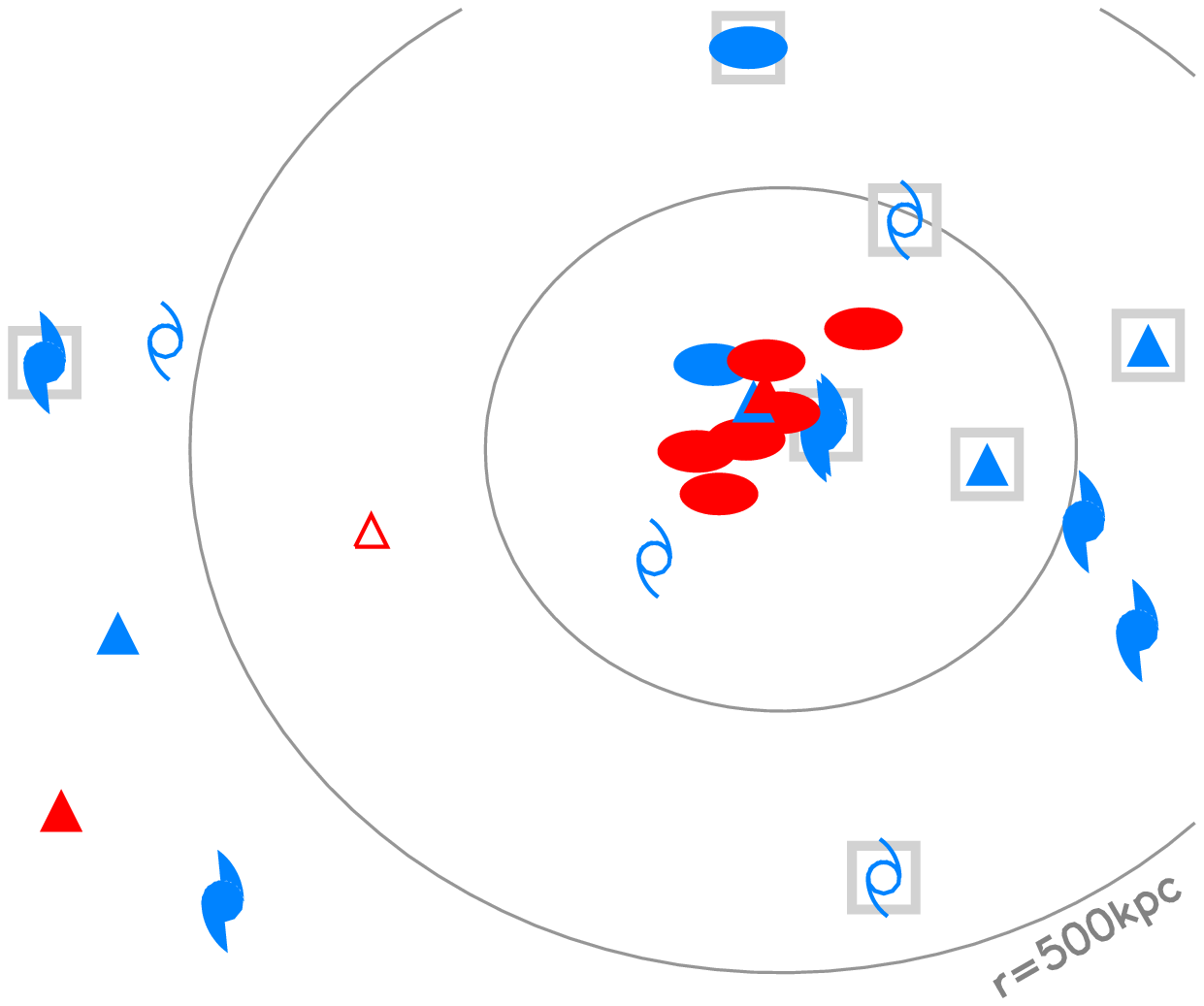}
\includegraphics[width=.48\textwidth,bb=87 374 544 696,clip]{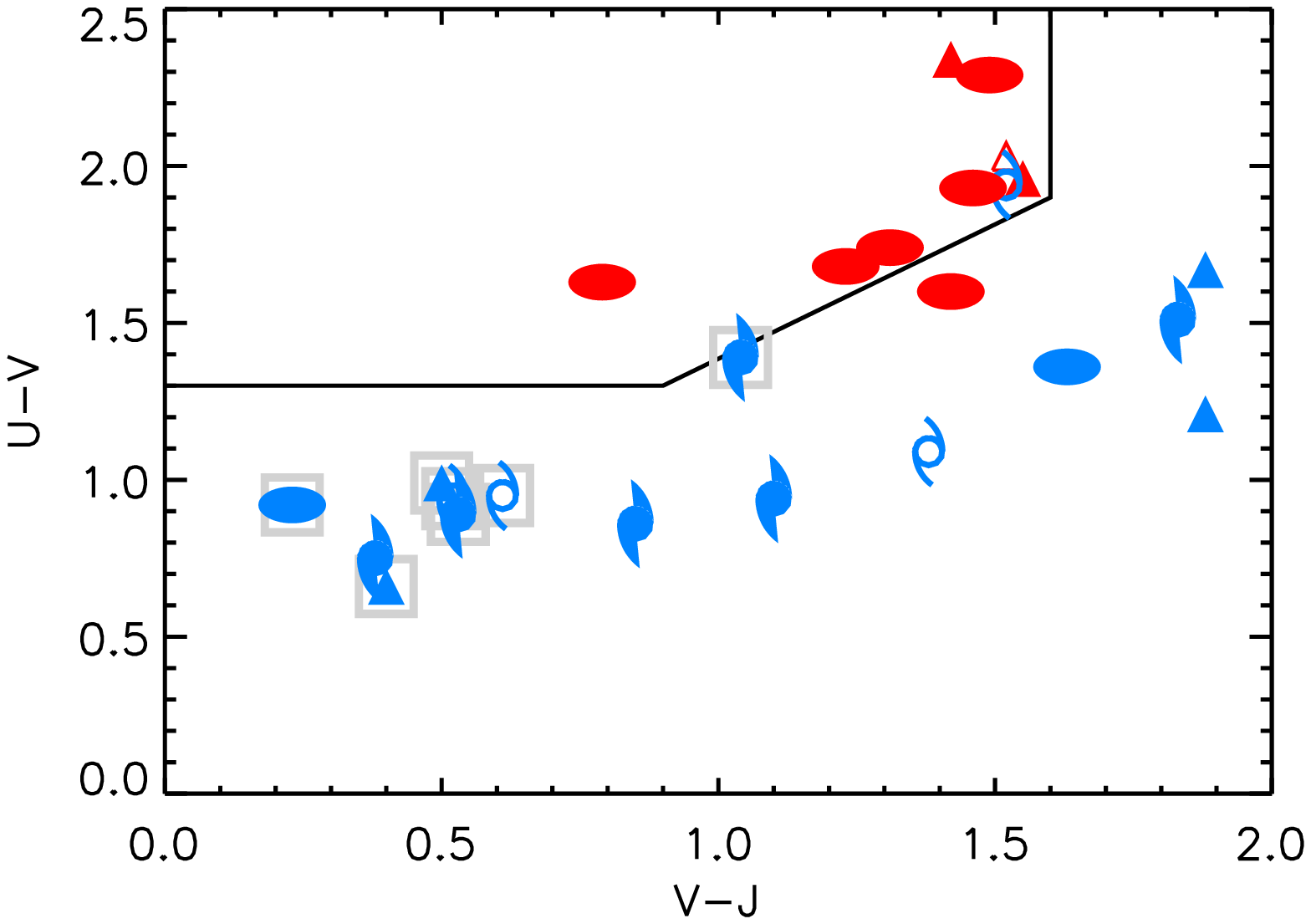}
\caption{The projected distribution (top panel, as in Figure
  \ref{fig:panoramixall}) and the UVJ restframe color-color plot
  (bottom panel) of the sample of candidate members brighter than the
  limit for morphological analysis $m_{140}<24.5$.  This sample is
  flux-limited, {\it not} mass complete: sources below the estimated
  mass completeness of logM/M$_{\odot}$=10.4 are highlighted with gray
  squares. Solid/empty symbols show likely/possible members,
  respectively. Galaxies classified as passive or star-forming are
  colored in red and blue, while galaxies with a $n_{Sersic}$ higher
  or below 2 are shown as ellipses and spirals, respectively. Galaxies
  for which no acceptable fit could be obtained are plotted as
  triangles: visual inspection shows that only one could be an
  early-type, a passive source very close to the cluster center.
\label{fig:uvjmorphmembers}}
\end{figure}

At least in the mass range probed by our morphological analysis
(logM/M$_{\odot}>$10.4), this segregation is also evident for
high-Sersic systems ($n>2$), which are all within a clustercentric
distance of $\sim$150~kpc, as shown in Figures \ref{fig:mapall} and
\ref{fig:uvjmorphmembers}.

This extends to a z=2 cluster previous results showing that, already
before $z\sim1$, the central regions of clusters and groups generally
exhibit a segregation of more massive, older, or morphologically
evolved galaxies \citep[e.g., among
  others][]{rosati2009,mei2012,muzzin2012}. In particular, the studies
of \citet{kurk2009,tanaka2012,tanaka2010,papovich2010,papovich2012} of
two X-ray detected low-mass clusters at $z\sim1.6$, suggest that at
least some overdense structures, even with relatively low masses and
already at $z>1.5$, host in their core galaxy populations which are
particularly evolved, in terms of their structure, stellar
populations, and assembled stellar mass, as compared to lower density
regions at the same epoch, and possibly in spite of the coexistence in
the same volume of a population of galaxies which are instead still
actively forming \citep[e.g.,][]{tran2010}. Indeed, we recall results
from several studies suggesting that, even at $z\gtrsim2$, some
proto-cluster environments may already host galaxies more massive,
with older stars, and more evolved structure, than their surroundings
\citep[e.g.,][]{steidel2005,kodama2007,tanaka2010b,hatch2011,zirm2012,spitler2012}.

With respect to the comparison of structural and stellar population
properties, we note the clear correlation in our sample of candidate
members (at least in the probed mass range) between a high-Sersic
profile and evolved host stellar populations (Figure
\ref{fig:uvjmorphmembers}), consistent with previous observations at
similar redshifts in both field and high-density environments
\citep[e.g.,
][]{cimatti2008,kurk2009,wuyts2011,cameron2011,bell2012,papovich2012,tanaka2012,patel2012}. As
discussed also in more detail below, 70$^{+10}_{-20}$\%\footnote{Here
  and in the following, errors on the fraction are calculated
  following \citet{cameron2011p}.}  of candidate members more massive
than the logM/M$_{\odot}>$10.4 threshold for morphological analysis,
and identified as passive, have n$_{Sersic}>$2. A
similar early-type fraction is found in the passive population of our sample of
interlopers at $1.5<z<2.5$. Conversely, only 10$^{+20}_{-4}$\% of
star-forming candidate members more massive than the same limit are
classified as morphological early-types.  For comparison,
\citet{papovich2012} finds that about 80\% of candidate members in the
cluster XMM–LSS J02182-05102 at $z\sim1.6$ have n$_{Sersic}>$ 2 (in a
mass range similar to ours).

In turn, $\sim$75$^{+9}_{-20}$\% of the logM/M$_{\odot}>$10.4
candidate members with an early-type morphology also appear to be
passive, with a similar fraction in our sample of interlopers at
$1.5<z<3$, although statistics are too poor to draw conclusions. For
comparison, \citet{bell2012} finds about 60\% of early-type galaxies
to be passive, at $z\sim2$, down to a stellar mass limit of
$5.5\times10^{10}$M$_{\odot}$.

\subsection{The environmental signature on galaxy populations}

\begin{figure}[ht]
\includegraphics[width=.48\textwidth,bb=42 360 400 721,clip]{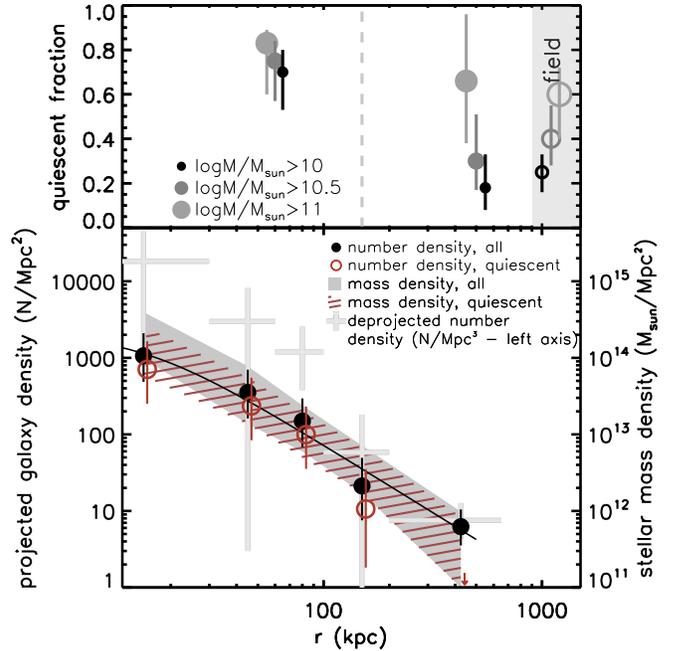}
\caption{{\it Bottom:} Number density profile of cluster galaxies as
  estimated from the sample of candidate members (see text for
  details). Black and red symbols show the whole population more
  massive than 10$^{10}$M$_{\odot}$, and the sub-sample of galaxies
  classified as passive, respectively (black and red symbols are
  slightly offset for clarity). Error bars include the Poisson error
  and the uncertainties in membership determination (see text). 
    The light gray crosses show a simplistic deprojection of the
    observed density profile assuming spherical symmetry (see text;
    units are galaxies/Mpc$^3$ read on the left-hand y-axis).  The
  gray-shaded and red-hatched areas show the inferred stellar mass
  density profiles (right-hand scale) for the same two samples
  ($>10^{10}$M$_{\odot}$, whole population and passive sources,
  respectively). {\it Top: } The fraction of passive galaxies in two
  radial bins of clustercentric radius ($r<150$~kpc and
  $150<r<700$~kpc, solid symbols) as estimated from the sample of
  candidate cluster members (see text for details).  The passive
  fraction as measured from these same data in the field (galaxies
  classified as interlopers at $1.5<z<2.5$) is shown with empty
  symbols. Black, dark-gray and light-gray symbols show mass-complete
  samples with logM/M$_{\odot}>$10,10.5, 11,
  respectively.  \label{fig:profiles}}
\end{figure}

In Figure \ref{fig:profiles} we show the projected number density
profile for the whole mass complete sample of candidate members more
massive than 10$^{10}$M$_{\odot}$, as well as for passive galaxies in
this sample, and the related stellar mass profiles. For the purpose of
this figure, the cluster center is taken at the center of the galaxy
overdensity, roughly located on a complex multi-component galaxy
system, with asymmetric halos and tails clearly suggestive of an
ongoing merging, that in \citet{gobat2011} was identified as the
possible proto-BCG still in a very active formation phase.  This is
offset by $\sim$50 kpc (in projection) from the estimated center of
the X-ray emission. Note that this offset is similar to what observed
in lower redshift clusters and groups, and is anyway comparable to the
uncertainty on the X-ray centroid position
\citep{fassbender2011b,george2012}.  All profiles shown take into
account the contamination by interlopers by resampling multiple times
the sample of candidate members, according to the contamination
estimates discussed above.  The errors shown on the number density
profiles are the largest between the Poisson error on the counts and
the scatter due to the resampling. For stellar mass profiles, an error
of a factor two on stellar mass is included.  At these masses the
impact of less-likely (``possible") members is marginal, and we verify
that the inclusion or exclusion of these galaxies does not affect the
profiles. The light gray crosses in Figure \ref{fig:profiles}
  show, as a simplistic illustration, the density profile obtained by
  deprojecting the observed profile (black points) assuming spherical
  symmetry, with a simple approach similar to \citet{mclaughlin1999},
  and assuming no significant contribution to the overdensity beyond
  650~kpc. Based on this estimate, the average volume number density
  of massive galaxies ($>10^{10}M_{\odot}$) within the region probed
  by this profile (650kpc from the cluster center) would be about
  $250\pm100$ times the density in the field at $z\sim2$ \citep[from wide
  field measurements, e.g. ][, see also section
  \ref{sec:complcont}]{muzzin2013}, reaching central densities 4-5 orders of
  magnitude larger than in the field within 100~kpc from the cluster
  center.  We stress that this is only a simplistic approximation for
  illustration purposes, and of course we have no proof - and likely
  no expectations - that this cluster is spherically simmetric.

The purpose of Figure \ref{fig:profiles} is to quantitatively show the
increased galaxy density of candidate members in the cluster central
region. A proper investigation of the shape of the galaxy number
density profile is beyond the scope of this work, but we show as a
reference the best-fitting projected $\beta$-model\footnote{A
  generalization of core profiles which is often used to describe
  cluster galaxy number density profiles \citep[e.g.,][]
  {girardi1998,lemze2009}. } \citep[$\Sigma(r) = \sigma_{0}
  (1+(\frac{r}{r_{core}})^{2})^{-\beta}$,][with $\sigma_{0}$ the
    central projected density, $r_{core}$ the core radius, and $\beta$
    the outer slope]{cavaliere1978} to the number density profile
(black points) as a black line. The profile suggests that, {\it if}
there is a core, it is very small (core radius $20^{+30}_{-10}$~kpc),
as also observed in low-redshift clusters
\citep[e.g.,][]{biviano2003}. The best-fit $\beta\sim0.9$ is close to
typical values observed in the nearby Universe
\citep[e.g.,][]{popesso2004}. Given the small offset between the X-ray
centroid and the center of the overdensity, we note that this figure
would be essentially the same if considering the X-ray centroid as the
cluster center, the only relevant effect being the increase of the
core size to $\sim50$~kpc.

The upper panel of Figure \ref{fig:profiles} shows the fraction of
candidate members which are classified as passive in two radial bins
(within and beyond a clustercentric distance of 150~kpc), for three
mass-limited samples (logM/M$_{\odot}>$10, 10.5, and 11). In spite of
the relatively poor statistics and of the contamination by field
galaxies, this Figure clearly shows that a larger fraction of galaxies
has already suppressed star formation in the cluster center,
corresponding to the high-density region shown by the profile in the
bottom panel. The effect is seen in all the mass-limited samples
shown, with the possible exception of the most massive systems
($>$10$^{11}$M$_{\odot}$). Statistics are too poor to draw any
conclusion, but the lack of a clear environmental effect for the most
massive galaxies would hint at a predominant role of mass-related
factors \citep[so-called ``mass quenching'',
  e.g.,][]{baldry2006,peng2010}, rather than local density, in
quenching galaxies at the highest masses, at this epoch and for this
kind of environment \citep[but see e.g.,][for clusters at
  $z\sim1$]{muzzin2012}. In the outer bin (between 150 and
$\sim700$~kpc) the passive fraction is consistent with the field value
(estimated from these same data using galaxies classified as
interlopers at $1.5<z<2.5$ and keeping into account the resampling of
candidate members). However, we remind the reader that where the
overdensity of cluster galaxies drops, dilution from contamination
dramatically affects the possibility to recover the properties of
cluster galaxies, which thus appear similar to those of the field
sample.  Statistically correcting for the contamination by
interlopers, as it is done here, is expected to give lower passive
fractions (if field galaxies have a higher star-forming fraction),
especially in the outer regions were field contamination is more
significant. Finally, we note that the passive fraction that we
measure in the field is consistent with previous determinations at
redshift two, for instance we find a passive fraction of $50\pm15$\%
at masses logM/M$_{\odot}=11.15\pm0.35$, close -- given the
uncertainties -- to the estimates of e.g.,
\citet{daddi2005b,brammer2011,patel2012b}.

Compared to $z\sim1$ clusters, we find as expected higher star-forming
fractions, but we recall the caveat just discussed above which might
bias high our estimates. For instance, in the range 10.25$\lesssim
$logM/M$_{\odot} \lesssim 10.95$, \citet{muzzin2012} finds about 20\%
of cluster galaxies with still active star formation within 200~kpc
from the cluster core, while we estimate about 50\%. However, at high
stellar masses (logM/M$_{\odot}>$10.95) 70$^{+10}_{-20}$\% of galaxies
within 200~kpc appear to be already passive, a result which is not
likely to be produced by field contamination, and is already similar
to the low ($\sim20\%$) star-forming fractions estimated by Muzzin et
al. in their $z\sim1$ sample.

\subsection{The high-mass tail}
\label{sec:massiveend}

At $z<1.5$, very massive galaxies are a characteristic population of
cluster cores, which generally exhibits the most evolved morphological
structures and stellar populations. However, studies of $z>1.5$
clusters often show significant activity (from both the star
formation and mass assembly points of view) even at the high-mass tail
of cluster galaxy populations, as discussed in the introduction.

\begin{figure}[ht]
\centering
\includegraphics[width=.48\textwidth]{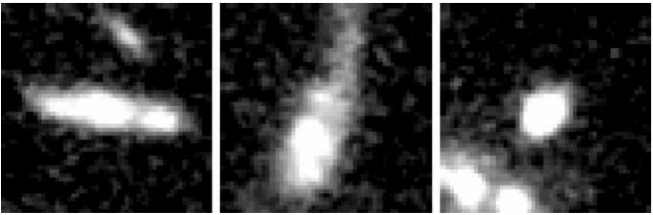}
\caption{WFC3 F140W cutouts of the three massive candidate members
  classified as passive (from both criteria described in section
  \ref{sec:stellarpop}), that do not have an early-type morphology
  (left and middle panel) or for which a reliable fit could not be
  obtained (right panel). Cutout size is 3'', or $\sim$26~kpc at
  z=2. \label{fig:cutoutsnoetgs}}
\end{figure}

Based on the results of SED fitting, in our sample there are nine
candidate members with stellar masses exceeding 10$^{11}$M$_{\odot}$,
six of which are spectroscopic members. These objects are mostly
concentrated close to the cluster center, with 5 of them within a
clustercentric distance $d_{cl}\lesssim$100~kpc. Only two are
classified as actively star-forming based on our criteria: one
candidate at $d_{cl}>600$~kpc, and one component of the proto-BCG
system that from recent analysis seems indeed to be associated with
the cluster (Gobat et al. 2013).  According to our criteria, its SED
is classified as star-forming. While the photometry of this source is
likely significantly affected by the presence of multiple components
and neighbors, a detection in the Herschel PACS imaging indeed
suggests a SFR of order $\sim$100M$_{\odot}$/yr (Gobat et
al. 2013). All the other seven $>$10$^{11}$M$_{\odot}$ candidate
members are classified as passive, four of them with an early-type
morphology, although one is embedded in a large asymmetric halo with
features suggestive of a recent interaction. The remaining three
galaxies classified as passive are {\it i)} a likely member of disky
morphology, {\it ii)} a possible member with distorted shape showing a
large tail, and {\it iii)} a galaxy very close ($\sim$1.5'') to the
star-forming component of the proto-BCG mentioned above, identified in
\citet{gobat2011} as a proto-BCG component itself, and for which we
were not able to obtain a reliable Sersic fit, likely due to its
  complicated surroundings.  The F140W images of these three galaxies
are shown in Figure \ref{fig:cutoutsnoetgs}.

In spite of the mentioned caveats, these observations would thus
picture the high-mass end galaxy population in this cluster as a mix
of passive galaxies with already established early-type
morphologies, and of galaxies which are instead still actively forming
their stars, assembling their mass, or reshaping their structure, in
some cases clearly through interactions.  We note that, with the
exception of the proto-BCG complex, the central region within
$<$150~kpc from the cluster center hosts the most evolved of these
very massive galaxies, while those star-forming or with disk or
irregular morphologies typically lie outside of the cluster core, at
$d_{cl}>350$~kpc.

As already mentioned in section \ref{generalgalaxyprop}, the most
massive galaxies in the core of this cluster already show a very high
passive fraction (83$^{+6}_{-20}$\% for the fully spectroscopically
confirmed sample of M$>$10$^{11}$M$_{\odot}$ members within 150~kpc from
the cluster center), close to the estimate by \citet{raichoor2012} for
very massive galaxies close to the central area of JKCS~041, assumed
to be at roughly similar redshift.

As also found in other studies at this redshift, the fraction of very
massive galaxies which have already attained an early-type morphology is
significantly lower than at lower redshift. Both in our cluster and
$1.5<z<2.5$ field samples, 40$\pm$15\% of galaxies more massive than
$10^{11}$M$_{\odot}$ are classified as morphological early-types
\citep[or 20$^{+20}_{-6}$\% for $\gtrsim2\times10^{11}$M$_{\odot}$, in
  agreement with][]{buitrago2011}. On the other hand, the fraction of
early-types could be larger for $>10^{11}$M$_{\odot}$ {\it passive}
galaxies (60$\pm$15\%, larger than the 35$\pm$15\% estimated by
\citet{vanderwel2011}, but still consistent given the significant
errors).

\begin{figure*}
\centering
\includegraphics[width=.98\textwidth,bb=10 506 677 836,clip]{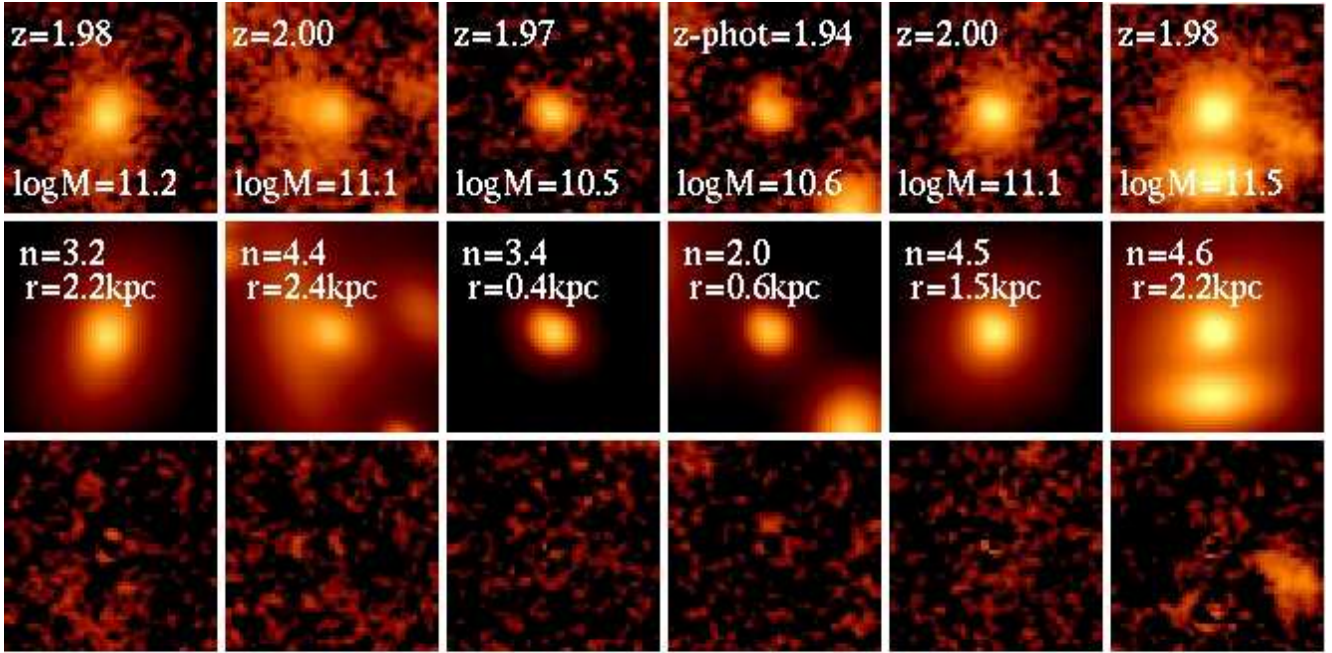}
\caption{WFC3 F140W cutouts of the 6 massive passive members with a
  $n_{Sersic}>2$. The top, middle, and bottom panels show the image,
  model, and residual for these six sources.  Cutout size is 3'', or $\sim$26~kpc at
  z=2. \label{fig:cutoutsetgs}}
\end{figure*}

\subsection{An estimate of the cluster mass from its host galaxies}

As reported in \citet{gobat2011}, the total mass of Cl~J1449+0856 as
inferred from its X-ray luminosity would be $M_{200} =
5.3\pm1$~10$^{13}$M$_{sun}$. \citet{gobat2011} also attempted an
independent estimate of the cluster mass based on the stellar mass
contained in the red galaxies in the very central (20'', $\sim$170
kpc) overdensity. We attempt here a refinement of this estimate based
on the sample of candidate members within a clustercentric radius of
500~kpc. We stress that this only gives a very rough indication of the
cluster mass, since besides the biases related to the selection of
candidate members, which are extensively discussed above, there are
many additional important uncertainties including our ignorance of the
cluster virial radius, and of the redshift evolution up to $z\sim2$ of
the relation between cluster total mass and stellar mass in galaxies.

Given the cluster redshift as well as the previous mass estimates, it
is reasonable to assume that the cluster virial radius is likely not
much larger than $\sim500$ kpc, and thus that a $r<500$ kpc area
accounts for most of the mass in galaxies in this system (as would be
also suggested by Figure \ref{fig:profiles}). We estimate the stellar
mass in galaxies within this area\footnote{We correct for a small
  fraction of uncovered area beyond 300~kpc, see e.g., Figure
  \ref{fig:mapall}.}, keeping into account contamination by
interlopers as discussed above, and extrapolating down to stellar
masses of $10^7$M$_{\odot}$ assuming that the shape of the mass
function is similar to what measured at $1.5<z<2$ by
\citet{ilbert2010}. The stellar mass calculated in this way is
2$\pm1\times$10$^{12}$M$_{\odot}$. Based on this, we then estimate the
cluster mass using its relation (in the nearby Universe) with stellar
mass in galaxies as determined by \citet{andreon2012}. Since as
discussed above we do not know the $r_{500}$ or
$r_{200}$\footnote{As for the usual definition, $r_{500}$ and
    $r_{200}$ are the radii within which the mean density of the
    cluster is 500 and 200 times, respectively, the critical density
    of the Universe at the cluster redshift.} of this cluster, we
apply both local calibrations based on stellar mass within $r_{500}$
and $r_{200}$, in the reasonable assumption that the 500 kpc radius we
use must be between or close to one of them. The two estimates,
$~4-5\times10^{13}$M$_{\odot}$, are consistent given the uncertainty
of at least 50\%. This would correspond to a stellar mass fraction
within the $r<500$ kpc area of $\sim$4-5\%, also in agreement with
other measurements up to $z\sim1$ \citep[][for the same
  IMF]{giodini2009,leauthaud2012}. On the other hand, while there is
currently no evidence for a significant evolution of the stellar mass
fraction in clusters up tp $z\sim1$, there might well be a stronger
evolution between $z\sim1$ and 2. The actual amount of such evolution
is difficult to quantify, and we note just for reference that the
\citet{bower2006} semi-analytic model would predict a slightly lower
stellar mass fraction for group/cluster-sized haloes at $z=2$
\citep{balogh2008}, which would thus mildly increase, by $\sim$30\%,
our estimate for the cluster mass.

In any case, our revised estimate of the cluster mass based on stellar
mass in galaxies is close to previous determinations. We stress
nonetheless once more that, given the significant assumptions and
uncertainties involved, this remains only a crude guess.

\section{Passive early-type galaxies}

According to the criteria described above, down to the completeness
mass limit of $8.5\times10^9$M$_{\odot}$ we identify 8 passive
``likely" members and 4 passive ``possible'' members. As discussed
above and shown in Figure \ref{fig:panoramixall}, these candidate
members -- and in particular those most likely associated with the
cluster -- tend to be located in the cluster core, at a clustercentric
radius of $<$150~kpc. As expected, the fraction of candidate members
classified as passive strongly depends on stellar mass. At masses
below logM/M$_{\odot}<10.5$ passive galaxies seem very rare: we have
only two in our mass-complete sample, making up $15^{+15}_{-5}$\% of
the $9.9<$logM/M$_{\odot}<10.5$ population of cluster
candidates. Statistics are poor, and the exact number could be
affected by contamination and photo-z uncertainties, but there seems
to be a paucity of passive candidate members at low stellar masses in
our sample \citep[see also e.g., among others, ][at lower to similar
  redshifts, and in different
  environments]{kodama2004,delucia2007,ilbert2010,rudnick2012}. The
passive fraction increases at higher masses, getting to
$30^{+20}_{-10}$\% at $10.5<$logM/M$_{\odot}<11$, and up to the
$\sim$80\% for logM/M$_{\odot}>11$ as quoted in section
\ref{sec:massiveend}.

Down to our limit for morphological analysis ($m_{140}<24.5$,
M$>2.5\times10^{10}$M$_{\odot}$), our passive sample contains 8
``likely'' and one ``possible" candidate members. While the surface
brightness distribution of most of these 9 sources may be described
with a $n>2$ Sersic profile and is overall suggestive of an early-type
structure, this sample also includes the three massive
($>10^{11}$M$_{\odot}$) passive systems with disk, distorted or
undetermined morphology, that were discussed in section
\ref{sec:massiveend} (Figure \ref{fig:cutoutsnoetgs}).  These three
galaxies are excluded from the following analysis. Cutouts of the six
remaining bona-fide passive cluster early-types (five spectroscopic
and one likely member) are shown in Figure \ref{fig:cutoutsetgs},
together with their Sersic models and residual maps\footnote{These
  include the early-type galaxy surrounded by a large asymmetric halo
  already mentioned in section \ref{sec:massiveend} -- this source
  will be highlighted below were relevant.}.

\begin{figure*}
\centering
\includegraphics[width=.98\textwidth,bb=10 177 677 506 ,clip]{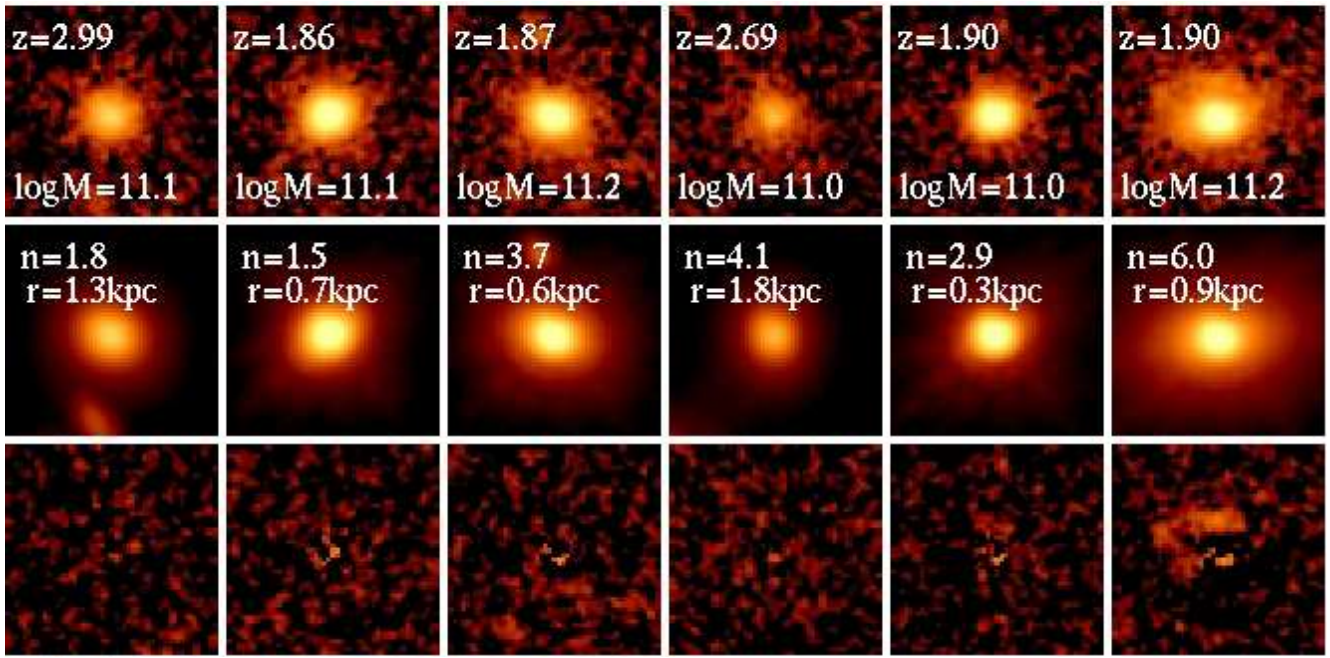}
\caption{WFC3 F140W cutouts of the 6 massive passive interlopers with a
  $n_{Sersic}>1.5$. The top, middle, and bottom panels show the image,
  model, and residual for these six sources.  Cutout size is
  3'', or $\sim$25~kpc in the considered redshift
  range. \label{fig:cutoutsinterlopers}}
\end{figure*}

Analogously, in the redshift range close to the cluster ($1.5<z<3$,
corresponding to $\pm1$Gyr around z=2) we identify 6 (all
spectroscopic) interlopers classified as passive and with a likely
early-type morphology\footnote{We note that, although four of these
  interlopers have a similar redshift $1.86<z<1.90$, the projected
  separation between any two of them is at least 200 kpc, and three
  out of four lie at more than 300 kpc from the cluster center.}.  The
passive nature inferred from their photometry is also confirmed by
their spectra.  Cutouts of the 6 passive interlopers are shown in
Figure \ref{fig:cutoutsinterlopers}, together with their Sersic models
and residual maps. We include in this sample two sources with
1.5$<$n$_{Sersic} <2$, which is below the n$_{Sersic}=2$ threshold
adopted in this work\footnote{There are no passive candidate members
  with a 1.5$<$n$_{Sersic} <2$.}, and one high-n$_{Sersic}$ source
which shows signatures of interaction\footnote{This source is shown in
  the right-hand panels of Figure \ref{fig:cutoutsinterlopers}, and
  appears in Figure \ref{fig:masssize} with a logM/M$_{\odot}$=11.2
  and an effective radius of 0.95~kpc. If fitting the faint component
  north of the source, its Sersic index and size are reduced by
  30-40\%. As Figure \ref{fig:masssize} shows, adopting the lower size
  estimate, or excluding this source (as well as the two n$_{Sersic}
  <2$ systems) from the sample, would not change our conclusions.}.

We note that this sample of interlopers might be biased, since due to
our membership criteria sources in this redshift range are likely to
be classified as candidate members unless they are bright enough, and
possibly compact enough given our grism data, that their redshift can
be measured discarding their association with the cluster. This sample
of field $z\sim2$ passive galaxies might thus be, in principle, not
completely representative of the logM/M$_{\odot}>10.4$ population of
passive galaxies at this redshift. On the other hand, we also note
that at the mass of these interlopers ($\gtrsim10^{11}$M$_{\odot}$)
all the passive cluster early-types are spectroscopically confirmed,
so at least at these masses there is no uncertainty due to membership
determination and we can make a meaningful, homogeneous comparison
between cluster and field passive early-type galaxies in our field.

\subsection{The mass-size relation of passive cluster early-types}

\begin{figure}[ht]
\includegraphics[width=.5\textwidth,bb= 1 310 418 730,clip]{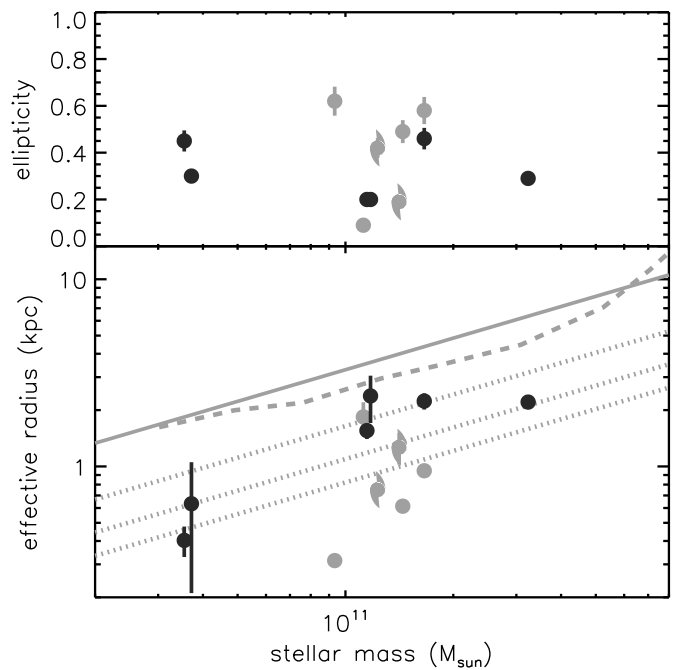}
\caption{The ellipticity and effective radius as a function of stellar
  mass, for cluster (black) and field (gray, $1.5<z<3$) passive
  early-type galaxies in the probed area.  Two field galaxies with
  $1.5<n_{Sersic}<2$ are included, and are shown as spiral symbols.
  In the bottom panel, the solid and dashed lines show, respectively,
  the local determination of the stellar-mass size relation by
  \citet{shen2003} for early-type galaxies, and by
  \citet{valentinuzzi2010} for nearby cluster early-types. The dotted
  lines show for reference the \citet{shen2003} relation scaled by a
  factor 2,3, and 4 in size.  \label{fig:masssize}}
\end{figure}

In Figure \ref{fig:masssize} we show the ellipticity and circularised
effective radius vs stellar mass for the passive candidate members
with $n_{Sersic}>2$, or in fact $>2.5$ for all but one of the plotted sources.
The six passive spectroscopic interlopers at $1.5<z<3$  are also shown, 
including those with $1.5<n_{Sersic}<2$  as spiral symbols.

While the average ellipticity of the cluster early-types tends to be
somewhat lower than for those in the field, statistics are too poor to
draw any significant conclusion concerning environmental dependence as
well as redshift evolution. With this important caveat, we just note
that the median ellipticity $\sim0.3$ of the cluster early-types seems
very similar to what is observed at low redshift
\citep[e.g.,][]{holden2009} in a similar mass range.

All galaxies in Figure \ref{fig:masssize} appear to be more compact
than similarly massive early-types in the nearby Universe, in
agreement with many previous studies at high redshift in clusters and
field (e.g., among many others,
\citet{daddi2005,trujillo2006b,zirm2007,cimatti2008,vanderwel2008,rettura2010,williams2010,cassata2011,cameron2011},
but see also e.g. \citet{saracco2009,onodera2010,mancini2010}).

For comparison, we show in Figure \ref{fig:masssize} the most commonly
used local reference relation \citep{shen2003}, and the determination
by \citet{valentinuzzi2010} for nearby cluster early-types. While the
\citet{shen2003} relation has been shown to be affected by some errors
and biases \citep[e.g.,][]{guo2009,taylor2010,valentinuzzi2010}, due
to its widespread use in previous work we use it as the z=0 reference
to compute size evolution factors, for ease of comparison with other
results. We remind the reader that, while our morphological analisys
is carried out in the restframe optical ($\sim$4700$\AA$), the
\citet{shen2003} sizes are still measured at longer wavelength (z
band), which might raise issues of morphological k-correction,
although up to now this does not seem to be a serious concern for the
kind of sources studied here
\citep{cassata2010,damjanov2011,cameron2011}.

While keeping in mind the small size of our sample, from Figure
\ref{fig:masssize} interlopers seem to have a larger spread in size,
and to be systematically more compact than candidate members of
similar mass\footnote{We note that this is not due to the large
  $1.5<z<3$ bin -- in fact, the four very compact galaxies are very
  close to the cluster redshift at $1.8<z<1.9$, while the two at
  $z>2.5$ have sizes closer to the cluster members.}. As compared to
the \citet{shen2003} relation, cluster early-types have sizes smaller
on average by a factor $r_{e}/r_{e,shen2003}=0.44\pm0.06$ (rms range
$\sim$ 0.2-0.7), while field early-types have an average
$r_{e}/r_{e,shen2003}=0.22\pm0.06$ (rms range $\sim$ 0.1-0.5). This
would support (at least at masses $\gtrsim 10^{11}$M$_{\odot}$) recent
claims on the typically larger sizes of early-types in high-redshift
dense environments
\citep{cooper2012,papovich2012,zirm2012,tanaka2012}.

\begin{figure}[b]
\includegraphics[width=.48\textwidth,bb= 83 366 546 696,clip]{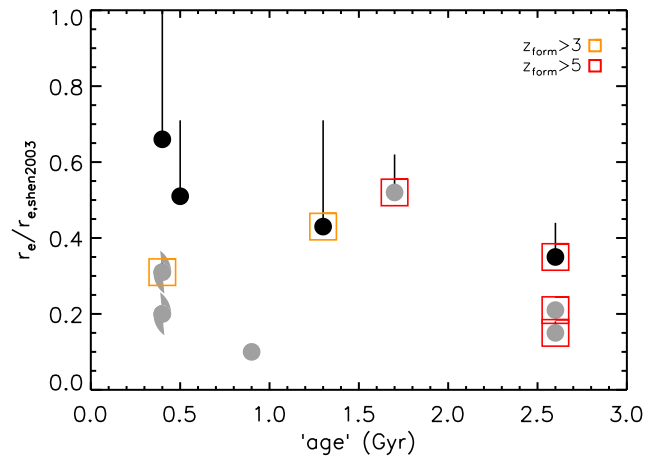}
\caption{The ``age'' vs size evolution factor for the
  $\gtrsim10^{11}$M$_{\sun}$ cluster (black symbols) and field (gray
  symbols) passive early-types (see Figure \ref{fig:masssize}, and the
  text for the definition of the plotted quantities). Black lines are
  not errors on the size determination (see Figure \ref{fig:masssize}),
  but show the offset in evolution factor if using M05 rather than
  BC03 stellar masses, for the same local
  reference \label{fig:agesize}. }
\end{figure}

Although still debated, a correlation has been claimed by several
studies across a broad redshift range, between the size of passive
early-types and the age of their stellar populations, with older
galaxies having smaller sizes, (e.g.,
\citet{bernardi2010,valentinuzzi2010,saracco2011} and references
therein, but see also results in
e.g., \citet{cimatti2012,onodera2012,whitaker2012}). In this respect,
we note that the difference in size between cluster and field
early-types in Figure \ref{fig:masssize} does not seem to reflect a
difference in age. We show in Figure \ref{fig:agesize} the size
evolution factor for the $\gtrsim$10$^{11}$M$_{\sun}$ sources with
respect to the \citet{shen2003} relation, vs. the ``age", defined as
the time when half of the stellar mass at the epoch of observations
was formed (based on the star formation history of the best-fitting
model SED). At M$\gtrsim$10$^{11}$M$_{\sun}$, the cluster and field
samples are comparable in stellar mass, sizes are relatively well
constrained, and both samples are fully spectroscopically
confirmed. Our poor statistics and very rough age estimates do not
allow us to draw any conclusion on the age vs size relation of
early-types in this work, but we cannot see any evidence of
segregation in this figure between cluster and field galaxies, besides
the larges sizes of cluster early-types, already shown in Figure
\ref{fig:masssize}.

Assuming a size evolution of the form (1+z)$^{\alpha}$, $\alpha$ would
be 0.75$\pm$0.15 for cluster early-types, and 1.4$\pm$0.2 for the
field. This amount of size evolution for field early-types at this
redshift would be in close agreement with previous estimates
\citep[e.g.,
][]{vandokkum2008,buitrago2008,vanderwel2008,cassata2011,damjanov2011,cimatti2012,patel2012b},
although other studies have found somewhat milder evolution
\citep[e.g.,][but see discussion below]{cimatti2008,papovich2012}. We
note that such comparison may be biased by the many systematics
affecting the measurement of the mass-size relation, especially at
different redshifts and on different data sets. Nonetheless, taken at
face value, our estimate of the size evolution factor for early-types
is consistent with the expectations from previous measurements.

Interpreting the difference in average size of field and cluster
early-types as evidence that structural evolution is accelerated in
the cluster environment, would suggest that cluster early-types reach
- on average - the observed (at $z\sim2$) size about 3 Gyr earlier
than early-types in the field, {\it assuming a smooth evolution} of
the form given above in the field, down to at least $z\sim0.8$. Note
that, while some work presented evidence for a smooth size evolution
in the $0<z<2$ range \citep{damjanov2011}, other studies suggest that
evolution could be faster before $z \sim2$
\citep[][]{cimatti2012}. With this important caveat, we note for
comparison that a difference in stellar populations has sometimes been
interpreted as a delay in the formation of the bulk of the stars in
field relative to cluster early-types ranging from $\sim0.4$ to 2 Gyr
\citep[e.g.,][all in the nearby
  Universe]{thomas2005,bernardi2005,clemens2006,vandokkum2007,rogers2010}. However,
several other studies, including work at higher redshifts, generally
concluded that {\it if} there is a delay it is small ($\sim0.5$~Gyr),
and often ascribed the slightly different stellar populations to a
more complex difference of star formation histories in different
environments rather than a delay in the bulk of the star formation
\citep[e.g.,][and references
  therein]{moran2005,gobat2008,thomas2010,rettura2011}.

In any case, Figure \ref{fig:masssize} excludes the presence of
extremely compact passive early-types in the cluster, at least in the
mass range and area probed here. We recall however that this
  sample does not include the passive component of the proto-BCG, for
  which we do not have a reliable estimate of morphological
  parameters, as well as the red, compact AGN host that we discarded
  from the spectroscopic member sample due to severe uncertainties in
  the determination of its properties, as discussed in section
  \ref{sec:candidatemembersample}. In principle, either or both might
  be examples of very compact early-types in the cluster core. 
Besides this caveat, cluster passive early-types seem to have sizes
typically a factor $\sim$2-3 smaller than similarly massive
early-types in the nearby Universe.  With the possible exception of
the massive source with asymmetric halo which, as mentioned above, may
suggest a post-interaction stage, there are essentially no passive
early-types within 1$\sigma$ of the local relation. This might be
linked to the still incomplete evolution of a massive, core galaxy
population at this epoch, at least in this cluster.

 On the other hand, the evolution of the mass-size relation does not
 necessarily imply an evolution in size of individual galaxies, and
 its interpretation is complicated by several biases and selection
 effects, as discussed in many studies including e.g.,
 \citet{franx2008,bernardi2010,hopkins2010,williams2010,saracco2009,saracco2010,saracco2011,poggianti2012}.
 In particular, the mismatch between samples of early-types at
 different redshifts is often considered as a significant contribution
 to the observed evolution of the mass size relation, as recently
 summarized by e.g., \citet{carollo2013} and \citet{cassata2013} with
 representative early-type samples up to $z\sim1$ and 3. In fact, as
 observed in the general field and, albeit with some differences, in
 all environments, continuous quenching of star forming galaxies
 through cosmic times significantly increases the number density of
 passive galaxies -- by about an order of magnitude in the field
 between $z\sim2$ and today, in the mass range of our passive sample
 \citep[e.g.][]{ilbert2013,muzzin2013}.  If galaxies quenched more
 recently have typically larger sizes (as for the age-size correlation
 discussed above), the observed mass-size relation evolves even if
 individual early-types in the high-redshift samples do not. Indeed,
 \citet{valentinuzzi2010} showed how early-types with sizes below the
 \citet{shen2003} relation by a factor 2-3 can be found also in nearby
 clusters, even at high stellar masses ($>10^{11}$M$_{\odot}$). Such
 compact galaxies tend to have older stellar populations than
 average-sized ones, and thus made it into the early-type samples at
 earlier times, shifting the average mass-size relation at higher
 redshifts to lower sizes. Attempts to model the effect of such kind
 of progenitor bias on the mass-size relation evolution
 \citep{vanderwel2009,valentinuzzi2010} suggested that, by comparing
 early-type samples at redshift 2 and 0, the observed z=2 mass-size
 relation could be shifted to lower sizes by a factor $\sim30\%$ even
 without any size evolution of the individual z=2 galaxies.  On the
 other hand, we recall that the analysis of size evolution in
 age-controlled samples by \citet{cimatti2012}, albeit affected by
 some caveats\footnote{Besides the intrinsic difficulties in
   estimating galaxy ages, \citet{cimatti2012} used a compilation of
   literature data, thus age measurements were not uniform across
   their sample - see original paper for details.}, might suggest that
 the effect of this bias could instead be relatively mild.  Although
 such amount of progenitor bias would be, in any case, insufficient to
 fully explain our observed size evolution, it still complicates the
 quantification of the relevance of size evolution for individual
 galaxies, especially when coupled to other biases and systematics on
 the determination of sizes \citep[e.g.,][]{pannella2009b,mancini2010}
 and stellar masses (IMF, stellar population models, etc.). In this
 respect, we note that stellar masses estimated with the M05 models
 for the sample of early-types in Figure \ref{fig:masssize} are lower
 by, on average, about a factor 2, thus decreasing the average
 evolution factor ($r_{e}/r_{e,shen2003} \sim 0.7$ rather than $\sim
 0.4$ with BC03 masses, see also Figure \ref{fig:agesize}).

\section{Summary}

We have studied galaxy populations in the field of the z=2 galaxy
cluster Cl J1449+0856, using samples of (candidate) members selected
through spectroscopic and photometric redshifts. Our mass completeness
limit is about 10$^{10}$M$_{\odot}$ (or
2.5$\times$10$^{10}$M$_{\odot}$ where morphological analysis is
involved) at the cluster redshift, thus probing the massive population
of cluster galaxies.

We summarise below our main results: 

$\bullet$ In spite of the residual contamination from field galaxies,
which is expected to be relevant especially at low masses, the cluster
clearly stands out as an overdensity both in the redshift distribution
and in the projected distribution of galaxies in the sky, close to the
center of the extended X-ray emission. In the central $r<100$~kpc
region, the projected number density of cluster galaxies more massive
than $\sim10^{10}$M$_{\odot}$ is estimated to exceed 100
galaxies/Mpc$^2$, for a stellar mass density exceeding
10$^{13}$M$_{\sun}$/Mpc$^2$.

$\bullet$ The highest density cluster core is already traced by a
population of massive, quiescent, early-type galaxies. On the other
hand, massive star forming galaxies, often significantly dust
reddened, also populate the cluster core, as observed in other $z>1.5$
clusters. It thus appears that the core of Cl J1449+0856 might be in a
transitional phase, where a population of already massive and passive
early-types coexists with galaxies still actively forming their stars,
and in some cases reshaping their structure through interactions or
merging.

$\bullet$ Besides the central overdensity which hosts the most massive
and evolved galaxy populations, a secondary galaxy concentration at
$\sim$250~kpc seems to host galaxies of clearly distinct nature, with
lower masses, on-going star formation, and late-type morphologies.

$\bullet$ Environmental signatures on galaxy populations are evident
within $\sim$200~kpc from the cluster center, where the great majority
of morphological early-types and of passive galaxies are
concentratred, resulting in a clear increase of the passive fraction
of massive galaxies. However, at the highest masses
($>10^{11}$M$_{\sun}$) the passive fraction is closer to (and
consistent with, given our uncertainties) the field level, which might
suggest a predominant role of mass over environment quenching for most
massive galaxies at this redshift and in this kind of environment.

$\bullet$ A first analysis of the spatial distribution of galaxies
around the cluster center would suggest a profile shape overall
similar to what observed in nearby clusters, consistent with a
small-core $\beta$ model with $\beta\sim0.9$.

$\bullet$ From the estimated stellar mass in galaxies, and using the
relation between stellar mass and total mass of groups and clusters in
the nearby Universe, we obtain an indicative estimate for the cluster mass
of about $5\times10^{13}$M$_{\sun}$, consistent with the mass inferred
from the X-ray emission.

$\bullet$ We observe a clear correlation between an early-type
morphology and passive stellar populations, as also observed in lower
density environments at similar redshift.

$\bullet$ Massive passive early-types in this cluster are smaller on
average by a factor 2-3 with respect to the \citet{shen2003}
determination of the $z=0$ stellar mass vs. size relation. However,
they appear typically larger by about a factor 2 than similarly
massive field galaxies at the same epoch. While statistics are still
very limited, this would lend support to recent claims of accelerated
structural evolution in high-redshift overdense environments.

This study pictures Cl 1449+0856 as a still-forming cluster which
retains some expected characteristics of low-mass systems at early
times, including massive galaxies still actively forming close to its
center, and likely infalling substructures accreting onto the central
regions lower mass, less evolved galaxies. On the other hand it shows
how, at the same time, early formed massive galaxies, quite evolved
both in their structure and in their stellar content, are a major
component of galaxy populations in cluster cores already 10 billion
years ago.

\acknowledgments We thank Maurilio Pannella, Anna Cibinel, Mark
Sargent, Matthieu B\'ethermin, Gabriella De Lucia, and Stefano Andreon
for valuable inputs, suggestions or comments at various stages of this
work. We also thank the referee for constructive feedback which
improved the presentation of this study.  VS, RG, and ED were
supported by grants ERC-StG UPGAL 240039 and ANR-08-JCJC-0008. AC
acknowledges the grants ASI n.I/023/12/0 "Attivit\`a relative alla
fase B2/C per la missione Euclid" and MIUR PRIN 2010-2011 "The dark
Universe and the cosmic evolution of baryons: from current surveys to
Euclid". Partly based on data collected at the Subaru Telescope,
operated by the National Astronomical Observatory of Japan, and at the
Very Large Telescope, operated by the European Southern
Observatory. Partly based on observations made under program GO-11648
with the NASA/ESA Hubble Space Telescope, which is operated by the
Association of Universities for Research in Astronomy, Inc., under
NASA contract NAS 5-26555. Partly based on observations made under
programs GTO-64 and GO-80103 with the Spitzer Space Telescope, which
is operated by the Jet Propulsion Laboratory, California Institute of
Technology under a contract with NASA."

\bibliography{stra1220}

\begin{thebibliography}{132}
\expandafter\ifx\csname natexlab\endcsname\relax\def\natexlab#1{#1}\fi

\bibitem[{{Andreon}(2008)}]{andreon2008}
{Andreon}, S. 2008, \mnras, 386, 1045

\bibitem[{{Andreon}(2012)}]{andreon2012}
{Andreon}, S. 2012, \aap, 548, A83

\bibitem[{{Baldry} {et~al.}(2006){Baldry}, {Balogh}, {Bower}, {Glazebrook},
  {Nichol}, {Bamford}, \& {Budavari}}]{baldry2006}
{Baldry}, I.~K., {Balogh}, M.~L., {Bower}, R.~G., {et~al.} 2006, \mnras, 373,
  469

\bibitem[{{Balogh} {et~al.}(2008){Balogh}, {McCarthy}, {Bower}, \&
  {Eke}}]{balogh2008}
{Balogh}, M.~L., {McCarthy}, I.~G., {Bower}, R.~G., \& {Eke}, V.~R. 2008,
  \mnras, 385, 1003

\bibitem[{{Bell} {et~al.}(2012){Bell}, {van der Wel}, {Papovich}, {Kocevski},
  {Lotz}, {McIntosh}, {Kartaltepe}, {Faber}, {Ferguson}, {Koekemoer}, {Grogin},
  {Wuyts}, {Cheung}, {Conselice}, {Dekel}, {Dunlop}, {Giavalisco},
  {Herrington}, {Koo}, {McGrath}, {de Mello}, {Rix}, {Robaina}, \&
  {Williams}}]{bell2012}
{Bell}, E.~F., {van der Wel}, A., {Papovich}, C., {et~al.} 2012, \apj, 753, 167

\bibitem[{{Bernardi} {et~al.}(2010){Bernardi}, {Shankar}, {Hyde}, {Mei},
  {Marulli}, \& {Sheth}}]{bernardi2010}
{Bernardi}, M., {Shankar}, F., {Hyde}, J.~B., {et~al.} 2010, \mnras, 436

\bibitem[{{Bernardi} {et~al.}(2005){Bernardi}, {Sheth}, {Nichol}, {Schneider},
  \& {Brinkmann}}]{bernardi2005}
{Bernardi}, M., {Sheth}, R.~K., {Nichol}, R.~C., {Schneider}, D.~P., \&
  {Brinkmann}, J. 2005, \aj, 129, 61

\bibitem[{{Bertin} \& {Arnouts}(1996)}]{sextractor}
{Bertin}, E. \& {Arnouts}, S. 1996, \aaps, 117, 393

\bibitem[{{Biviano} \& {Girardi}(2003)}]{biviano2003}
{Biviano}, A. \& {Girardi}, M. 2003, \apj, 585, 205

\bibitem[{{Bower} {et~al.}(2006){Bower}, {Benson}, {Malbon}, {Helly}, {Frenk},
  {Baugh}, {Cole}, \& {Lacey}}]{bower2006}
{Bower}, R.~G., {Benson}, A.~J., {Malbon}, R., {et~al.} 2006, \mnras, 370, 645

\bibitem[{{Brainerd} {et~al.}(1996){Brainerd}, {Blandford}, \&
  {Smail}}]{brainerd1996}
{Brainerd}, T.~G., {Blandford}, R.~D., \& {Smail}, I. 1996, \apj, 466, 623

\bibitem[{{Brammer} {et~al.}(2008){Brammer}, {van Dokkum}, \& {Coppi}}]{eazy}
{Brammer}, G.~B., {van Dokkum}, P.~G., \& {Coppi}, P. 2008, \apj, 686, 1503

\bibitem[{{Brammer} {et~al.}(2011){Brammer}, {Whitaker}, {van Dokkum},
  {Marchesini}, {Franx}, {Kriek}, {Labb{\'e}}, {Lee}, {Muzzin}, {Quadri},
  {Rudnick}, \& {Williams}}]{brammer2011}
{Brammer}, G.~B., {Whitaker}, K.~E., {van Dokkum}, P.~G., {et~al.} 2011, \apj,
  739, 24

\bibitem[{{Bruzual} \& {Charlot}(2003)}]{bc03}
{Bruzual}, G. \& {Charlot}, S. 2003, \mnras, 344, 1000

\bibitem[{{Buitrago} {et~al.}(2008){Buitrago}, {Trujillo}, {Conselice},
  {Bouwens}, {Dickinson}, \& {Yan}}]{buitrago2008}
{Buitrago}, F., {Trujillo}, I., {Conselice}, C.~J., {et~al.} 2008, \apjl, 687,
  L61

\bibitem[{{Buitrago} {et~al.}(2011){Buitrago}, {Trujillo}, {Conselice}, \&
  {Haeussler}}]{buitrago2011}
{Buitrago}, F., {Trujillo}, I., {Conselice}, C.~J., \& {Haeussler}, B. 2011,
  ArXiv e-prints

\bibitem[{{Calzetti} {et~al.}(2000){Calzetti}, {Armus}, {Bohlin}, {Kinney},
  {Koornneef}, \& {Storchi-Bergmann}}]{calzetti2000}
{Calzetti}, D., {Armus}, L., {Bohlin}, R.~C., {et~al.} 2000, \apj, 533, 682

\bibitem[{{Cameron}(2011)}]{cameron2011p}
{Cameron}, E. 2011, PASA, 28, 128

\bibitem[{{Cameron} {et~al.}(2011){Cameron}, {Carollo}, {Oesch}, {Bouwens},
  {Illingworth}, {Trenti}, {Labb{\'e}}, \& {Magee}}]{cameron2011}
{Cameron}, E., {Carollo}, C.~M., {Oesch}, P.~A., {et~al.} 2011, \apj, 743, 146

\bibitem[{{Capak} {et~al.}(2007){Capak}, {Aussel}, {Ajiki}, {McCracken},
  {Mobasher}, {Scoville}, {Shopbell}, {Taniguchi}, {Thompson}, {Tribiano},
  {Sasaki}, {Blain}, {Brusa}, {Carilli}, {Comastri}, {Carollo}, {Cassata},
  {Colbert}, {Ellis}, {Elvis}, {Giavalisco}, {Green}, {Guzzo}, {Hasinger},
  {Ilbert}, {Impey}, {Jahnke}, {Kartaltepe}, {Kneib}, {Koda}, {Koekemoer},
  {Komiyama}, {Leauthaud}, {Le Fevre}, {Lilly}, {Liu}, {Massey}, {Miyazaki},
  {Murayama}, {Nagao}, {Peacock}, {Pickles}, {Porciani}, {Renzini}, {Rhodes},
  {Rich}, {Salvato}, {Sanders}, {Scarlata}, {Schiminovich}, {Schinnerer},
  {Scodeggio}, {Sheth}, {Shioya}, {Tasca}, {Taylor}, {Yan}, \&
  {Zamorani}}]{capak2007}
{Capak}, P., {Aussel}, H., {Ajiki}, M., {et~al.} 2007, \apjs, 172, 99

\bibitem[{{Carollo} {et~al.}(2013){Carollo}, {Bschorr}, {Renzini}, {Lilly},
  {Capak}, {Cibinel}, {Ilbert}, {Onodera}, {Scoville}, {Cameron}, {Mobasher},
  {Sanders}, \& {Taniguchi}}]{carollo2013}
{Carollo}, C.~M., {Bschorr}, T.~J., {Renzini}, A., {et~al.} 2013, ArXiv
  e-prints

\bibitem[{{Cassata} {et~al.}(2010){Cassata}, {Giavalisco}, {Guo}, {Ferguson},
  {Koekemoer}, {Renzini}, {Fontana}, {Salimbeni}, {Dickinson}, {Casertano},
  {Conselice}, {Grogin}, {Lotz}, {Papovich}, {Lucas}, {Straughn}, {Gardner}, \&
  {Moustakas}}]{cassata2010}
{Cassata}, P., {Giavalisco}, M., {Guo}, Y., {et~al.} 2010, \apjl, 714, L79

\bibitem[{{Cassata} {et~al.}(2011){Cassata}, {Giavalisco}, {Guo}, {Renzini},
  {Ferguson}, {Koekemoer}, {Salimbeni}, {Scarlata}, {Grogin}, {Conselice},
  {Dahlen}, {Lotz}, {Dickinson}, \& {Lin}}]{cassata2011}
{Cassata}, P., {Giavalisco}, M., {Guo}, Y., {et~al.} 2011, \apj, 743, 96

\bibitem[{{Cassata} {et~al.}(2013){Cassata}, {Giavalisco}, {Williams}, {Guo},
  {Lee}, {Renzini}, {Ferguson}, {Faber}, {Barro}, {McIntosh}, {Lu}, {Bell},
  {Koo}, {Papovich}, {Ryan}, {Conselice}, {Grogin}, {Koekemoer}, \&
  {Hathi}}]{cassata2013}
{Cassata}, P., {Giavalisco}, M., {Williams}, C.~C., {et~al.} 2013, ArXiv
  e-prints

\bibitem[{{Cavaliere} \& {Fusco-Femiano}(1978)}]{cavaliere1978}
{Cavaliere}, A. \& {Fusco-Femiano}, R. 1978, \aap, 70, 677

\bibitem[{{Cimatti} {et~al.}(2008){Cimatti}, {Cassata}, {Pozzetti}, {Kurk},
  {Mignoli}, {Renzini}, {Daddi}, {Bolzonella}, {Brusa}, {Rodighiero},
  {Dickinson}, {Franceschini}, {Zamorani}, {Berta}, {Rosati}, \&
  {Halliday}}]{cimatti2008}
{Cimatti}, A., {Cassata}, P., {Pozzetti}, L., {et~al.} 2008, \aap, 482, 21

\bibitem[{{Cimatti} {et~al.}(2012){Cimatti}, {Nipoti}, \&
  {Cassata}}]{cimatti2012}
{Cimatti}, A., {Nipoti}, C., \& {Cassata}, P. 2012, \mnras, 422, L62

\bibitem[{{Clemens} {et~al.}(2006){Clemens}, {Bressan}, {Nikolic}, {Alexander},
  {Annibali}, \& {Rampazzo}}]{clemens2006}
{Clemens}, M.~S., {Bressan}, A., {Nikolic}, B., {et~al.} 2006, \mnras, 370, 702

\bibitem[{{Cooper} {et~al.}(2012){Cooper}, {Griffith}, {Newman}, {Coil},
  {Davis}, {Dutton}, {Faber}, {Guhathakurta}, {Koo}, {Lotz}, {Weiner},
  {Willmer}, \& {Yan}}]{cooper2012}
{Cooper}, M.~C., {Griffith}, R.~L., {Newman}, J.~A., {et~al.} 2012, \mnras,
  419, 3018

\bibitem[{{Daddi} {et~al.}(2004){Daddi}, {Cimatti}, {Renzini}, {Fontana},
  {Mignoli}, {Pozzetti}, {Tozzi}, \& {Zamorani}}]{daddi2004}
{Daddi}, E., {Cimatti}, A., {Renzini}, A., {et~al.} 2004, \apj, 617, 746

\bibitem[{{Daddi} {et~al.}(2005{\natexlab{a}}){Daddi}, {Dickinson}, {Chary},
  {Pope}, {Morrison}, {Alexander}, {Bauer}, {Brandt}, {Giavalisco}, {Ferguson},
  {Lee}, {Lehmer}, {Papovich}, \& {Renzini}}]{daddi2005b}
{Daddi}, E., {Dickinson}, M., {Chary}, R., {et~al.} 2005{\natexlab{a}}, \apjl,
  631, L13

\bibitem[{{Daddi} {et~al.}(2005{\natexlab{b}}){Daddi}, {Renzini}, {Pirzkal},
  {Cimatti}, {Malhotra}, {Stiavelli}, {Xu}, {Pasquali}, {Rhoads}, {Brusa}, {di
  Serego Alighieri}, {Ferguson}, {Koekemoer}, {Moustakas}, {Panagia}, \&
  {Windhorst}}]{daddi2005}
{Daddi}, E., {Renzini}, A., {Pirzkal}, N., {et~al.} 2005{\natexlab{b}}, \apj,
  626, 680

\bibitem[{{Damjanov} {et~al.}(2011){Damjanov}, {Abraham}, {Glazebrook},
  {McCarthy}, {Caris}, {Carlberg}, {Chen}, {Crampton}, {Green}, {J{\o}rgensen},
  {Juneau}, {Le Borgne}, {Marzke}, {Mentuch}, {Murowinski}, {Roth}, {Savaglio},
  \& {Yan}}]{damjanov2011}
{Damjanov}, I., {Abraham}, R.~G., {Glazebrook}, K., {et~al.} 2011, \apjl, 739,
  L44

\bibitem[{{De Lucia} {et~al.}(2007){De Lucia}, {Poggianti},
  {Arag{\'o}n-Salamanca}, {White}, {Zaritsky}, {Clowe}, {Halliday}, {Jablonka},
  {von der Linden}, {Milvang-Jensen}, {Pell{\'o}}, {Rudnick}, {Saglia}, \&
  {Simard}}]{delucia2007}
{De Lucia}, G., {Poggianti}, B.~M., {Arag{\'o}n-Salamanca}, A., {et~al.} 2007,
  \mnras, 374, 809

\bibitem[{{De Lucia} {et~al.}(2006){De Lucia}, {Springel}, {White}, {Croton},
  \& {Kauffmann}}]{delucia2006}
{De Lucia}, G., {Springel}, V., {White}, S.~D.~M., {Croton}, D., \&
  {Kauffmann}, G. 2006, \mnras, 366, 499

\bibitem[{{De Propris} {et~al.}(2007){De Propris}, {Stanford}, {Eisenhardt},
  {Holden}, \& {Rosati}}]{depropris2007}
{De Propris}, R., {Stanford}, S.~A., {Eisenhardt}, P.~R., {Holden}, B.~P., \&
  {Rosati}, P. 2007, \aj, 133, 2209

\bibitem[{{Dressler}(1980)}]{dressler1980}
{Dressler}, A. 1980, \apj, 236, 351

\bibitem[{{Eisenhardt} {et~al.}(2008){Eisenhardt}, {Brodwin}, {Gonzalez},
  {Stanford}, {Stern}, {Barmby}, {Brown}, {Dawson}, {Dey}, {Doi}, {Galametz},
  {Jannuzi}, {Kochanek}, {Meyers}, {Morokuma}, \& {Moustakas}}]{eisenhardt2008}
{Eisenhardt}, P.~R.~M., {Brodwin}, M., {Gonzalez}, A.~H., {et~al.} 2008, \apj,
  684, 905

\bibitem[{{Fassbender} {et~al.}(2011{\natexlab{a}}){Fassbender},
  {B{\"o}hringer}, {Nastasi}, {{\v S}uhada}, {M{\"u}hlegger}, {de Hoon},
  {Kohnert}, {Lamer}, {Mohr}, {Pierini}, {Pratt}, {Quintana}, {Rosati},
  {Santos}, \& {Schwope}}]{fassbender2011b}
{Fassbender}, R., {B{\"o}hringer}, H., {Nastasi}, A., {et~al.}
  2011{\natexlab{a}}, New Journal of Physics, 13, 125014

\bibitem[{{Fassbender} {et~al.}(2011{\natexlab{b}}){Fassbender}, {Nastasi},
  {B{\"o}hringer}, {{\v S}uhada}, {Santos}, {Rosati}, {Pierini},
  {M{\"u}hlegger}, {Quintana}, {Schwope}, {Lamer}, {de Hoon}, {Kohnert},
  {Pratt}, \& {Mohr}}]{fassbender2011}
{Fassbender}, R., {Nastasi}, A., {B{\"o}hringer}, H., {et~al.}
  2011{\natexlab{b}}, \aap, 527, L10

\bibitem[{{Franx} {et~al.}(2008){Franx}, {van Dokkum}, {Schreiber}, {Wuyts},
  {Labb{\'e}}, \& {Toft}}]{franx2008}
{Franx}, M., {van Dokkum}, P.~G., {Schreiber}, N.~M.~F., {et~al.} 2008, \apj,
  688, 770

\bibitem[{{George} {et~al.}(2012){George}, {Leauthaud}, {Bundy}, {Finoguenov},
  {Ma}, {Rykoff}, {Tinker}, {Wechsler}, {Massey}, \& {Mei}}]{george2012}
{George}, M.~R., {Leauthaud}, A., {Bundy}, K., {et~al.} 2012, \apj, 757, 2

\bibitem[{{Giodini} {et~al.}(2009){Giodini}, {Pierini}, {Finoguenov}, {Pratt},
  {Boehringer}, {Leauthaud}, {Guzzo}, {Aussel}, {Bolzonella}, {Capak}, {Elvis},
  {Hasinger}, {Ilbert}, {Kartaltepe}, {Koekemoer}, {Lilly}, {Massey},
  {McCracken}, {Rhodes}, {Salvato}, {Sanders}, {Scoville}, {Sasaki}, {Smolcic},
  {Taniguchi}, {Thompson}, \& {the COSMOS Collaboration}}]{giodini2009}
{Giodini}, S., {Pierini}, D., {Finoguenov}, A., {et~al.} 2009, \apj, 703, 982

\bibitem[{{Girardi} {et~al.}(1998){Girardi}, {Giuricin}, {Mardirossian},
  {Mezzetti}, \& {Boschin}}]{girardi1998}
{Girardi}, M., {Giuricin}, G., {Mardirossian}, F., {Mezzetti}, M., \&
  {Boschin}, W. 1998, \apj, 505, 74

\bibitem[{{Gobat} {et~al.}(2011){Gobat}, {Daddi}, {Onodera}, {Finoguenov},
  {Renzini}, {Arimoto}, {Bouwens}, {Brusa}, {Chary}, {Cimatti}, {Dickinson},
  {Kong}, \& {Mignoli}}]{gobat2011}
{Gobat}, R., {Daddi}, E., {Onodera}, M., {et~al.} 2011, \aap, 526, A133

\bibitem[{{Gobat} {et~al.}(2008){Gobat}, {Rosati}, {Strazzullo}, {Rettura},
  {Demarco}, \& {Nonino}}]{gobat2008}
{Gobat}, R., {Rosati}, P., {Strazzullo}, V., {et~al.} 2008, \aap, 488, 853

\bibitem[{{Guo} {et~al.}(2009){Guo}, {McIntosh}, {Mo}, {Katz}, {van den Bosch},
  {Weinberg}, {Weinmann}, {Pasquali}, \& {Yang}}]{guo2009}
{Guo}, Y., {McIntosh}, D.~H., {Mo}, H.~J., {et~al.} 2009, \mnras, 398, 1129

\bibitem[{{Hatch} {et~al.}(2011){Hatch}, {Kurk}, {Pentericci}, {Venemans},
  {Kuiper}, {Miley}, \& {R{\"o}ttgering}}]{hatch2011}
{Hatch}, N.~A., {Kurk}, J.~D., {Pentericci}, L., {et~al.} 2011, \mnras, 415,
  2993

\bibitem[{{Hayashi} {et~al.}(2011){Hayashi}, {Kodama}, {Koyama}, {Tadaki}, \&
  {Tanaka}}]{hayashi2011}
{Hayashi}, M., {Kodama}, T., {Koyama}, Y., {Tadaki}, K.-I., \& {Tanaka}, I.
  2011, \mnras, 415, 2670

\bibitem[{{Hayashi} {et~al.}(2010){Hayashi}, {Kodama}, {Koyama}, {Tanaka},
  {Shimasaku}, \& {Okamura}}]{hayashi2010}
{Hayashi}, M., {Kodama}, T., {Koyama}, Y., {et~al.} 2010, \mnras, 402, 1980

\bibitem[{{Hilton} {et~al.}(2010){Hilton}, {Lloyd-Davies}, {Stanford}, {Stott},
  {Collins}, {Romer}, {Hosmer}, {Hoyle}, {Kay}, {Liddle}, {Mehrtens}, {Miller},
  {Sahl{\'e}n}, \& {Viana}}]{hilton2010}
{Hilton}, M., {Lloyd-Davies}, E., {Stanford}, S.~A., {et~al.} 2010, \apj, 718,
  133

\bibitem[{{Hoaglin} {et~al.}(1983){Hoaglin}, {Mosteller}, \&
  {Tukey}}]{hoaglin1983}
{Hoaglin}, D.~C., {Mosteller}, F., \& {Tukey}, J.~W. 1983, {Understanding
  robust and exploratory data anlysis} (Wiley Series in Probability and
  Mathematical Statistics, New York: Wiley, 1983, edited by Hoaglin, David C.;
  Mosteller, Frederick; Tukey, John W.)

\bibitem[{{Holden} {et~al.}(2009){Holden}, {Franx}, {Illingworth}, {Postman},
  {van der Wel}, {Kelson}, {Blakeslee}, {Ford}, {Demarco}, \&
  {Mei}}]{holden2009}
{Holden}, B.~P., {Franx}, M., {Illingworth}, G.~D., {et~al.} 2009, \apj, 693,
  617

\bibitem[{{Hopkins} {et~al.}(2010){Hopkins}, {Bundy}, {Hernquist}, {Wuyts}, \&
  {Cox}}]{hopkins2010}
{Hopkins}, P.~F., {Bundy}, K., {Hernquist}, L., {Wuyts}, S., \& {Cox}, T.~J.
  2010, \mnras, 401, 1099

\bibitem[{{Ilbert} {et~al.}(2009){Ilbert}, {Capak}, {Salvato}, {Aussel},
  {McCracken}, {Sanders}, {Scoville}, {Kartaltepe}, {Arnouts}, {LeFloc'h},
  {Mobasher}, {Taniguchi}, {Lamareille}, {Leauthaud}, {Sasaki}, {Thompson},
  {Zamojski}, {Zamorani}, {Bardelli}, {Bolzonella}, {Bongiorno}, {Brusa},
  {Caputi}, {Carollo}, {Contini}, {Cook}, {Coppa}, {Cucciati}, {de la Torre},
  {de Ravel}, {Franzetti}, {Garilli}, {Hasinger}, {Iovino}, {Kampczyk},
  {Kneib}, {Knobel}, {Kovac}, {LeBorgne}, {LeBrun}, {LeF{\`e}vre}, {Lilly},
  {Looper}, {Maier}, {Mainieri}, {Mellier}, {Mignoli}, {Murayama}, {Pell{\`o}},
  {Peng}, {P{\'e}rez-Montero}, {Renzini}, {Ricciardelli}, {Schiminovich},
  {Scodeggio}, {Shioya}, {Silverman}, {Surace}, {Tanaka}, {Tasca}, {Tresse},
  {Vergani}, \& {Zucca}}]{ilbert2009}
{Ilbert}, O., {Capak}, P., {Salvato}, M., {et~al.} 2009, \apj, 690, 1236

\bibitem[{{Ilbert} {et~al.}(2013){Ilbert}, {McCracken}, {Le Fevre}, {Capak},
  {Dunlop}, {Karim}, {Renzini}, {Caputi}, {Boissier}, {Arnouts}, {Aussel},
  {Comparat}, {Guo}, {Hudelot}, {Kartaltepe}, {Kneib}, {Krogager}, {Le Floc'h},
  {Lilly}, {Mellier}, {Milvang-Jensen}, {Moutard}, {Onodera}, {Richard},
  {Salvato}, {Sanders}, {Scoville}, {Silverman}, {Taniguchi}, {Tasca},
  {Thomas}, {Toft}, {Tresse}, {Vergani}, {Wolk}, \& {Zirm}}]{ilbert2013}
{Ilbert}, O., {McCracken}, H.~J., {Le Fevre}, O., {et~al.} 2013, ArXiv e-prints

\bibitem[{{Ilbert} {et~al.}(2010){Ilbert}, {Salvato}, {Le Floc'h}, {Aussel},
  {Capak}, {McCracken}, {Mobasher}, {Kartaltepe}, {Scoville}, {Sanders},
  {Arnouts}, {Bundy}, {Cassata}, {Kneib}, {Koekemoer}, {Le F{\`e}vre}, {Lilly},
  {Surace}, {Taniguchi}, {Tasca}, {Thompson}, {Tresse}, {Zamojski}, {Zamorani},
  \& {Zucca}}]{ilbert2010}
{Ilbert}, O., {Salvato}, M., {Le Floc'h}, E., {et~al.} 2010, \apj, 709, 644

\bibitem[{{Johansson} {et~al.}(2012){Johansson}, {Naab}, \&
  {Ostriker}}]{johansson2012}
{Johansson}, P.~H., {Naab}, T., \& {Ostriker}, J.~P. 2012, \apj, 754, 115

\bibitem[{{Kodama} {et~al.}(2007){Kodama}, {Tanaka}, {Kajisawa}, {Kurk},
  {Venemans}, {De Breuck}, {Vernet}, \& {Lidman}}]{kodama2007}
{Kodama}, T., {Tanaka}, I., {Kajisawa}, M., {et~al.} 2007, \mnras, 377, 1717

\bibitem[{{Kodama} {et~al.}(2004){Kodama}, {Yamada}, {Akiyama}, {Aoki}, {Doi},
  {Furusawa}, {Fuse}, {Imanishi}, {Ishida}, {Iye}, {Kajisawa}, {Karoji},
  {Kobayashi}, {Komiyama}, {Kosugi}, {Maeda}, {Miyazaki}, {Mizumoto},
  {Morokuma}, {Nakata}, {Noumaru}, {Ogasawara}, {Ouchi}, {Sasaki}, {Sekiguchi},
  {Shimasaku}, {Simpson}, {Takata}, {Tanaka}, {Ueda}, {Yasuda}, \&
  {Yoshida}}]{kodama2004}
{Kodama}, T., {Yamada}, T., {Akiyama}, M., {et~al.} 2004, \mnras, 350, 1005

\bibitem[{{Kriek} {et~al.}(2009){Kriek}, {van Dokkum}, {Labb{\'e}}, {Franx},
  {Illingworth}, {Marchesini}, \& {Quadri}}]{fast}
{Kriek}, M., {van Dokkum}, P.~G., {Labb{\'e}}, I., {et~al.} 2009, \apj, 700,
  221

\bibitem[{{Kurk} {et~al.}(2009){Kurk}, {Cimatti}, {Zamorani}, {Halliday},
  {Mignoli}, {Pozzetti}, {Daddi}, {Rosati}, {Dickinson}, {Bolzonella},
  {Cassata}, {Renzini}, {Franceschini}, {Rodighiero}, \& {Berta}}]{kurk2009}
{Kurk}, J., {Cimatti}, A., {Zamorani}, G., {et~al.} 2009, \aap, 504, 331

\bibitem[{{Leauthaud} {et~al.}(2012){Leauthaud}, {George}, {Behroozi}, {Bundy},
  {Tinker}, {Wechsler}, {Conroy}, {Finoguenov}, \& {Tanaka}}]{leauthaud2012}
{Leauthaud}, A., {George}, M.~R., {Behroozi}, P.~S., {et~al.} 2012, \apj, 746,
  95

\bibitem[{{Lemze} {et~al.}(2009){Lemze}, {Broadhurst}, {Rephaeli}, {Barkana},
  \& {Umetsu}}]{lemze2009}
{Lemze}, D., {Broadhurst}, T., {Rephaeli}, Y., {Barkana}, R., \& {Umetsu}, K.
  2009, \apj, 701, 1336

\bibitem[{{Lidman} {et~al.}(2008){Lidman}, {Rosati}, {Tanaka}, {Strazzullo},
  {Demarco}, {Mullis}, {Ageorges}, {Kissler-Patig}, {Petr-Gotzens}, \&
  {Selman}}]{lidman2008}
{Lidman}, C., {Rosati}, P., {Tanaka}, M., {et~al.} 2008, \aap, 489, 981

\bibitem[{{Mancini} {et~al.}(2010){Mancini}, {Daddi}, {Renzini}, {Salmi},
  {McCracken}, {Cimatti}, {Onodera}, {Salvato}, {Koekemoer}, {Aussel}, {Le
  Floc'h}, {Willott}, \& {Capak}}]{mancini2010}
{Mancini}, C., {Daddi}, E., {Renzini}, A., {et~al.} 2010, \mnras, 401, 933

\bibitem[{{Mancone} {et~al.}(2010){Mancone}, {Gonzalez}, {Brodwin}, {Stanford},
  {Eisenhardt}, {Stern}, \& {Jones}}]{mancone2010}
{Mancone}, C.~L., {Gonzalez}, A.~H., {Brodwin}, M., {et~al.} 2010, \apj, 720,
  284

\bibitem[{{Maraston}(2005)}]{maraston2005}
{Maraston}, C. 2005, \mnras, 362, 799

\bibitem[{{Maraston} {et~al.}(2010){Maraston}, {Pforr}, {Renzini}, {Daddi},
  {Dickinson}, {Cimatti}, \& {Tonini}}]{maraston2010}
{Maraston}, C., {Pforr}, J., {Renzini}, A., {et~al.} 2010, \mnras, 407, 830

\bibitem[{{McLaughlin}(1999)}]{mclaughlin1999}
{McLaughlin}, D.~E. 1999, \aj, 117, 2398

\bibitem[{{Mei} {et~al.}(2009){Mei}, {Holden}, {Blakeslee}, {Ford}, {Franx},
  {Homeier}, {Illingworth}, {Jee}, {Overzier}, {Postman}, {Rosati}, {Van der
  Wel}, \& {Bartlett}}]{mei2009}
{Mei}, S., {Holden}, B.~P., {Blakeslee}, J.~P., {et~al.} 2009, \apj, 690, 42

\bibitem[{{Mei} {et~al.}(2012){Mei}, {Stanford}, {Holden}, {Raichoor},
  {Postman}, {Nakata}, {Finoguenov}, {Ford}, {Illingworth}, {Kodama}, {Rosati},
  {Tanaka}, {Huertas-Company}, {Rettura}, {Shankar}, {Carrasco}, {Demarco},
  {Eisenhardt}, {Jee}, {Koyama}, \& {White}}]{mei2012}
{Mei}, S., {Stanford}, S.~A., {Holden}, B.~P., {et~al.} 2012, \apj, 754, 141

\bibitem[{{Moran} {et~al.}(2005){Moran}, {Ellis}, {Treu}, {Smail}, {Dressler},
  {Coil}, \& {Smith}}]{moran2005}
{Moran}, S.~M., {Ellis}, R.~S., {Treu}, T., {et~al.} 2005, \apj, 634, 977

\bibitem[{{Muzzin} {et~al.}(2013){Muzzin}, {Marchesini}, {Stefanon}, {Franx},
  {McCracken}, {Milvang-Jensen}, {Dunlop}, {Fynbo}, {Le Fevre}, {Brammer}, \&
  {Labbe}}]{muzzin2013}
{Muzzin}, A., {Marchesini}, D., {Stefanon}, M., {et~al.} 2013, ArXiv e-prints

\bibitem[{{Muzzin} {et~al.}(2012){Muzzin}, {Wilson}, {Yee}, {Gilbank},
  {Hoekstra}, {Demarco}, {Balogh}, {van Dokkum}, {Franx}, {Ellingson}, {Hicks},
  {Nantais}, {Noble}, {Lacy}, {Lidman}, {Rettura}, {Surace}, \&
  {Webb}}]{muzzin2012}
{Muzzin}, A., {Wilson}, G., {Yee}, H.~K.~C., {et~al.} 2012, \apj, 746, 188

\bibitem[{{Onodera} {et~al.}(2010){Onodera}, {Daddi}, {Gobat}, {Cappellari},
  {Arimoto}, {Renzini}, {Yamada}, {McCracken}, {Mancini}, {Capak}, {Carollo},
  {Cimatti}, {Giavalisco}, {Ilbert}, {Kong}, {Lilly}, {Motohara}, {Ohta},
  {Sanders}, {Scoville}, {Tamura}, \& {Taniguchi}}]{onodera2010}
{Onodera}, M., {Daddi}, E., {Gobat}, R., {et~al.} 2010, \apjl, 715, L6

\bibitem[{{Onodera} {et~al.}(2012){Onodera}, {Renzini}, {Carollo},
  {Cappellari}, {Mancini}, {Strazzullo}, {Daddi}, {Arimoto}, {Gobat}, {Yamada},
  {McCracken}, {Ilbert}, {Capak}, {Cimatti}, {Giavalisco}, {Koekemoer}, {Kong},
  {Lilly}, {Motohara}, {Ohta}, {Sanders}, {Scoville}, {Tamura}, \&
  {Taniguchi}}]{onodera2012}
{Onodera}, M., {Renzini}, A., {Carollo}, M., {et~al.} 2012, \apj, 755, 26

\bibitem[{{Pannella} {et~al.}(2009){Pannella}, {Gabasch}, {Goranova}, {Drory},
  {Hopp}, {Noll}, {Saglia}, {Strazzullo}, \& {Bender}}]{pannella2009b}
{Pannella}, M., {Gabasch}, A., {Goranova}, Y., {et~al.} 2009, \apj, 701, 787

\bibitem[{{Papovich}(2008)}]{papovich2008}
{Papovich}, C. 2008, \apj, 676, 206

\bibitem[{{Papovich} {et~al.}(2012){Papovich}, {Bassett}, {Lotz}, {van der
  Wel}, {Tran}, {Finkelstein}, {Bell}, {Conselice}, {Dekel}, {Dunlop}, {Guo},
  {Faber}, {Farrah}, {Ferguson}, {Finkelstein}, {H{\"a}ussler}, {Kocevski},
  {Koekemoer}, {Koo}, {McGrath}, {McLure}, {McIntosh}, {Momcheva}, {Newman},
  {Rudnick}, {Weiner}, {Willmer}, \& {Wuyts}}]{papovich2012}
{Papovich}, C., {Bassett}, R., {Lotz}, J.~M., {et~al.} 2012, \apj, 750, 93

\bibitem[{{Papovich} {et~al.}(2011){Papovich}, {Finkelstein}, {Ferguson},
  {Lotz}, \& {Giavalisco}}]{papovich2011}
{Papovich}, C., {Finkelstein}, S.~L., {Ferguson}, H.~C., {Lotz}, J.~M., \&
  {Giavalisco}, M. 2011, \mnras, 412, 1123

\bibitem[{{Papovich} {et~al.}(2010){Papovich}, {Momcheva}, {Willmer},
  {Finkelstein}, {Finkelstein}, {Tran}, {Brodwin}, {Dunlop}, {Farrah}, {Khan},
  {Lotz}, {McCarthy}, {McLure}, {Rieke}, {Rudnick}, {Sivanandam}, {Pacaud}, \&
  {Pierre}}]{papovich2010}
{Papovich}, C., {Momcheva}, I., {Willmer}, C.~N.~A., {et~al.} 2010, \apj, 716,
  1503

\bibitem[{{Patel} {et~al.}(2012{\natexlab{a}}){Patel}, {Holden}, {Kelson},
  {Franx}, {van der Wel}, \& {Illingworth}}]{patel2012}
{Patel}, S.~G., {Holden}, B.~P., {Kelson}, D.~D., {et~al.} 2012{\natexlab{a}},
  \apjl, 748, L27

\bibitem[{{Patel} {et~al.}(2009){Patel}, {Kelson}, {Holden}, {Illingworth},
  {Franx}, {van der Wel}, \& {Ford}}]{patel2009}
{Patel}, S.~G., {Kelson}, D.~D., {Holden}, B.~P., {et~al.} 2009, \apj, 694,
  1349

\bibitem[{{Patel} {et~al.}(2012{\natexlab{b}}){Patel}, {van Dokkum}, {Franx},
  {Quadri}, {Muzzin}, {Marchesini}, {Williams}, {Holden}, \&
  {Stefanon}}]{patel2012b}
{Patel}, S.~G., {van Dokkum}, P.~G., {Franx}, M., {et~al.} 2012{\natexlab{b}},
  ArXiv e-prints

\bibitem[{{Peng} {et~al.}(2002){Peng}, {Ho}, {Impey}, \& {Rix}}]{galfit1}
{Peng}, C.~Y., {Ho}, L.~C., {Impey}, C.~D., \& {Rix}, H.-W. 2002, \aj, 124, 266

\bibitem[{{Peng} {et~al.}(2010{\natexlab{a}}){Peng}, {Ho}, {Impey}, \&
  {Rix}}]{galfit2}
{Peng}, C.~Y., {Ho}, L.~C., {Impey}, C.~D., \& {Rix}, H.-W. 2010{\natexlab{a}},
  \aj, 139, 2097

\bibitem[{{Peng} {et~al.}(2010{\natexlab{b}}){Peng}, {Lilly}, {Kova{\v c}},
  {Bolzonella}, {Pozzetti}, {Renzini}, {Zamorani}, {Ilbert}, {Knobel},
  {Iovino}, {Maier}, {Cucciati}, {Tasca}, {Carollo}, {Silverman}, {Kampczyk},
  {de Ravel}, {Sanders}, {Scoville}, {Contini}, {Mainieri}, {Scodeggio},
  {Kneib}, {Le F{\`e}vre}, {Bardelli}, {Bongiorno}, {Caputi}, {Coppa}, {de la
  Torre}, {Franzetti}, {Garilli}, {Lamareille}, {Le Borgne}, {Le Brun},
  {Mignoli}, {Perez Montero}, {Pello}, {Ricciardelli}, {Tanaka}, {Tresse},
  {Vergani}, {Welikala}, {Zucca}, {Oesch}, {Abbas}, {Barnes}, {Bordoloi},
  {Bottini}, {Cappi}, {Cassata}, {Cimatti}, {Fumana}, {Hasinger}, {Koekemoer},
  {Leauthaud}, {Maccagni}, {Marinoni}, {McCracken}, {Memeo}, {Meneux}, {Nair},
  {Porciani}, {Presotto}, \& {Scaramella}}]{peng2010}
{Peng}, Y.-j., {Lilly}, S.~J., {Kova{\v c}}, K., {et~al.} 2010{\natexlab{b}},
  \apj, 721, 193

\bibitem[{{Poggianti} {et~al.}(2012){Poggianti}, {Calvi}, {Bindoni},
  {D'Onofrio}, {Moretti}, {Valentinuzzi}, {Fasano}, {Fritz}, {De Lucia},
  {Vulcani}, {Bettoni}, {Gullieuszik}, \& {Omizzolo}}]{poggianti2012}
{Poggianti}, B., {Calvi}, R., {Bindoni}, D., {et~al.} 2012, ArXiv e-prints

\bibitem[{{Popesso} {et~al.}(2004){Popesso}, {B{\"o}hringer}, {Brinkmann},
  {Voges}, \& {York}}]{popesso2004}
{Popesso}, P., {B{\"o}hringer}, H., {Brinkmann}, J., {Voges}, W., \& {York},
  D.~G. 2004, \aap, 423, 449

\bibitem[{{Postman} {et~al.}(2005){Postman}, {Franx}, {Cross}, {Holden},
  {Ford}, {Illingworth}, {Goto}, {Demarco}, {Rosati}, {Blakeslee}, {Tran},
  {Ben{\'{\i}}tez}, {Clampin}, {Hartig}, {Homeier}, {Ardila}, {Bartko},
  {Bouwens}, {Bradley}, {Broadhurst}, {Brown}, {Burrows}, {Cheng}, {Feldman},
  {Golimowski}, {Gronwall}, {Infante}, {Kimble}, {Krist}, {Lesser}, {Martel},
  {Mei}, {Menanteau}, {Meurer}, {Miley}, {Motta}, {Sirianni}, {Sparks}, {Tran},
  {Tsvetanov}, {White}, \& {Zheng}}]{postman2005}
{Postman}, M., {Franx}, M., {Cross}, N.~J.~G., {et~al.} 2005, \apj, 623, 721

\bibitem[{{Raichoor} \& {Andreon}(2012)}]{raichoor2012}
{Raichoor}, A. \& {Andreon}, S. 2012, \aap, 537, A88

\bibitem[{{Rettura} {et~al.}(2011){Rettura}, {Mei}, {Stanford}, {Raichoor},
  {Moran}, {Holden}, {Rosati}, {Ellis}, {Nakata}, {Nonino}, {Treu},
  {Blakeslee}, {Demarco}, {Eisenhardt}, {Ford}, {Fosbury}, {Illingworth},
  {Huertas-Company}, {Jee}, {Kodama}, {Postman}, {Tanaka}, \&
  {White}}]{rettura2011}
{Rettura}, A., {Mei}, S., {Stanford}, S.~A., {et~al.} 2011, \apj, 732, 94

\bibitem[{{Rettura} {et~al.}(2010){Rettura}, {Rosati}, {Nonino}, {Fosbury},
  {Gobat}, {Menci}, {Strazzullo}, {Mei}, {Demarco}, \& {Ford}}]{rettura2010}
{Rettura}, A., {Rosati}, P., {Nonino}, M., {et~al.} 2010, \apj, 709, 512

\bibitem[{{Rogers} {et~al.}(2010){Rogers}, {Ferreras}, {Pasquali}, {Bernardi},
  {Lahav}, \& {Kaviraj}}]{rogers2010}
{Rogers}, B., {Ferreras}, I., {Pasquali}, A., {et~al.} 2010, \mnras, 405, 329

\bibitem[{{Rosati} {et~al.}(2009){Rosati}, {Tozzi}, {Gobat}, {Santos},
  {Nonino}, {Demarco}, {Lidman}, {Mullis}, {Strazzullo}, {B{\"o}hringer},
  {Fassbender}, {Dawson}, {Tanaka}, {Jee}, {Ford}, {Lamer}, \&
  {Schwope}}]{rosati2009}
{Rosati}, P., {Tozzi}, P., {Gobat}, R., {et~al.} 2009, \aap, 508, 583

\bibitem[{{Rudnick} {et~al.}(2012){Rudnick}, {Tran}, {Papovich}, {Momcheva}, \&
  {Willmer}}]{rudnick2012}
{Rudnick}, G.~H., {Tran}, K.-V., {Papovich}, C., {Momcheva}, I., \& {Willmer},
  C. 2012, \apj, 755, 14

\bibitem[{{Salpeter}(1955)}]{salpeter1955}
{Salpeter}, E.~E. 1955, \apj, 121, 161

\bibitem[{{Santos} {et~al.}(2011){Santos}, {Fassbender}, {Nastasi},
  {B{\"o}hringer}, {Rosati}, {{\v S}uhada}, {Pierini}, {Nonino},
  {M{\"u}hlegger}, {Quintana}, {Schwope}, {Lamer}, {de Hoon}, \&
  {Strazzullo}}]{santos2011}
{Santos}, J.~S., {Fassbender}, R., {Nastasi}, A., {et~al.} 2011, \aap, 531, L15

\bibitem[{{Saracco} {et~al.}(2009){Saracco}, {Longhetti}, \&
  {Andreon}}]{saracco2009}
{Saracco}, P., {Longhetti}, M., \& {Andreon}, S. 2009, \mnras, 392, 718

\bibitem[{{Saracco} {et~al.}(2010){Saracco}, {Longhetti}, \&
  {Gargiulo}}]{saracco2010}
{Saracco}, P., {Longhetti}, M., \& {Gargiulo}, A. 2010, \mnras, L115+

\bibitem[{{Saracco} {et~al.}(2011){Saracco}, {Longhetti}, \&
  {Gargiulo}}]{saracco2011}
{Saracco}, P., {Longhetti}, M., \& {Gargiulo}, A. 2011, \mnras, 412, 2707

\bibitem[{{Sargent} {et~al.}(2007){Sargent}, {Carollo}, {Lilly}, {Scarlata},
  {Feldmann}, {Kampczyk}, {Koekemoer}, {Scoville}, {Kneib}, {Leauthaud},
  {Massey}, {Rhodes}, {Tasca}, {Capak}, {McCracken}, {Porciani}, {Renzini},
  {Taniguchi}, {Thompson}, \& {Sheth}}]{sargent2007}
{Sargent}, M.~T., {Carollo}, C.~M., {Lilly}, S.~J., {et~al.} 2007, \apjs, 172,
  434

\bibitem[{{Shen} {et~al.}(2003){Shen}, {Mo}, {White}, {Blanton}, {Kauffmann},
  {Voges}, {Brinkmann}, \& {Csabai}}]{shen2003}
{Shen}, S., {Mo}, H.~J., {White}, S.~D.~M., {et~al.} 2003, \mnras, 343, 978

\bibitem[{{Spitler} {et~al.}(2012){Spitler}, {Labb{\'e}}, {Glazebrook},
  {Persson}, {Monson}, {Papovich}, {Tran}, {Poole}, {Quadri}, {van Dokkum},
  {Kelson}, {Kacprzak}, {McCarthy}, {Murphy}, {Straatman}, \&
  {Tilvi}}]{spitler2012}
{Spitler}, L.~R., {Labb{\'e}}, I., {Glazebrook}, K., {et~al.} 2012, \apjl, 748,
  L21

\bibitem[{{Steidel} {et~al.}(2005){Steidel}, {Adelberger}, {Shapley}, {Erb},
  {Reddy}, \& {Pettini}}]{steidel2005}
{Steidel}, C.~C., {Adelberger}, K.~L., {Shapley}, A.~E., {et~al.} 2005, \apj,
  626, 44

\bibitem[{{Strazzullo} {et~al.}(2010){Strazzullo}, {Rosati}, {Pannella},
  {Gobat}, {Santos}, {Nonino}, {Demarco}, {Lidman}, {Tanaka}, {Mullis},
  {Nu{\~n}ez}, {Rettura}, {Jee}, {B{\"o}hringer}, {Bender}, {Bouwens},
  {Dawson}, {Fassbender}, {Franx}, {Perlmutter}, \& {Postman}}]{strazzullo2010}
{Strazzullo}, V., {Rosati}, P., {Pannella}, M., {et~al.} 2010, \aap, 524, A17

\bibitem[{{Tanaka} {et~al.}(2010{\natexlab{a}}){Tanaka}, {De Breuck},
  {Venemans}, \& {Kurk}}]{tanaka2010b}
{Tanaka}, M., {De Breuck}, C., {Venemans}, B., \& {Kurk}, J.
  2010{\natexlab{a}}, \aap, 518, A18

\bibitem[{{Tanaka} {et~al.}(2012){Tanaka}, {Finoguenov}, {Mirkazemi}, {Wilman},
  {Mulchaey}, {Ueda}, {Xue}, {Brandt}, \& {Cappelluti}}]{tanaka2012}
{Tanaka}, M., {Finoguenov}, A., {Mirkazemi}, M., {et~al.} 2012, ArXiv e-prints

\bibitem[{{Tanaka} {et~al.}(2010{\natexlab{b}}){Tanaka}, {Finoguenov}, \&
  {Ueda}}]{tanaka2010}
{Tanaka}, M., {Finoguenov}, A., \& {Ueda}, Y. 2010{\natexlab{b}}, \apjl, 716,
  L152

\bibitem[{{Taylor} {et~al.}(2010){Taylor}, {Franx}, {Glazebrook}, {Brinchmann},
  {van der Wel}, \& {van Dokkum}}]{taylor2010}
{Taylor}, E.~N., {Franx}, M., {Glazebrook}, K., {et~al.} 2010, \apj, 720, 723

\bibitem[{{Thomas} {et~al.}(2005){Thomas}, {Maraston}, {Bender}, \& {de
  Oliveira}}]{thomas2005}
{Thomas}, D., {Maraston}, C., {Bender}, R., \& {de Oliveira}, C.~M. 2005, \apj,
  621, 673

\bibitem[{{Thomas} {et~al.}(2010){Thomas}, {Maraston}, {Schawinski}, {Sarzi},
  \& {Silk}}]{thomas2010}
{Thomas}, D., {Maraston}, C., {Schawinski}, K., {Sarzi}, M., \& {Silk}, J.
  2010, \mnras, 404, 1775

\bibitem[{{Tran} {et~al.}(2010){Tran}, {Papovich}, {Saintonge}, {Brodwin},
  {Dunlop}, {Farrah}, {Finkelstein}, {Finkelstein}, {Lotz}, {McLure},
  {Momcheva}, \& {Willmer}}]{tran2010}
{Tran}, K., {Papovich}, C., {Saintonge}, A., {et~al.} 2010, \apjl, 719, L126

\bibitem[{{Trujillo} {et~al.}(2006{\natexlab{a}}){Trujillo}, {Feulner},
  {Goranova}, {Hopp}, {Longhetti}, {Saracco}, {Bender}, {Braito}, {Della Ceca},
  {Drory}, {Mannucci}, \& {Severgnini}}]{trujillo2006b}
{Trujillo}, I., {Feulner}, G., {Goranova}, Y., {et~al.} 2006{\natexlab{a}},
  \mnras, 373, L36

\bibitem[{{Trujillo} {et~al.}(2006{\natexlab{b}}){Trujillo}, {F{\"o}rster
  Schreiber}, {Rudnick}, {Barden}, {Franx}, {Rix}, {Caldwell}, {McIntosh},
  {Toft}, {H{\"a}ussler}, {Zirm}, {van Dokkum}, {Labb{\'e}}, {Moorwood},
  {R{\"o}ttgering}, {van der Wel}, {van der Werf}, \& {van
  Starkenburg}}]{trujillo2006a}
{Trujillo}, I., {F{\"o}rster Schreiber}, N.~M., {Rudnick}, G., {et~al.}
  2006{\natexlab{b}}, \apj, 650, 18

\bibitem[{{Valentinuzzi} {et~al.}(2010){Valentinuzzi}, {Fritz}, {Poggianti},
  {Cava}, {Bettoni}, {Fasano}, {D'Onofrio}, {Couch}, {Dressler}, {Moles},
  {Moretti}, {Omizzolo}, {Kj{\ae}rgaard}, {Vanzella}, \&
  {Varela}}]{valentinuzzi2010}
{Valentinuzzi}, T., {Fritz}, J., {Poggianti}, B.~M., {et~al.} 2010, \apj, 712,
  226

\bibitem[{{van der Wel} {et~al.}(2009){van der Wel}, {Bell}, {van den Bosch},
  {Gallazzi}, \& {Rix}}]{vanderwel2009}
{van der Wel}, A., {Bell}, E.~F., {van den Bosch}, F.~C., {Gallazzi}, A., \&
  {Rix}, H. 2009, \apj, 698, 1232

\bibitem[{{van der Wel} {et~al.}(2007){van der Wel}, {Holden}, {Franx},
  {Illingworth}, {Postman}, {Kelson}, {Labb{\'e}}, {Wuyts}, {Blakeslee}, \&
  {Ford}}]{vanderwel2007}
{van der Wel}, A., {Holden}, B.~P., {Franx}, M., {et~al.} 2007, \apj, 670, 206

\bibitem[{{van der Wel} {et~al.}(2008){van der Wel}, {Holden}, {Zirm}, {Franx},
  {Rettura}, {Illingworth}, \& {Ford}}]{vanderwel2008}
{van der Wel}, A., {Holden}, B.~P., {Zirm}, A.~W., {et~al.} 2008, \apj, 688, 48

\bibitem[{{van der Wel} {et~al.}(2011){van der Wel}, {Rix}, {Wuyts}, {McGrath},
  {Koekemoer}, {Bell}, {Holden}, {Robaina}, \& {McIntosh}}]{vanderwel2011}
{van der Wel}, A., {Rix}, H.-W., {Wuyts}, S., {et~al.} 2011, \apj, 730, 38

\bibitem[{{van Dokkum} {et~al.}(2008){van Dokkum}, {Franx}, {Kriek}, {Holden},
  {Illingworth}, {Magee}, {Bouwens}, {Marchesini}, {Quadri}, {Rudnick},
  {Taylor}, \& {Toft}}]{vandokkum2008}
{van Dokkum}, P.~G., {Franx}, M., {Kriek}, M., {et~al.} 2008, \apjl, 677, L5

\bibitem[{{van Dokkum} \& {van der Marel}(2007)}]{vandokkum2007}
{van Dokkum}, P.~G. \& {van der Marel}, R.~P. 2007, \apj, 655, 30

\bibitem[{{Wetzel} {et~al.}(2012){Wetzel}, {Tinker}, \& {Conroy}}]{wetzel2012}
{Wetzel}, A.~R., {Tinker}, J.~L., \& {Conroy}, C. 2012, \mnras, 424, 232

\bibitem[{{Whitaker} {et~al.}(2012){Whitaker}, {Kriek}, {van Dokkum},
  {Bezanson}, {Brammer}, {Franx}, \& {Labb{\'e}}}]{whitaker2012}
{Whitaker}, K.~E., {Kriek}, M., {van Dokkum}, P.~G., {et~al.} 2012, \apj, 745,
  179

\bibitem[{{Whitaker} {et~al.}(2011){Whitaker}, {Labb{\'e}}, {van Dokkum},
  {Brammer}, {Kriek}, {Marchesini}, {Quadri}, {Franx}, {Muzzin}, {Williams},
  {Bezanson}, {Illingworth}, {Lee}, {Lundgren}, {Nelson}, {Rudnick}, {Tal}, \&
  {Wake}}]{whitaker2011}
{Whitaker}, K.~E., {Labb{\'e}}, I., {van Dokkum}, P.~G., {et~al.} 2011, \apj,
  735, 86

\bibitem[{{Williams} {et~al.}(2009){Williams}, {Quadri}, {Franx}, {van Dokkum},
  \& {Labb{\'e}}}]{williams2009}
{Williams}, R.~J., {Quadri}, R.~F., {Franx}, M., {van Dokkum}, P., \&
  {Labb{\'e}}, I. 2009, \apj, 691, 1879

\bibitem[{{Williams} {et~al.}(2010){Williams}, {Quadri}, {Franx}, {van Dokkum},
  {Toft}, {Kriek}, \& {Labb{\'e}}}]{williams2010}
{Williams}, R.~J., {Quadri}, R.~F., {Franx}, M., {et~al.} 2010, \apj, 713, 738

\bibitem[{{Wuyts} {et~al.}(2011){Wuyts}, {F{\"o}rster Schreiber}, {van der
  Wel}, {Magnelli}, {Guo}, {Genzel}, {Lutz}, {Aussel}, {Barro}, {Berta},
  {Cava}, {Graci{\'a}-Carpio}, {Hathi}, {Huang}, {Kocevski}, {Koekemoer},
  {Lee}, {Le Floc'h}, {McGrath}, {Nordon}, {Popesso}, {Pozzi}, {Riguccini},
  {Rodighiero}, {Saintonge}, \& {Tacconi}}]{wuyts2011}
{Wuyts}, S., {F{\"o}rster Schreiber}, N.~M., {van der Wel}, A., {et~al.} 2011,
  \apj, 742, 96

\bibitem[{{Wuyts} {et~al.}(2007){Wuyts}, {Labb{\'e}}, {Franx}, {Rudnick}, {van
  Dokkum}, {Fazio}, {F{\"o}rster Schreiber}, {Huang}, {Moorwood}, {Rix},
  {R{\"o}ttgering}, \& {van der Werf}}]{wuyts2007}
{Wuyts}, S., {Labb{\'e}}, I., {Franx}, M., {et~al.} 2007, \apj, 655, 51

\bibitem[{{Zirm} {et~al.}(2012){Zirm}, {Toft}, \& {Tanaka}}]{zirm2012}
{Zirm}, A.~W., {Toft}, S., \& {Tanaka}, M. 2012, \apj, 744, 181

\bibitem[{{Zirm} {et~al.}(2007){Zirm}, {van der Wel}, {Franx}, {Labb{\'e}},
  {Trujillo}, {van Dokkum}, {Toft}, {Daddi}, {Rudnick}, {Rix},
  {R{\"o}ttgering}, \& {van der Werf}}]{zirm2007}
{Zirm}, A.~W., {van der Wel}, A., {Franx}, M., {et~al.} 2007, \apj, 656, 66

\end{thebibliography}

\end{document}